\pdfoutput=1

\documentclass[11pt]{article}
\usepackage{jheppub}
\usepackage[cp1251]{inputenc}
\usepackage{amsmath}
\usepackage{amssymb}
\usepackage{array}
\usepackage{graphicx,xcolor,slashed}
\usepackage{enumerate}
\usepackage{ifpdf}
\usepackage{mathtools}
\usepackage[english]{babel}
\usepackage{url}
\usepackage{booktabs}
\usepackage{xspace}
\usepackage{xinttools}
\usepackage{mathrsfs}
\usepackage{tensor}
\usepackage{cancel}
\usepackage{cleveref}
\usepackage{subfig}
\notoc

\definecolor{darkgreen}{rgb}{0.0, 0.26, 0.15}
\definecolor{darkred}{rgb}{0.65,0.15,0}

\usepackage{float}

\hypersetup{bookmarksnumbered,bookmarksdepth=3}

\allowdisplaybreaks[1]

\DeclareFontFamily{U}{mathx}{\hyphenchar\font45}
\DeclareFontShape{U}{mathx}{m}{n}{
      <5> <6> <7> <8> <9> <10>
      <10.95> <12> <14.4> <17.28> <20.74> <24.88>
      mathx10
      }{}
\DeclareSymbolFont{mathx}{U}{mathx}{m}{n}
\DeclareFontSubstitution{U}{mathx}{m}{n}
\DeclareMathAccent{\widecheck}{0}{mathx}{"71}

\restylefloat{figure}
\definecolor{dgreen}{rgb}{0,0.70,0.30}
\definecolor{gold}{rgb}{0.85,.66,0}
\definecolor{purple}{rgb}{1.0,0.3,0.6}

\usepackage{tikz}
\usetikzlibrary{calc}
\usetikzlibrary{patterns}
\usetikzlibrary{decorations.pathreplacing}
\usetikzlibrary{decorations.markings}
\usetikzlibrary{decorations.pathmorphing}
\usetikzlibrary{positioning}
\usetikzlibrary{arrows}
\usetikzlibrary{arrows.meta}

\newcommand{\bea}{\begin{eqnarray}}
\newcommand{\eea}{\end{eqnarray}}
\def\beq{\begin{equation}}
\def\eeq{\end{equation}}

\newcommand{\mN}{\mathfrak{N}}

\let\Re\relax
\let\Im\relax
\DeclareMathOperator{\Re}{Re}
\DeclareMathOperator{\Im}{Im}

\newcommand{\dd}{\mathrm{d}}
\newcommand{\te}{\textrm}
\newcommand{\ap}{{\alpha'}}

\DeclareMathOperator\sign{sgn}

\newcommand{\RR}{\mathbb R}

\DeclareMathOperator{\sgn}{sgn}

\crefname{equation}{}{}

\title{Pinching rules in the chiral-splitting description of one-loop string amplitudes}

\author[a,b]{Filippo Maria Balli,}
\author[c]{Alex Edison,}
\author[d,e]{and Oliver Schlotterer}

\affiliation[a]{Dipartimento di Scienze Fisiche, Informatiche e Matematiche, Universit\`a degli Studi di Modena e Reggio Emilia, Via Campi 213/A, I-41125 Modena, Italy}
\affiliation[b]{INFN, Sezione di Bologna, Via Irnerio 46, I-40126 Bologna, Italy}
\affiliation[c]{Department of Physics and Astronomy, Northwestern University, Evanston, Illinois 60208, USA}
\affiliation[d]{Department of Physics and Astronomy, Uppsala University, Box 516, 75120 Uppsala, Sweden}
\affiliation[e]{Department of Mathematics, Centre for Geometry and Physics, Uppsala University, Box 480, 75106 Uppsala, Sweden}

\emailAdd{filippo.balli@unimore.it}
\emailAdd{alexander.edison@northwestern.edu}
\emailAdd{oliver.schlotterer@physics.uu.se}

\date{\today}

\abstract{Loop amplitudes in string theories reduce to those of gauge theories and (super)gravity
in their worldline description as the inverse string tension $\alpha'$ tends to zero. The appearance
of reducible diagrams in these $\alpha' \rightarrow 0$ limits is determined through so-called pinching rules in 
the worldline literature. In this work, we extend these pinching rules to the chiral-splitting description of
one-loop superstring amplitudes where left- and right-moving degrees of freedom decouple at fixed 
loop momentum.
Starting from six points, the Kronecker-Eisenstein integrands of chiral amplitudes introduce subtleties
into the pinching rules and integration-by-parts simplifications. Resolutions of these subtleties 
are presented and applied
to produce a new superspace representation of the six-point one-loop amplitude of 
type IIA/B supergravity. The worldline computations and their subtleties are compared with the 
ambitwistor-string approach to one-loop field-theory amplitudes where integration-by-parts manipulations
are shown to be more flexible. Throughout this work, the homology invariance of loop-momentum
dependent correlation functions on the torus is highlighted as a consistency condition
of $\alpha' \rightarrow 0$ limits and their comparison with ambitwistor methods.}

\preprint{UUITP--30/24}

\begin{document}

\maketitle{}

\newpage

\setcounter{page}{1}
\pagenumbering{roman}

\setcounter{tocdepth}{2}

\tableofcontents

\numberwithin{equation}{section}

\setcounter{page}{1}
\pagenumbering{arabic}

\newpage

%%%%%%%%%%%%%%%%%%%%%%%%%%%%%%%%%%%%%%%%%%%%%%%% 
%%%%%%%%%%%%%%%%%%%%%%%%%%%%%%%%%%%%%%%%%%%%%%%% 
%%%%%%%%%%%%%%%%%%%%%%%%%%%%%%%%%%%%%%%%%%%%%%%% 

\section{Introduction}
\label{sec:intro}
The point-particle limit of string amplitudes offers a wealth of perspectives on and tools for studying scattering amplitudes in gauge theories and (super)gravity. Besides taming the spurious
combinatorial complexity of traditional Feynman rules, string amplitudes naturally expose
the double-copy relation between gauge theory, gravity and a growing web of different field-
and string theories. Moreover, superstring loop amplitudes 
elegantly manifest the cancellations between bosonic and fermionic states in the loops of supersymmetric 
field-theory amplitudes.

At a practical level, field-theory amplitudes are obtained from the infinite-tension limit $\alpha' \rightarrow 0$
of their string-theory ancestors while degenerating their two-dimensional worldsheets to 
one-dimensional worldlines. In this way, string-theory methods fruitfully draw from and extend the techniques
of the worldline literature. 

One rather technical step in the transition from worldsheets to worldlines at loop level
is the extraction of \textit{reducible diagrams}, namely worldline configurations with external tree-level diagrams.
The systematic identification of such reducible diagrams has been studied over several decades and is known
under the name of \textit{pinching rules} \cite{Bern:1990cu, Bern:1990ux, Bern:1991aq, Strassler:1992zr, Bern:1993wt, Dunbar:1994bn}\textemdash see \cite{Schubert:2001he} for a review. In both the string-theory literature \cite{Ochirov:2013xba, towardsOne, He:2015wgf, Berg:2016fui, Casali:2020knc, DHoker:2020prr} and the worldline literature \cite{Ahmadiniaz:2021fey,Bastianelli:2021rbt,Ahmadiniaz:2021ayd}, the pinching rules lead to explicit realizations of the color-kinematics duality
at the heart of the double copy \cite{BCJ, Bern:2010yg, loopBCJ, Bern:2019prr, Bern:2022wqg, Adamo:2022dcm}. 
 
The string-inspired worldline techniques and pinching rules typically recover the Schwinger-parameter
representation of Feynman integrals where the integration variables are worldline quantities descending
from the moduli of string worldsheets instead of loop momenta. On the one hand, 
Schwinger parameters inherit valuable structures from the moduli space of worldsheets
which is particularly conveniently described by tropical geometry \cite{Tourkine:2013rda, Lam:2024vqs}.
On the other hand, the departure from loop-momentum space obscures the
Bern-Carrasco-Johansson double-copy structure of (super)gravity amplitudes: their assembly from bilinears of gauge-theory building blocks
is performed at the level of loop integrands
 \cite{BCJ, Bern:2010yg, loopBCJ}, see
\cite{Bern:2019prr, Bern:2022wqg, Adamo:2022dcm} for reviews.

The double-copy relations between loop integrands naturally originate from string theory
because gravitational and gauge multiplets are vibration modes of closed and open strings, 
respectively. At loop level, the worldsheet description of string amplitudes requires
\textit{chiral splitting} \cite{Verlinde:1987sd, DHoker:1988pdl, DHoker:1989cxq} to manifest 
the double-copy structure of closed strings: the moduli-space integrand of closed-string
amplitude factorizes into two chiral amplitudes involving open-string degrees of freedom upon introducing
string-theory ancestors of loop momenta (certain zero modes of worldsheet fields).
Accordingly, it is desirable to perform the field-theory limit of loop-level string amplitudes
within the chiral-splitting formulation such as to preserve manifest double-copy structures
of gravitational loop integrands.

However, the vast body of literature on the pinching rules for the field-theory limits
usually starts by integrating out the string loop momentum. The goal of this work is
to extend the pinching rules to the chiral-splitting formulation, with an account and 
resolution of subtleties specific to multiparticle amplitudes starting from six points.
These extra subtleties within chiral splitting can be understood from the
fact that reducible diagrams are inferred from poles in the worldsheet moduli
in the integrands of string amplitudes. In case of chiral splitting, the central building 
blocks of worldsheet integrands are meromorphic and thereby have a different pole
structure than the non-meromorphic outcome of the loop integration. 

We describe the appearance of additional reducible diagrams in the pinching rules of chiral splitting
which for maximally supersymmetric open- and closed-string amplitudes kick in at six points.\footnote{By the parallels between 
one-loop $n$-point superstring amplitudes with maximal supersymmetry and $(n{+}2)$-point
amplitudes with reduced supersymmetry \cite{Berg:2016wux, Berg:2016fui}, the subtleties discussed and
addressed in this work already concern four-point one-loop amplitudes in K3
or Calabi-Yau compactifications.} The new pinching rules 
specific to chiral splitting can be traced back to the meromorphic Kronecker-Eisenstein
coefficients $g^{(k)}$ with $k\in \mathbb N$ that furnish a natural language for multiparticle
one-loop string amplitudes \cite{Dolan:2007eh, Broedel:2014vla, Berg:2016wux, Tsuchiya:2017joo, 
Lee:2017ujn, Gerken:2018jrq, Mafra:2018pll, oneloopIII}. Starting from $k=2$, the multivalued $g^{(k)}$ have
subtle poles away from the origin that are determined by their monodromies and absent for their doubly-periodic
counterparts $f^{(k)}$ that arise from loop integration. The same monodromies pose
restrictions on the integration-by-parts relations between meromorphic (as opposed to doubly-periodic)
Kronecker-Eisenstein coefficients, and we showcase their resolution at the six-point level. Hence, a key 
result of this work is a discussion and constructive fix of two related pitfalls arising in the field-theory limit
of multiparticle string amplitudes in the chiral-splitting formulation.

As a concrete application of the extended pinching rules, we derive a new representation
of the six-point one-loop amplitude of type-IIA/B supergravity from chiral splitting. The result
is written in pure-spinor superspace \cite{Berkovits:2000fe, Berkovits:2006ik, Berkovits:2022fth, Mafra:2022wml}, 
manifesting spacetime supersymmetry and encoding all component amplitudes for arbitrary 
combinations of external bosons and fermions. For all the one-loop box-, pentagon- and hexagon diagrams,
the kinematic numerator factors are written as double copies of gauge-theory polarizations. Moreover, in 
contrast to the outcome of the one-loop KLT formula in field theory \cite{He:2016mzd, He:2017spx}, 
the propagators in our six-point amplitude representation take the conventional form quadratic in the 
loop momentum. However, the gauge-theory numerators in our supergravity loop integrand
do not obey the kinematic Jacobi relations required by the color-kinematics duality. Relatedly, 
not all of the terms in the supergravity integrand correspond to graphs with solely cubic vertices -- we
also encounter double-copy contributions associated with contact diagrams akin to the
generalized double copy \cite{Bern:2017yxu}.

Many important facets of double-copy structures in gravitational loop amplitudes
were informed by ambitwistor string theories \cite{Mason:2013sva, Berkovits:2013xba, Adamo:2013tsa, Adamo:2015hoa, Geyer:2022cey}. The ambitwistor construction of
loop integrands in double-copy form from forward limits of tree amplitudes \cite{Geyer:2015bja, Baadsgaard:2015hia, He:2015yua, Geyer:2015jch, Cachazo:2015aol, Cardona:2016bpi, 
Cardona:2016wcr, Geyer:2017ela, Edison:2020uzf} gave rise to all-multiplicity realizations 
of the color-kinematics duality in a modified
setting \cite{He:2017spx}: the bookkeeping of cubic one-loop graphs includes all single-cut diagrams,
where most of their inverse propagators are linearized in the loop momentum. In converting back
to quadratic propagators (which can be attained using a variety of approaches 
\cite{Gomez:2016cqb, Gomez:2017lhy, Gomez:2017cpe, Ahmadiniaz:2018nvr, 
Agerskov:2019ryp, Farrow:2020voh, Feng:2022wee, Dong:2023stt, Xie:2024pro}), the color-kinematics
duality is not guaranteed to be preserved. Since one of the counterexamples is the one-loop six-point
amplitude of type-IIA/B supergravity \cite{towardsOne, He:2017spx}, this work provides its first
superspace representation written in terms of quadratic propagators.

Our extended pinching rules are compared with the results of ambitwistor-string methods
converted to quadratic propagators. Apart from the expected matching of the full
one-loop amplitudes, the results of the two different formalisms are found to match
at a refined level, namely at the level of individual \textit{homology invariants} within the 
underlying correlation functions. Homology invariance generalizes the notion of double periodicity
for correlation functions on a torus where shifts of loop momenta compensate for certain types of
monodromies \cite{DHoker:1989cxq, Geyer:2017ela, Mafra:2018pll, DHoker:2020prr}. 
Hence, each of the numerous homology invariants contributing to
multiparticle one-loop amplitudes admits the combined use of the
complementary advantages of pinching rules from conventional string theories and ambitwistor-string techniques.

Integration-by-parts relations in chiral splitting are restricted by the fact that total derivatives
can only be discarded when they act on homology invariants. We show that this restriction
does not apply to the integrands of ambitwistor-string formulae for one-loop amplitudes,
at least when following the evaluation strategy of \cite{Geyer:2015bja, Geyer:2015jch} via forward limits of trees.

In summary, this work closes several gaps in the literature related to the interplay of different
formalisms for one-loop amplitudes of gauge theories and (super)gravity, with the
one-loop six-point supergravity amplitude as a key illustration and application:
\begin{itemize}
\item the pinching rules in the worldline approach to the field-theory limit of one-loop string
amplitudes are extended to account for chiral splitting and the Kronecker-Eisenstein coefficients
in the associated chiral amplitudes
\item the comparison between conventional $\alpha'$-dependent strings and ambitwistor strings
is refined to smaller subsectors of amplitude computations (individual homology {invariants}), and 
the restrictions on integration-by-parts relations in both approaches is clarified
\end{itemize}
\textbf{Note Added:} The closely related work \cite{Geyer:2024} reverses the traditional information flow 
from string theory to field theory and constructs chiral one-loop amplitudes at $n\leq 7$ points directly from
color-kinematics dual representations of super-Yang-Mills amplitudes and homology invariance.
The absence of Kroencker-Eisenstein derivatives $\partial_z g^{(k)}(z)$ in the chiral amplitudes of the
reference resonates with our resolution of the restriction on integrations by parts from monodromies in 
chiral splitting. We are grateful to the authors of \cite{Geyer:2024} for notifying us of their results, 
and for coordinating the submission.

%%%%%%%%%%%%%%%%%%%%%%%%%%%%%%%%%%%%%%%%%%%%%%%% 
%%%%%%%%%%%%%%%%%%%%%%%%%%%%%%%%%%%%%%%%%%%%%%%% 

\subsection{Outline}

This work is organized as follows: we set the stage for the main results through a
detailed review of pinching rules and chiral splitting of string amplitudes in section \ref{sec:2}
which also fixes our notation.
In section \ref{sec:3}, the pinching rules are extended to chiral splitting, describing and overcoming
subtleties in the pole structure of Kronecker-Eisenstein coefficients and integration by parts.
Our findings are then applied to present a new superspace representation of the one-loop 
six-point supergravity amplitude in section \ref{sec:5}. Finally, the comparison and synergies
with ambitwistor strings are discussed in section \ref{sec:6}. Three appendices provide further
details on superspace kinematic factors, alternative derivations of reducible diagrams
and explicit formula for field-theory limits at six points.

%%%%%%%%%%%%%%%%%%%%%%%%%%%%%%%%%%%%%%%%%%%%%%%% 
%%%%%%%%%%%%%%%%%%%%%%%%%%%%%%%%%%%%%%%%%%%%%%%% 

\section{Review}
\label{sec:2}

%%%%%%%%%%%%%%%%%%%%%%%%%%%%%%%%%%%%%%%%%%%%%%%% 
%%%%%%%%%%%%%%%%%%%%%%%%%%%%%%%%%%%%%%%%%%%%%%%% 

\subsection{Basics of tree-level pinching rules}
\label{sec:2.1}

The pole structure of string amplitudes is determined by the singularities in
the integration over punctures on the worldsheet. In particular, the massless
propagators given by inverses of Mandelstam invariants
\beq
s_{ij} = (k_i{+}k_j)^2 = 2 k_i \cdot k_j \, , \ \ \ \ \ \ 
s_{i_1 i_2 \ldots i_p} = (k_{i_1}{+}k_{i_2}{+}\ldots{+}k_{i_p})^2 
\label{baspin.01}
\eeq
arise from a region in moduli space where some of the punctures
approach each other: $z_i \rightarrow z_j$ in case of two-particle propagators
$s^{-1}_{ij}$ and a cascade of limits that bring all of $z_{i_1} ,z_{i_2},\ldots,z_{i_p}$
close to each other in case of $p$-particle propagators $s^{-1}_{i_1 i_2 \ldots i_p}$.
However,
the kinematic poles are only realized when the moduli-space integrand is ``sufficiently
singular'' in the associated limits for the marked point. The purpose of {\it pinching rules} 
is to give a precise characterization of a ``sufficiently singular'' behavior needed
to encounter a given kinematic pole in (\ref{baspin.01}).

The short-distance behavior of the moduli-space integrands for open- and closed-string amplitudes
is governed by the universal \textit{Koba-Nielsen factor} which is
\beq
{\cal J}^{\rm tree}_n(\alpha') = \prod_{1\leq i < j}^n (z_{ij})^{\alpha' s_{ij}} \, , \ \ \ \ z_{ij} = z_i {-} z_j
\label{baspin.02}
\eeq
in open-string tree-level amplitudes, $|{\cal J}^{\rm tree}_n(\alpha'/4)|^{2}$ in closed-string 
tree amplitudes and
more generally determined by correlation functions of plane waves $e^{ik\cdot X(z)}$
in the conformal-field-theory description of string amplitudes \cite{Polchinski:1998rq, Polchinski:1998rr, Blumenhagen:2013fgp}. Even though the Green functions
on worldsheets of genus $\geq 1$ take a more involved form than $\log | z_{ij}|$, the
leading $z_i \rightarrow z_j$ behavior $ | z_{ij}|^{\alpha' s_{ij}}$ and $ | z_{ij}|^{\alpha' s_{ij}/2}$
is universal to Koba-Nielsen factors at any loop order. Throughout this work, we shall rescale $\alpha' \rightarrow 4\alpha'$ in closed-string amplitudes, resulting for instance in a tree-level Koba-Nielsen factor $ | z_{ij}|^{2\alpha' s_{ij}}$ instead of $ | z_{ij}|^{\alpha' s_{ij}/2}$. This rescaling ensures that the meromorphic $z_i$-dependence in their integrand matches that of open-string amplitudes and unifies various pairs of later formulae. Our main results concern the $\alpha' \rightarrow 0$ limit of string amplitudes and are therefore unaffected by this rescaling of $\alpha' $.

The nuances of pinching rules are different for
open- and closed-string amplitudes. As will be illustrated via tree-level examples, the
organization of open-string amplitudes by cyclic orderings of punctures on worldsheet
boundaries leads to pinching rules where the neighboring legs $\ldots < z_i < z_j < z_k < \ldots$
in integration regions are compared with the integrand and its poles in $z_{ij}, z_{ik}, z_{jk}$.
For closed strings in turn, the integration region for punctures is the entire worldsheet -- without any
notion of neighbors in a cyclic ordering -- and pinching rules amount to comparing the singularity
structure in the holomorphic and antiholomorphic $z_i$-dependence of the integrand.

%%%%%%%%%%%%%%%%%%%%%%%%%%%%%%%%%%%%%%%%%%%%%%%% 
%%%%%%%%%%%%%%%%%%%%%%%%%%%%%%%%%%%%%%%%%%%%%%%% 

\subsubsection{Open-string examples at tree level}
\label{sec:2.1.1}

As our first illustrations of pinching rules, we review field-theory limits
of selected contributions to open-string tree amplitudes
derived from integrating punctures over the boundary of a disk worldsheet.
More specifically, color-ordered open-string amplitudes $A^{\rm tree}(1,2,\ldots,n;\alpha')$
descend from the cyclic ordering of punctures that corresponds to
 $z_1 < z_2 < \ldots < z_n$ when parametrizing the disk boundary through the real line.

In our first example at $n=4$ points, we pick the
${\rm SL}(2,\mathbb R)$ frame where $(z_1,z_3,z_4) \rightarrow (0,1,\infty)$
and focus on the disk integral
\begin{align}
\int^1_0 \frac{\dd z_2}{z_2} \, z_2^{\alpha' s_{12}} ( 1{-}z_2)^{\alpha' s_{23}} = \frac{1}{\alpha' s_{12}} + {\cal O(\alpha')}
\label{baspin.03}
\end{align}
while ignoring the accompanying polarization dependent factors.\footnote{The 
polarization dependence of $n$-point tree amplitudes in supersymmetric, heterotic
and bosonic string theories can for instance be found in the recent review \cite{Mafra:2022wml}.}
The pole in $s_{12}$ reflects the fact that a naive $\alpha' \rightarrow 0$ limit at the
integrand level would introduce an endpoint divergence from integrating $\frac{\dd z_2}{z_2}$
close to $z_2 \rightarrow 0$. There is no pole in $s_{23}$ since $\frac{\dd z_2}{z_2}$
in absence of additional rational factors $(1{-}z_2)^{-1}$ can be smoothly integrated up 
to $z_2 \rightarrow 1$.
The pole of (\ref{baspin.03}) is tied to the integration region $z_2 \in (0,1)$ reflecting the color-ordering
of $A^{\rm tree}(1,2,3,4;\alpha')$ with cyclic ordering $z_1 < z_2 < z_3 < z_4$ on the boundary of the disk worldsheet.

At $n\geq 5$ points, the $(n{-}3)$-dimensional moduli-space integrals for punctured disk 
worldsheets and integrands with logarithmic singularities
yield products of $n{-}3$ propagators $s_{i\ldots j}$ in the field-theory limit.
The three examples in the ${\rm SL}(2,\mathbb R)$ frame $(z_1,z_4,z_5) \rightarrow (0,1,\infty)$
\begin{align}
&\lim_{\ap \rightarrow 0 } (\alpha')^2\int^1_0  \dd z_3 \int^{z_3}_0  \dd z_2 \, 
{\cal J}^{\rm tree}_5(\alpha')
 \left[ \begin{array}{c}
(  z_{12}z_{13} )^{-1}  \\ ( z_{13}z_{23} )^{-1}  \\ ( z_{12}z_{23} )^{-1}  \end{array}
\right] =  \left[ \begin{array}{c}
(  s_{12}s_{123} )^{-1}  \\ (  s_{23}s_{123} )^{-1}  \\ (  s_{12}s_{123} )^{-1} +(  s_{23}s_{123} )^{-1}  \end{array}
\right]
\label{baspin.04}
\end{align}
illustrate how two-particle propagators $s_{ij}^{-1}$ correlate with simple poles
$z_{ij}^{-1}$ in the integrand provided that $z_i$ and $z_j$ are neighbors in the
integration region (i.e.\ there is no pole in $s_{13}$ from integrals over $z_{13}^{-1}$). Similarly,
three-particle propagators $s_{i-1,i,i+1}$ involving three neighboring legs arise from
integrals over two factors out of $\{z_{i-1,i}^{-1},z_{i-1,i+1}^{-1},z_{i,i+1}^{-1}\}$ as seen in
(\ref{baspin.04}) for $i=2$. 
After earlier studies of five- and six-point examples \cite{Barreiro:2005hv, Stieberger:2006te, 6ptOprisa}, the 
systematics of field-theory limits of $n$-point disk integrals was discussed from several perspectives \cite{nptStringII, Zfunctions, Schlotterer:2018zce, Brown:2019wna} and can be neatly encoded in doubly-partial amplitudes of bi-adjoint scalars \cite{DPellis, FTlimit}.

%%%%%%%%%%%%%%%%%%%%%%%%%%%%%%%%%%%%%%%%%%%%%%%% 
%%%%%%%%%%%%%%%%%%%%%%%%%%%%%%%%%%%%%%%%%%%%%%%% 

\subsubsection{Closed-string examples at tree level}
\label{sec:2.1.2}

Closed-string tree amplitudes arise from sphere worldsheets where the
punctures are independently integrated over $\mathbb C$ and do not 
follow any analogue of the ordering along the disk boundary of the open-string case. 
In this case, the appearance of massless propagators is controlled by the alignment
of poles $z_{ij}^{-1}$ and $\bar z_{pq}^{-1}$ in the holomorphic and antiholomorphic moduli dependence.
In the four-point example (with ${\rm SL}(2,\mathbb C)$ frame $(z_1,z_3,z_4) \rightarrow (0,1,\infty)$
and $\dd^2 z = \dd x \, \dd y$ for $z= x{+}iy$ with $x,y \in \mathbb R$)
\begin{align}
\frac{1}{\pi}\int_{\mathbb C} \frac{\dd^2 z_2 }{z_2 \bar z_2 (1{-}\bar z_2)} \, 
|z_2|^{2\alpha' s_{12}} |1{-}z_2|^{2\alpha' s_{23}}  = \frac{1}{\alpha' s_{12}} + {\cal O}(\alpha'^2)
\label{baspin.06}
\end{align}
the pole in $s_{12}$ can be understood from the fact that a naive limit $\alpha' \rightarrow 0$ in the integrand
yields a divergent integral over $\frac{\dd^2 z_2}{|z_2|^2}$ from a disk
around $z_2 = 0$. Integrating $\frac{\dd^2 z_2}{(1{-}\bar z_2)}$ over
the analogous disk centered at $z_2=1$ yields a finite result as one
can for instance check in polar coordinates $z_2{-}1= r e^{i \varphi}$ 
for $\varphi \in [0,2\pi)$ and  small $r\geq 0$, explaining the absence of propagators 
$s_{23}^{-1}$ in (\ref{baspin.06}). Hence, the pole in (\ref{baspin.06}) is tied
to the matching of the holomorphic and antiholomorphic singularity 
in $z_{ij}$ and $\bar z_{ij}$ at $(i,j)=(1,2)$.\footnote{Note that changing the rational function
$(z_2 \bar z_2 (1{-}\bar z_2))^{-1}$ in (\ref{baspin.06}) to one of $(z_2 \bar z_2)^{-1}$,
$((1{-}z_2 )(1{-}\bar z_2))^{-1}$, $(z_2(1{-}\bar z_2))^{-1}$ or $((1{-}z_2 ) \bar z_2)^{-1}$
would yield poles in $s_{13}=s_{24}$ from the region $z_2 \rightarrow \infty$
(with additional poles $s_{12}^{-1}$ and $s_{23}^{-1}$ in the case of $(z_2 \bar z_2)^{-1}$ and $((1{-}z_2 )(1{-}\bar z_2))^{-1}$, respectively). 
These poles in $s_{13}$ can be traced back to singularities in $z_{13},\bar z_{13}$ or $z_{24},\bar z_{24}$ by undoing the choice
of ${\rm SL}(2,\mathbb C)$ frame in (\ref{baspin.06}), i.e.\ recognizing
$(z_{12}z_{24} z_{43} z_{31})^{-1}$ and $(z_{14}z_{42} z_{23} z_{31})^{-1}$ as the ${\rm SL}(2,\mathbb C)$-covariant uplift of $z_2^{-1}$ and $(1{-}z_2)^{-1}$, respectively.}

Similarly, the $(n{-}3)$-dimensional moduli-space of $n$-point closed-string
amplitudes yields $(n{-}3)$ simultaneous poles in the field-theory limit of sphere integrals over
functions with logarithmic singularities. Suitable closed-string analogues of the five-point open-string examples
(\ref{baspin.04}) (in the ${\rm SL}(2,\mathbb C)$ frame $(z_1,z_4,z_5) \rightarrow (0,1,\infty)$) are
\begin{align}
&\lim_{\ap \rightarrow 0 } \bigg( \frac{ \alpha' }{\pi} \bigg)^2\int_{\mathbb C^2} \frac{  \dd^2 z_2 \, \dd^2 z_3 \, |{\cal J}^{\rm tree}_5(\alpha')|^{2}  }{ \bar z_{43} \bar z_{32}  \bar z_{21}} \, 
 \left[ \begin{array}{c}
(  z_{12}z_{13} )^{-1}  \\ ( z_{13}z_{23} )^{-1}  \\ ( z_{12}z_{23} )^{-1}  \end{array}
\right] =  \left[ \begin{array}{c}
(  s_{12}s_{123} )^{-1}  \\ (  s_{23}s_{123} )^{-1}  \\ (  s_{12}s_{123} )^{-1} +(  s_{23}s_{123} )^{-1}  \end{array}
\right]
\label{baspin.07}
\end{align}
Here and in fact at arbitrary multiplicity \cite{Stieberger:2014hba}, the two-particle propagators $s_{ij}^{-1}$ 
arise from matching singularities $z_{ij}\bar z_{ij}$ for arbitrary pairs $1\leq i < j\leq n$ (by the permutation
symmetry of the integration region in the punctures, there is no restriction to $j=i{\pm}1$ here).
The field-theory limits (\ref{baspin.07}) furthermore illustrate that three-particle propagators
$s_{ijk}$ in closed-string amplitudes arise from integrands involving two out of three factors
$\{z_{ij}^{-1},z_{ik}^{-1},z_{jk}^{-1} \}$ and at the same time two out of $\{\bar z_{ij}^{-1}, \bar z_{ik}^{-1},\bar z_{jk}^{-1}\}$, for arbitrary $1\leq i < j< k\leq n$.

Most of the literature on the sphere integrals of closed-string tree amplitudes and their field-theory limits makes use of the Kawai-Lewellen-Tye relations \cite{KLTpaper} or the single-valued map between open- and closed-string $\alpha'$-expansions\footnote{The denominator $ (\bar z_{43} \bar z_{32}  \bar z_{21})^{-1}$ is naturally introduced by taking the single-valued map of open-string integrals over the integration region for $z_2,z_3$ bounded by $z_1<z_2 <z_3<z_4$, see for instance \cite{Schlotterer:2018zce}.} \cite{Schlotterer:2012ny,Stieberger:2013wea,Stieberger:2014hba,Schlotterer:2018zce,Vanhove:2018elu,Brown:2019wna}. Direct computations of propagators $(s_{ij} s_{pq})^{-1}$ in five-point sphere integrals at $\alpha' \rightarrow 0$ can for instance be found in~\cite{Ochirov:2013xba}.

%%%%%%%%%%%%%%%%%%%%%%%%%%%%%%%%%%%%%%%%%%%%%%%% 
%%%%%%%%%%%%%%%%%%%%%%%%%%%%%%%%%%%%%%%%%%%%%%%% 

\subsubsection{Summary of tree-level pinching rules}
\label{sec:2.1.3}

We shall summarize the examples of open- and closed-string pinching rules at tree level for one and two 
simultaneous propagators at $n=4$ and $n=5$ points as follows
\begin{align}
\begin{array}{c | c | c}
&\te{open strings} &\te{closed strings}
 \\\hline
n=4 \ \ & \frac{1}{z_{21}} \rightarrow \frac{1}{s_{12}} \bigg.&
 \frac{1 }{ \bar z_{32} \bar z_{21} \times z_{21}} \rightarrow \frac{1}{s_{12}}
 \\\hline
 &\bigg. \frac{ 1 }{ z_{12}z_{13}}
\rightarrow \frac{1}{s_{12}s_{123} }  \bigg.
 &\frac{ 1 }{ \bar z_{43} \bar z_{32}  \bar z_{21} \times z_{12}z_{13}}
\rightarrow \frac{1}{s_{12}s_{123} } \\
 n=5 \ \ &\frac{ 1 }{ z_{13}z_{23}}
\rightarrow \frac{1}{ s_{23}s_{123}}  &\frac{ 1 }{ \bar z_{43} \bar z_{32}  \bar z_{21} \times z_{13}z_{23}}
\rightarrow \frac{1}{ s_{23}s_{123}} \\
&\ \bigg.\frac{ 1 }{ z_{12}z_{23}}
\rightarrow \frac{1}{s_{12}s_{123} } +  \frac{1}{ s_{23}s_{123}}  \bigg. \ \,
&\ \frac{ 1 }{ \bar z_{43} \bar z_{32}  \bar z_{21} \times  z_{12}z_{23}}
\rightarrow \frac{1}{s_{12}s_{123} } +  \frac{1}{ s_{23}s_{123}} \
\end{array}
\label{baspin.05}
\end{align}
The above cases are adapted to the frame where $z_n \rightarrow \infty$, and there is
no analogue of this choice in the one-loop pinching rules below. Throughout this work, 
open-string pinching rules at tree level and one loop are given for color-ordered
amplitudes associated with the cyclic ordering ${\rm Tr}(t^{a_1}t^{a_2}\ldots t^{a_n})$.

%%%%%%%%%%%%%%%%%%%%%%%%%%%%%%%%%%%%%%%%%%%%%%%% 
%%%%%%%%%%%%%%%%%%%%%%%%%%%%%%%%%%%%%%%%%%%%%%%% 

\subsection{Basics of one-loop pinching rules}
\label{sec:2.2}

One-loop string amplitudes are derived from moduli-space integrals over
genus-one worldsheets involving modular parameters $\tau \in \mathbb H$
in addition to the punctures $z_i$ (with $ \mathbb H$ denoting the upper half plane 
in $\mathbb C$). In the open-string case, the worldsheets of
cylinder and M\"obius-strip topologies are described by $\tau \in i \mathbb R_+$
and $\tau \in i \mathbb R_+ {+} \tfrac{1}{2}$, respectively. The modular
invariance of closed-string one-loop amplitudes restricts the modular parameter
of the torus worldsheet to the fundamental domain ${\cal F}$ of ${\rm SL}(2,\mathbb Z)$
defined by $|\Re(\tau) | < \frac{1}{2}$ and $|\tau|>1$ 
\cite{Polchinski:1998rq, Blumenhagen:2013fgp, DHoker:2022dxx}.
Our parametrizations of cylinder and torus worldsheets ${\cal C}_\tau$ and ${\cal T}_\tau$ 
are depicted in figure \ref{figWS} below, with a single periodic direction $z_i \cong z_i{+}\tau$ 
for the cylinder as well as two periodic directions $z_i \cong z_i{+}1$ and 
$z_i \cong z_i{+}\tau$ for the torus. In order to account for the cyclic orderings of
the punctures on the cylinder boundary, we shall use the following shorthand
\beq
\partial {\cal C}^{12\ldots n}_\tau = \{ z_j =  \tau u_j , \ 0=u_1 <  u_2< \ldots <  u_n < 1 \}
\label{baspin.11}
\eeq
for the integration domain of planar one-loop open-string amplitudes $A ^{\textrm{1-loop}}(1,2,\ldots,n)$
associated with the color factor ${\rm Tr}(t^{a_1}t^{a_2}\ldots t^{a_n})$. In other
words, we have introduced real moduli $u_i \in (0,1)$ to encode the insertion
points of open-string states on the cylinder boundary. We do
not consider the M\"obius-strip contribution in this work since the cylinder 
amplitudes already degenerate to a basis of color-ordered
one-loop amplitudes in gauge theories through their $\alpha' \rightarrow 0$ limits.\footnote{Non-planar one-loop
gauge-theory amplitudes associated with double-trace color factors
${\rm Tr}(t^{a_1}t^{a_2}\ldots t^{a_r}) {\rm Tr}(t^{a_{r+1}}\ldots t^{a_n})$
can be obtained from linear combinations of the planar ones accompanying
${\rm Tr}(t^{a_1}t^{a_2}\ldots t^{a_n})$ with coefficients $0,\pm1$ through the all-multiplicity
relations of \cite{Bern:1994zx} and will not be discussed in this work.} 

\begin{figure}[t]
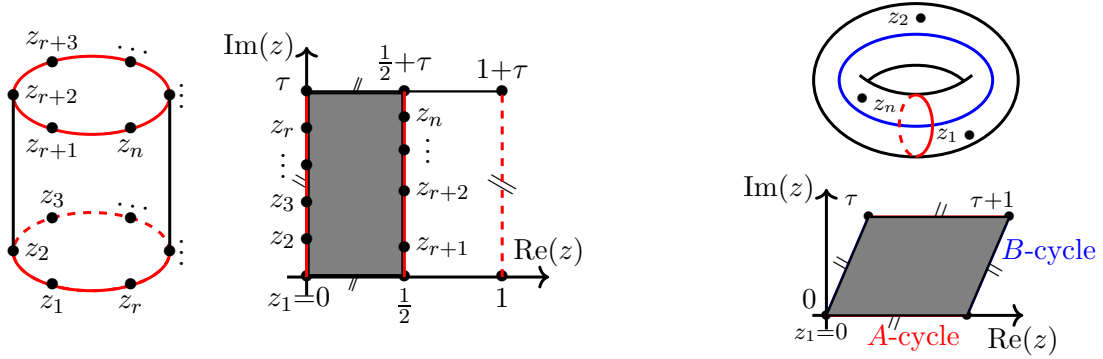

\begin{center}
\tikzpicture
\scope[scale=0.34, line width=0.40mm]
\draw(0,0) ellipse  (4cm and 3cm);
\draw(-2.2,0.2) .. controls (-1,-0.8) and (1,-0.8) .. (2.2,0.2);
\draw(-1.9,-0.05) .. controls (-1,0.8) and (1,0.8) .. (1.9,-0.05);
\draw[blue](0,0) ellipse  (3cm and 1.8cm);
\draw[red] (0,-2.975) arc (-90:90:0.65cm and 1.2cm);
\draw[red,dashed] (0,-0.575) arc (90:270:0.65cm and 1.2cm);
\draw(2.1,-2.15)node{\footnotesize $\bullet$}node[left]{\footnotesize $z_1$};
\draw(0.2,2.4)node{\footnotesize $\bullet$}node[left]{\footnotesize $z_2$};
\draw(-2.1,-0.7)node{\footnotesize $\bullet$};
\draw(-1.2,-1)node{\footnotesize $z_n$};
\scope[xshift=-3.5cm,yshift=-9.2cm,scale=1.1]
\draw[->](-0.5,0) -- (7,0) node[below]{${\rm Re}(z)$};
\draw[->](0,-0.5) -- (0,4.5) node[left]{${\rm Im}(z)$};
\draw[red](3.1,-0.8)node{$A$-cycle};
\draw[blue](7.9,2.25)node{$B$-cycle};
\draw[blue](0,0) -- (1.5,3.5);
\draw[red](0,0) -- (5,0);
\draw[red](1.5,3.5) -- (6.5,3.5);
\draw[blue](5,0) -- (6.5,3.5);
\draw(0,0)node{\footnotesize $\bullet$};
\draw(-0.6,0.6)node{$0$};
\draw(-0.2,-0.6)node{\footnotesize $z_1{=}0$};
\draw(0.75,1.75)node[rotate=60]{$| \; \! \!|$};
\draw (1.5,3.5)node{\footnotesize $\bullet$} ;
\draw(0.9,4.1)node{\footnotesize $\tau$};
\draw(2.5,0)node[rotate=-20]{$| \; \! \! |$};
\draw (5,0)node{\footnotesize $\bullet$};
\draw (4.4,0.6)node{\footnotesize $1$};
\draw(4,3.5)node[rotate=-20]{$| \; \! \! |$};
\draw(5.75,1.75)node[rotate=60]{$| \; \! \! |$};
\draw(6.5,3.5)node{\footnotesize $\bullet$};
\draw(5.9,4.1)node{\footnotesize $\tau{+}1$};
\draw(1.3,1)node{\footnotesize $\bullet$}node[right]{\footnotesize $z_n$};
\draw(5,2.25)node{\footnotesize $\bullet$}node[left]{\footnotesize $z_2$};
\draw(5.9/2,4.1/2)node{\footnotesize $\ddots$};
\draw[fill=gray, opacity=0.2, line width=0.3mm] (0, 0) -- (1.5,3.5) -- (6.5,3.5) -- (5,0) -- cycle;
\endscope
\endscope
%%%%%%%%%%%%
%%%%%%%%%%%%
\scope [xshift=-12cm, yshift=-0.2cm, scale=0.52, line width=0.40mm]
\draw[red] (2,-4) ellipse (2cm and 1cm);
\draw[white,fill=white] (0,-2.8) rectangle (4,-3.9);
\draw[red,dashed] (2,-4) ellipse (2cm and 1cm);
\draw[red] (2,0) ellipse (2cm and 1cm);
\draw (0,0) -- (0,-4);
\draw (4,0) -- (4,-4);
%%%%%%
\draw (1,0.85)node{$\bullet$}node[above]{$z_{r{+}3}$};
\draw (3,0.85)node{$\bullet$}node[above,rotate=-10]{$\ldots$};
\draw (4,0) node{$\bullet$};
\draw (4.3,0.18) node{$\vdots$};
\draw (0,0) node{$\bullet$}node[right]{$z_{r{+}2}$};
\draw (1,-0.85) node{$\bullet$}  node[below]{$z_{r{+}1}$};
\draw (3,-0.85) node{$\bullet$}  node[below]{$z_n$};
\scope[yshift=-4cm]
\draw (1,0.85)node{$\bullet$}node[above]{$z_3$};
\draw (3,0.85)node{$\bullet$}node[above,rotate=-10]{$\ldots$};
\draw (4,0) node{$\bullet$} ;
\draw (4.3,0.18) node{$\vdots$};
\draw (0,0) node{$\bullet$}node[right]{$z_2$};
\draw (1,-0.85) node{$\bullet$}  node[below]{$z_1$};
\draw (3,-0.85) node{$\bullet$}  node[below]{$z_r$};
\endscope
%%%%%%%%%%%%%%%
%%%%%%%%%%%%%%%
\scope[xshift=7.5cm,yshift=-4.65cm,yscale=0.95]
\draw[->](0,-0.5) -- (0,6.2) node[left]{${\rm Im}(z)$};
\draw[->](-0.5,0) -- (6.2,0) node[above]{${\rm Re}(z)$};
\draw[line width=0.3mm](0,5) -- (5,5) ;
\draw(0,0)node{$\bullet$};
\draw(-0.2,-0.6)node{$z_1{=}0$};
\draw(1.25,0)node[rotate=-20]{$| \! |$};
\draw(1.25,5)node[rotate=-20]{$| \! |$};
\draw(0,2.5)node[rotate=60]{$| \; \! \! |$};
\draw(5,2.5)node[rotate=60]{$| \; \! \! |$};
\draw (5,0)node{$\bullet$}node[below]{$1$};
\draw (2.5,0)node{$\bullet$}node[below]{$\tfrac{1}{2}$};
\draw (2.5,5)node{$\bullet$}node[above]{$\tfrac{1}{2}{+}\tau$};
\draw (5,5)node{$\bullet$}node[above]{$1{+}\tau$};
\draw[fill=gray, opacity=0.2,line width=0.3mm] (0.05,0.05) rectangle (2.45, 4.95);
\draw[red](0,0) -- (0,5);
\draw[red](2.5,0) -- (2.5,5);
\draw[red,dashed](5,0) -- (5,5);
\draw (0,5)node{$\bullet$};
\draw (-0.6,5.2)node{$\tau$};
\draw(0,1)node{$\bullet$}node[left]{$z_2$};
\draw(0,2)node{$\bullet$}node[left]{$z_3$};
\draw(0,3)node{$\bullet$};
\draw(-0.6,3.2)node{$\vdots$};
\draw(0,4)node{$\bullet$}node[left]{$z_r$};
\draw(2.5,0.8)node{$\bullet$}node[right]{$z_{r+1}$};
\draw(2.5,2.3)node{$\bullet$}node[right]{$z_{r+2}$};
\draw(2.5,3.4)node{$\bullet$};
\draw(3.1,3.6)node{$\vdots$};
\draw(2.5,4.3)node{$\bullet$}node[right]{$z_n$};
\endscope
\endscope
\endtikzpicture
\caption{Parametrization of the cylinder worldsheet ${\cal C}_\tau$ (left panel) and the torus worldsheet~${\cal T}_\tau$ (right panel) in one-loop $n$-point open- and closed-string amplitudes (referring to the color factor ${\rm Tr}(t^{a_1}t^{a_2}\ldots t^{a_r}) {\rm Tr}(t^{a_{r+1}}\ldots t^{a_n})$ in the open-string case) after fixing $z_1=0$ by translation invariance.}
\label{figWS}
\end{center}
\end{figure}

%%%%%%%%%%%%%%%%%%%%%%%%%%%%%%%%%%%%%%%%%%%%%%%% 
%%%%%%%%%%%%%%%%%%%%%%%%%%%%%%%%%%%%%%%%%%%%%%%% 

\subsubsection{Chiral splitting}
\label{sec:chsp}

In the chiral-splitting formulation of one-loop string amplitudes \cite{DHoker:1988pdl, DHoker:1989cxq},
the moduli-space integrands for open and closed strings are governed by the same
meromorphic functions of $z_i,\tau$ known as {\it chiral amplitudes}. Moreover,
chiral amplitudes are considered at fixed loop momentum $\ell^m$ -- the shared zero 
mode of the left-and right-moving worldsheet fields $\partial_z X^m(z)$ and $\partial_{\bar z} X^m(z)$ 
in $D$ spacetime dimensions, with Lorentz index $m=0,1,\ldots,D{-}1$. The loop integral  in
string amplitudes is Gaussian and straightforward to perform in later steps.
We organize chiral amplitudes into a theory-dependent function ${\cal K}_n$ of the polarizations
to be referred to as {\it chiral correlators} (see section \ref{sec:2.3.4} below for examples) and the universal Koba-Nielsen factor\footnote{We use the following conventions for the odd Jacobi theta function subject to $ \theta_1(z,\tau)= z \theta_1'(0,\tau)+ {\cal O}(z^3)$:
\[
\theta_1(z,\tau)= 2 q^{1/8} \sin(\pi z) \prod_{n=1}^{\infty} (1{-}e^{2\pi i z}q^n) (1{-}e^{-2\pi i z}q^n) (1{-}q^n)
\]} 
\begin{align}
{\cal J}_n(\alpha')&= \exp\bigg( \alpha' \bigg[ 2\pi i \tau \ell^2 +4\pi i \sum_{j=1}^n z_j(\ell \cdot k_j) + \sum_{1\leq i<j}^n s_{ij} \log \theta_1(z_{ij},\tau) \bigg] \bigg) 
\label{baspin.12}
\end{align}
generalizing its tree-level counterpart (\ref{baspin.02}). We will keep on rescaling
$\alpha' \rightarrow  4\alpha'$ in closed-string amplitudes (the field-theory limits under investigation arise at
the leading order in~$\alpha'$) and stop displaying the $\alpha'$-dependence in the ${\cal J}_n$ notation.
In this setting, the (planar) cylinder and torus contributions to $n$-point 
open- and closed-string 
amplitudes $A^{\textrm{1-loop}}$ and $M^{\textrm{1-loop}}$ are given by 
\begin{align}
A^{\textrm{1-loop}}(1,2,\ldots,n) &= (2\pi)^4(\alpha')^n \int_{\RR^D} \dd^{D} \ell \int_0^{i\infty} \dd \tau  \int_{\partial {\cal C}^{12\ldots n}_\tau} \dd z_2\ldots \dd z_n \, |{\cal J}_n |\, {\cal K}_n
\label{baspin.13} \\
M^{\textrm{1-loop}}_n &= (2\pi)^8\left(\frac{\alpha'}{\pi}\right)^n \int_{\RR^D} \dd^{D} \ell \int_{\cal F} \dd^2 \tau  \int_{ ({\cal T}_\tau)^{n-1}  } \dd^2 z_2\ldots \dd^2 z_n \, |{\cal J}_n  \, {\cal K}_n|^2 
\notag
\end{align}
where we fix translation in variance at genus one by setting
$z_1=0$ in both cases and in fact throughout this work. As a key virtue of chiral splitting, the closed-string integrand
$\sim |{\cal J}_n  \, {\cal K}_n|^2 $ at fixed loop momentum is an absolute-value square
and referred to as a \textit{double-copy} of the chiral amplitude. In particular, the polarization
dependence of closed-string amplitudes is a double copy of that in the chiral correlators 
${\cal K}_n$ and independent polarization degrees of freedom\footnote{The polarization
vector $e_i^m$ of the $i^{\rm th}$ gluon encountered in the meromorphic ${\cal K}_n$ of open-string amplitudes
is taken to be independent on the polarization vector $\tilde e^p_i$ in the anti-meromorphic $\overline{{\cal K}_n}$. The notation $\overline{{\cal K}_n}$ within the closed-string integrand $| {\cal K}_n |^2 = {\cal K}_n\overline{{\cal K}_n}$ is understood to take the complex conjugate of the worldsheet moduli
and to replace $e^m_i \rightarrow \tilde e^m_i$ without requiring that $\tilde e^m_i$ is the complex conjugate of $e^m_i$.}
in the complex conjugate $\overline{{\cal K}_n}$.
On these grounds, chiral splitting can be viewed as the origin
of the BCJ double-copy structure of gravity amplitudes
\cite{Bern:2019prr, Bern:2022wqg, Adamo:2022dcm} and explains
why double copy typically occurs at the level of the loop integrand
(as opposed to after loop integration).

%%%%%%%%%%%%%%%%%%%%%%%%%%%%%%%%%%%%%%%%%%%%%%%% 
%%%%%%%%%%%%%%%%%%%%%%%%%%%%%%%%%%%%%%%%%%%%%%%% 

\subsubsection{Field-theory limit at one loop}
\label{sec:2.2.1}

One-loop string amplitudes have a long history in offering streamlined 
derivations of one-loop amplitudes in supergravity and gauge theories, see
for instance \cite{Green:1982sw, Minahan:1987ha,Kaplunovsky:1987rp,Bern:1987tw}
and more recently \cite{Bjerrum-Bohr:2008vag, Bjerrum-Bohr:2008qoa, Tourkine:2012vx, Magnea:2013lna, Tourkine:2013rda, Ochirov:2013xba, towardsOne, Magnea:2015fsa, He:2015wgf, Berg:2016fui, Casali:2020knc, DHoker:2020prr}.
The key idea is that the $\alpha' \rightarrow 0$ limits are supported on the boundary
$\tau \rightarrow i\infty$ of moduli space where the cylinder and torus worldsheets
both degenerate to a worldline. These computations of field-theory limits are traditionally
performed in the worldline formalism where the loop momentum is integrated out
and renders the worldline Green function quadratic in proper times \cite{Bern:1990cu, Bern:1991aq, Strassler:1992zr, Bern:1993mq,Bern:1993wt,Dunbar:1994bn,Schmidt:1994aq,Schubert:2001he, Bastianelli:2002fv,Bastianelli:2002qw,Bastianelli:2005vk,Bastianelli:2004zp, Bjerrum-Bohr:2008qoa, Tourkine:2013rda}.

In order to investigate double-copy representations of loop integrands of gauge theory and supergravity in momentum space, one can even bypass the evaluation of the Gaussian loop integral
in performing the simultaneous limits $\alpha' \rightarrow 0$ and $\tau \rightarrow i\infty$ of the string amplitudes (\ref{baspin.13}). This is particularly relevant at $n\geq 5$ points where the chiral correlators
${\cal K}_n$ are non-trivial polynomials in loop momentum and where the Gaussian loop integral in
presence of the closed-string integrand $| {\cal K}_n |^2$ introduces a growing number of
Wick contractions. The latter contract the Lorentz-vector indices of left- and right-moving polarizations
and illustrate the well-known departures from the BCJ double-copy in passing from momentum space to
Schwinger parameters \cite{Bern:1993wt, Dunbar:1994bn, Ochirov:2013xba}.

Starting from the chiral-splitting representation (\ref{baspin.13}) of one-loop string
amplitudes, the field-theory limit of a constant chiral correlator ${\cal K}_n \rightarrow 1$ takes the
particularly simple form\footnote{\label{reffn}The $(n{-}1)!$-term permutation sum in the closed-string case of (\ref{baspin.14}) double-counts the scalar $n$-gons since reflections $(k_2,k_3,\ldots,k_n) \rightarrow (k_n,\ldots,k_3,k_2)$ leave these loop integrals invariant. Nevertheless, we keep all pairs of identical scalar $n$-gons separated to avoid the tedious bookkeeping efforts of relating the tensor integrals in later sections (with additional factors of $\ell^m$) to their reflection images.}
\begin{align}
 \int \limits_{\RR^D} \dd^{D} \ell \int \limits_0^{i\infty} \dd \tau 
 \! \! \! \! \!  \int \limits_{\partial {\cal C}^{12\ldots n}_\tau} \! \! \! \! \!
 \dd z_2\ldots \dd z_n \, |{\cal J}_n | &\rightarrow   \frac{1}{(2\pi i)^n(\alpha')^n}\int_{\RR^D}  \frac{ \dd^{D} \ell }{\ell^2 (\ell{-} k_1)^2 (\ell{-} k_{12})^2\ldots  (\ell{-} k_{12\ldots n-1})^2}
\notag \\
\int \limits_{\RR^D} \dd^{D} \ell \int \limits_{\cal F} \dd^2 \tau  
\! \! \! \! \! \int \limits_{ ({\cal T}_\tau)^{n-1} } \! \! \! \! \!
\dd^2 z_2\ldots \dd^2 z_n \, |{\cal J}_n |^2 
 &\rightarrow   \frac{1}{(2\pi)^{2n}}\left(\frac{\pi}{\alpha'}\right)^n \int_{\RR^D}  \frac{ \dd^{D} \ell }{\ell^2 (\ell{-} k_1)^2 (\ell{-} k_{12})^2\ldots  (\ell{-} k_{12\ldots n-1})^2}  \notag \\
 &\quad \quad \quad \quad \quad\quad \quad\quad   + {\rm perm}(2,3,\ldots,n)
\label{baspin.14}
\end{align}
of a single $n$-gon Feynman integral in the open-string case and a permutation sum
of $(n{-}1)!$ such $n$-gons in the closed-string case. Here and below, we are using the
notation 
\beq
k_{i_1 i_2 \ldots i_p} = k_{i_1}{+}k_{i_2}{+}\ldots{+}k_{i_p}
\label{mpmom}
\eeq
for multiparticle momenta such that $s_{i_1 i_2 \ldots i_p}=k_{i_1 i_2 \ldots i_p}^2$
by (\ref{baspin.01}). 

The field-theory limits (\ref{baspin.14}) can be understood from
the degeneration limit of the chiral Green function $\log \theta_1(u \tau {+} v,\tau)$
in the Koba-Nielsen factor (\ref{baspin.12}) at constant
\textit{comoving coordinates} $u,v \in [0,1)$,\footnote{The $z_i$ independent shift of the chiral Green function in $\log( \frac{\theta_1(u \tau {+} v,\tau)}{2 q^{1/8}} ) = \log( \theta_1(u \tau {+} v,\tau) ) - \log(2 q^{1/8})$ drops out from the Koba-Nielsen factor (\ref{baspin.12}) by momentum-conservation $\sum_{1\leq i< j}^n s_{ij}=0$.}
\beq
 \log\bigg( \frac{\theta_1(u \tau {+} v,\tau)}{2 q^{1/8}} \bigg) 
 \stackrel{ \tau \rightarrow i \infty }{\longrightarrow} - i \pi \tau |u|  
\label{baspin.15}
\eeq
In other words, the second comoving coordinate $v_i$ of $z_i= u_i \tau {+} v_i$ contributing solely
to the real part drops out at the boundary $\tau\rightarrow i\infty$ of moduli space. The independent
integrations over $0< u_i <1$ from the  integration region
$({\cal T}_\tau)^{n-1} $ of the closed-string amplitude (\ref{baspin.13}) then decomposes
into a sum over $(n{-}1)!$ permutations $\partial {\cal C}^{1\rho(2\ldots n)}_\tau $ 
with $\rho \in S_{n-1}$ covering all cyclically inequivalent 
open-string integration regions in (\ref{baspin.11}). Each integration simplex
$\partial {\cal C}^{1\rho(2\ldots n)}_\tau $ for the $u_i$ in (\ref{baspin.11})
realizes the Schwinger parametrization of an $n$-gon integral
\begin{align}
&\int \frac{\dd^D \ell}{\pi^{D/2}} \frac{1}{\ell^2 (\ell{-}K_1)^2 (\ell{-}K_{12})^2\ldots  (\ell{-}K_{12\ldots n-1})^2}
= \int^\infty_0 \frac{\dd t}{t} t^{n-D/2} 
\label{swpara.01} \\
&\quad \times \int_{0<u_2<u_3<\ldots <u_n<1} \dd u_2\, \dd u_3\,\ldots \, \dd u_n \, \exp \bigg[ {-}t \sum_{1\leq i<j}^n K_i \cdot K_j \big( u_{ij}^2 - |u_{ij}| \big)\bigg]
\notag
\end{align}
with $u_1=0$ which is also valid for massive external momenta $K_i = k_{ab\ldots}$, i.e.\ if $K_i^2 \neq~0$.
Further information
on the worldsheet origin of the Feynman integrals (\ref{swpara.01}) from the combined
limit $\alpha' \rightarrow 0$ and $\tau \rightarrow i\infty$ of open- and
closed-string integrals can be found in appendix \ref{bdyWL}.

%%%%%%%%%%%%%%%%%%%%%%%%%%%%%%%%%%%%%%%%%%%%%%%% 
%%%%%%%%%%%%%%%%%%%%%%%%%%%%%%%%%%%%%%%%%%%%%%%% 

\subsubsection{One-loop pinching rules in presence of simple poles}
\label{sec:2.2.2}

The field-theory limits (\ref{baspin.14}) only cover constant chiral correlators
and need to be extended to accommodate the non-trivial $z$-dependence of the
actual ${\cal K}_n$ of the superstring. The short-distance singularities are well-known to be expressible
through derivatives of bosonic Green functions \cite{Tsuchiya:1988va, Stieberger:2002wk, Berkovits:2004px, Bjerrum-Bohr:2008vag, Bjerrum-Bohr:2008qoa, Broedel:2014vla}  which in the chiral-splitting context reduces to
\beq
g^{(1)}_{ij} = \partial_{z_i} \log \theta(z_{ij},\tau) 
\label{baspin.16}
\eeq
without the doubly-periodic completion by $2\pi i \frac{\Im z_{ij}}{\Im \tau}$. By the simple
pole $g^{(1)}_{ij} = z_{ij}^{-1} + {\cal O}(z_{ij})$ and the universal short distance behavior of the
Koba-Nielsen factor ${\cal J}_n= z_{ij}^{\alpha' s_{ij}} (1+ {\cal O}(z_{ij}))$ at any genus,
integrating $g^{(1)}_{ij} {\cal J}_n$ over the region $z_j \rightarrow z_i$ gives rise to 
massless propagators $s_{ij}^{-1}$ as in the tree-level case. Hence, the open- and closed-string
pinching rules at genus zero summarized in section \ref{sec:2.1.3} have an immediate
genus-one uplift with almost identical combinatorial rules to deal with multiple simultaneous poles
such as $(s_{ijk} s_{ij})^{-1}$.

In presence of a single factor of $g^{(1)}_{ij} $ in ${\cal K}_n$ (resulting 
in a single pair $g^{(1)}_{ij}  \overline{g^{(1)}_{pq} }$ from the double copy $|{\cal K}_n|^2$ in the
closed-string integrand), the $n$-gon integrals in the field-theory limit (\ref{baspin.14})
may get augmented by $(n{-}1)$-gons. However, while the $\alpha' \rightarrow 0$
limits at tree level reduce to the terms with the maximum number of external propagators $s_{i\ldots j}^{-1}$, the
field-theory limits at one loop retain the irreducible $n$-gon contributions in addition to
the $(n{-}1)$-gons due to the pinch caused by $g^{(1)}_{ij} $.
More generally, terms with a total of $r$ factors $g^{(1)}_{ij} $ 
in ${\cal K}_n$ lead to $m$-gon Feynman integrals in the range of $m \in \{n{-}r, \, n{-}r{+}1,\,
\ldots,n{-}1,n\}$ corresponding to $0,1,\ldots,r$ simultaneous pinches.

From the derivative of the degeneration limit (\ref{baspin.15}) of the Green function, we obtain
\beq
\lim_{\tau \rightarrow i \infty} g^{(1)}_{ij} =- i\pi \, \sgn(u_i{-}u_j)
\label{baspin.17}
\eeq
which introduces prefactors to the higher-gons for each $g^{(1)}_{ij} $ that does not 
contribute a pinch. In particular, both individual $ g^{(1)}_{ij}$ and pairs of $g^{(1)}_{ij}  \overline{g^{(1)}_{pq} }$ introduce 
alternating signs into permutation sums of field-theory limits\footnote{Also
for non-constant chiral correlators ${\cal K}_n$, the
permutation sums over different orderings of the external $m$-gon legs arise from
the decomposition of the closed-string integration region $0<u_i<1$ for
each $i=2,3,\ldots,n$ into the simplices (\ref{baspin.11}) from open-string integration
regions. Since the signum function in the degeneration limit (\ref{baspin.17}) takes 
different constant values on these simplices which individually yield a single
Feynman integral, each factor of $ g^{(1)}_{ij} $ or $\overline{ g^{(1)}_{ij} }$
translates into $\sgn^\rho_{ij}$  in (\ref{baspin.18}) when it does not contribute a pinch.} such as (\ref{baspin.14}).

%%%%%%%%%%%%%%%%%%%%%%%%%%%%%%%%%%%%%%%%%%%%%%%% 
%%%%%%%%%%%%%%%%%%%%%%%%%%%%%%%%%%%%%%%%%%%%%%%% 

\subsubsection{Notation for one-loop pinching rules}
\label{sec:2.2.9}

\begin{figure}[t]
\begin{center}
\begin{tikzpicture} [scale=1.5, line width=0.30mm]
\draw(-0.866,0.5) -- (-0.866,-0.5);
\draw[dashed](0.866,0.5) -- (0,1);
\draw(-0.866,-0.5) -- (0,-1);
\draw[dashed](0.866,0.5) -- (0.866,-0.5);
\draw(-0.866,0.5) -- (-0.433,0.75);
\draw[dashed](-0.433,0.75) -- (0,1);
\draw[dashed](0.866,-0.5) -- (0.433,-0.75);
\draw(0.433,-0.75) -- (0,-1);
\draw[->] (0,-1) -- (-0.433,-0.75)node[above]{$\ell$};
\draw[->] (-0.866,-0.01) -- (-0.866,0)node[right]{$\ell{-}k_{1A_1}$};
\draw(0,-1) -- (0.2,-1.4);
\draw(0,-1) -- (-0.2,-1.4);
\draw[dotted](-0.2,-1.4) arc (243.435: 296.565 :0.447cm);
\draw(0,-1.65)node{$A_m$};
\scope[rotate =-60]
\draw(0,-1) -- (0.2,-1.4) node[below]{$1$};
\draw(0,-1) -- (-0.2,-1.4);
\draw(0,-1) -- (0,-1.4472);
\draw[dotted](-0.2,-1.4) arc (243.435: 270 :0.447cm);
\draw(-0.2,-1.6)node{$A_1$};
\endscope
\scope[rotate =-120]
\draw(0,-1) -- (0.2,-1.4);
\draw(0,-1) -- (-0.2,-1.4);
\draw[dotted](-0.2,-1.4) arc (243.435: 296.565 :0.447cm);
\draw(0,-1.7)node{$A_2$};
\endscope
%%%%
%%%%
%%%%
\scope[xshift=5cm]
\draw(-0.866,0.5) -- (-0.866,-0.5);
\draw[dashed](0.866,0.5) -- (0,1);
\draw(-0.866,-0.5) -- (0,-1);
\draw[dashed](0.866,0.5) -- (0.866,-0.5);
\draw(-0.866,0.5) -- (-0.433,0.75);
\draw[dashed](-0.433,0.75) -- (0,1);
\draw[dashed](0.866,-0.5) -- (0.433,-0.75);
\draw(0.433,-0.75) -- (0,-1);
\draw[->] (0,-1) -- (-0.433,-0.75)node[above]{$\ \ \ \ell{+}k_B$};
\draw[->] (-0.866,-0.01) -- (-0.866,0)node[right]{$\ell{-}k_{1C}$};
\draw(0,-1) -- (0.2,-1.4);
\draw(0,-1) -- (-0.2,-1.4);
\draw[dotted](-0.2,-1.4) arc (243.435: 296.565 :0.447cm);
\draw(0,-1.65)node{$A_m$};
\scope[rotate =-60]
\draw(0,-1) -- (0.2,-1.4) ;
\draw(0,-1) -- (-0.2,-1.4);
\draw(0,-1) -- (0,-1.4472)node[left]{$1$};
\draw(0,-1) -- (-0.357768, -1.26833);
\draw(0,-1) -- (0.357768, -1.26833);
\draw[dotted](-0.2,-1.4) arc (243.435: 216.87 :0.447cm);
\draw(-0.38,-1.5)node{$C$};
\draw[dotted](0.2,-1.4) arc (296.565: 323.13 :0.447cm);
\draw(0.38,-1.5)node{$B$};
\endscope
\scope[rotate =-120]
\draw(0,-1) -- (0.2,-1.4);
\draw(0,-1) -- (-0.2,-1.4);
\draw[dotted](-0.2,-1.4) arc (243.435: 296.565 :0.447cm);
\draw(0,-1.7)node{$A_2$};
\endscope
\endscope
\end{tikzpicture}
\caption{Routing of internal and external momenta for the propagators of (\ref{baspin.19}) (left panel, leg 1 in the first position of a massive corner) and (\ref{baspin.20}) (right panel, leg 1 in generic~positions).}
\label{figngon}
\end{center}
\end{figure}
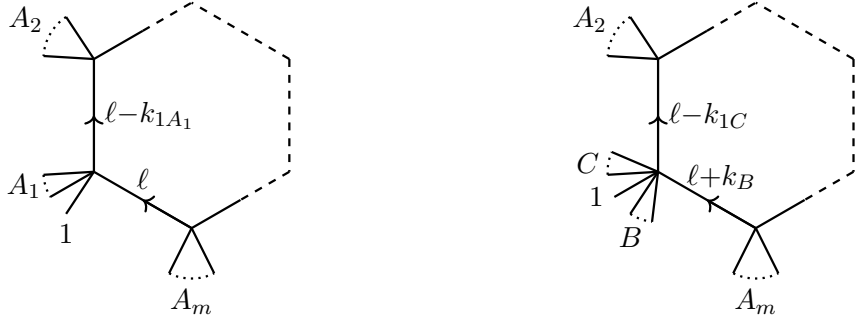

In order to compactly represent the multitude of Feynman integrals arising from some
of the subsequent field-theory limits, we shall here introduce several shorthand notations.
First, the alternating signs due to the signum function in (\ref{baspin.17}) are represented via
\beq
\sgn^\rho_{ij} = \left\{ \begin{array}{rl}
+1 &: \ i \ \te{is right of} \ j \ {\rm in} \ \rho(\ldots i \ldots j \ldots)  \\
-1 &: \ i \ \te{is left of} \ j \ {\rm in} \ \rho(\ldots i \ldots j \ldots)  
\end{array} \right. 
\label{baspin.18}
\eeq
Second, the $m$ loop-momentum dependent propagators of an $m$-gon
(i.e.\ {\em internal} propagators as opposed to {\em external} ones
$s_{i\ldots j}$) will be gathered in the notation
\begin{equation}
	I^{(m)}_{1A_1,A_2,\ldots,A_{m-1},A_m}
	= \frac{1}{\ell^2 (\ell-k_{1A_1})^2(\ell-k_{1A_1 A_2})^2 \ldots
	 (\ell-k_{1A_1 A_2 \ldots  A_{m-1}})^2}
	 \label{baspin.19}
\end{equation}
where commas separate the ordered sets $A_i= a^1_i a^2_i \ldots$ of particle 
labels $a_i^j$ referring to the sums (\ref{mpmom}) of external momenta in the $i^{\rm th}$ corner of the $m$-gon. 
We are assigning loop momentum $\ell$
to the edge of the $m$-gon adjacent to external leg 1 in counterclockwise direction as depicted in the left panel of figure \ref{figngon}. This way of singling out leg $1$ corresponds to our choice of fixing $z_1=0$ in the parental string amplitudes (\ref{baspin.13}). We will also encounter cases where leg 1 is not in the first position of a massive corner
\begin{equation}
	I^{(m)}_{B\underline{1}C,A_2,\ldots,A_{m-1},A_m}
	= \frac{1}{ (\ell-k_{1C})^2(\ell-k_{1C A_2})^2 \ldots
	 (\ell-k_{1C A_2 \ldots  A_{m}})^2}
	 \label{baspin.20}
\end{equation}
and emphasize through the underscore notation of $\underline{1}$ that leg 1 governs 
our choices of assigning $\ell$ as in the right panel of figure \ref{figngon}.

Finally, we shall use the following shorthands for the field-theory limits in the
open- and closed-string setting of (\ref{baspin.13}): 
\begin{align}
  \int_{\RR^D} \dd^{D} \ell  \ {\rm FT}^{\rm op}_{12\ldots n}\big[h(z_i,\tau,\ell) \big] &= 
 (2\pi)^4 \lim_{\alpha' \rightarrow 0} (\alpha')^n 
   \int_{\RR^D} \dd^{D} \ell \int_0^{i\infty} \dd \tau \notag \\
   &\quad  \int_{\partial {\cal C}^{12\ldots n}_\tau} \dd z_2\ldots \dd z_n \, |{\cal J}_n | \, h(z_i,\tau,\ell)
\notag \\
  \int_{\RR^D} \dd^{D} \ell \ {\rm FT}^{\rm cl}_n\big[h(z_i,\tau,\ell) \big]  &= 
  (2\pi)^8 \lim_{\alpha' \rightarrow 0} \left(\frac{\alpha'}{\pi}\right)^n  \int_{\RR^D} \dd^{D} \ell \int_{\cal F} \dd^2 \tau  \notag \\
  &\quad \int_{ ({\cal T}_\tau)^{n-1} } \dd^2 z_2\ldots \dd^2 z_n \, |{\cal J}_n |^2 \, h(z_i,\tau,\ell)
  \label{attempt1.0}
\end{align}
In other words, the ${\rm FT}^{\rm op}_{12\ldots n}[\ldots]$- and $ {\rm FT}^{\rm cl}_n[\ldots]$ notations
isolate the field-theory limit of the loop integrands of one-loop string amplitudes in the ellipsis -- up to shifts of
loop momentum that drop out under the $  \int_{\RR^D} \dd^{D} \ell $ operation on both sides, e.g.
\begin{align}
 {\rm FT}^{\rm op}_{12\ldots n}\big[h(z_i,\tau,\ell) \big] &= 
 (2\pi)^4   \lim_{\alpha' \rightarrow 0} (\alpha')^n
\int \limits_0^{i\infty} \dd \tau \! \! \!  \int \limits_{\partial {\cal C}^{12\ldots n}_\tau}  \! \! \! \dd z_2\ldots \dd z_n \, |{\cal J}_n | \, h(z_i,\tau,\ell) \ {\rm mod} \ (\ell \rightarrow \ell{\pm} k_j)
  \label{attempt1.1}
\end{align}
In the open-string case, the placeholder functions $h(z_i,\tau,\ell) $ in (\ref{attempt1.0}) of
interest are meromorphic combinations of Kronecker-Eisenstein
coefficients $g^{(m)}$ (and their $z$-derivatives) to be reviewed in section \ref{sec:2.3} below
with $k_i$- and $\ell$-dependent coefficients. 
For closed strings, the placeholders $h(z_i,\tau,\ell) $  in the second line of (\ref{attempt1.0}) 
additionally incorporate complex
conjugate Kroencker-Eisenstein coefficients.

%%%%%%%%%%%%%%%%%%%%%%%%%%%%%%%%%%%%%%%%%%%%%%%% 
%%%%%%%%%%%%%%%%%%%%%%%%%%%%%%%%%%%%%%%%%%%%%%%% 

\subsubsection{Examples of one-loop pinching rules}
\label{sec:2.2.3}
We shall now spell out examples of the one-loop pinching rules
that apply to the chiral correlators ${\cal K}_n$ of type-I and type-II
superstring amplitudes which introduce a maximum of $n{-}4$ factors of~$g^{(1)}_{ij}$ at $n$ points. 
Specializing the FT notations (\ref{attempt1.0}) to $z_i$-independent terms
$h(z_i,\tau,\ell)\rightarrow 1$ of chiral correlators casts the $n$-gon examples in (\ref{baspin.14}) into the
compact form 
\begin{align}
{\rm FT}^{\rm op}_{12\ldots n}\big[1\big] &= (2\pi i)^{4-n} I^{(n)}_{1,2,\ldots,n}
\, , \ \ \ \ \ \
{\rm FT}^{\rm cl}_n\big[1 \big]  = (2\pi)^{2(4-n)} \! \! \sum_{\rho \in S_{\{2,3,\ldots,n \}}}  \! \! I^{(n)}_{1,\rho(2,3,\ldots,n)}
 \label{baspin.22}
\end{align}
where the commas in the subscripts on the right-hand sides reduce all the ordered 
sets $A_i$ in (\ref{baspin.19}) to length one. Here and below, the 
$p!$-element group of permutations of $A_1,A_2,\ldots,A_p$~is denoted by
$S_{\{A_1,A_2,\ldots,A_p \}}$ (instead of $S_p$), and we tolerate the double-counting in the
$(n{-}1)!$ permutation sum over $\rho \in S_{\{2,3,\ldots,n \}}$ from
$ \int_{\RR^D} \dd^{D} \ell \, I^{(n)}_{1,2,3,\ldots,n}=  \int_{\RR^D} \dd^{D} \ell \, I^{(n)}_{1,n,\ldots,3,2}$
by the comments in footnote~\ref{reffn}. 

The chiral correlators of five-point superstring amplitudes involve a single factor of $g^{(1)}_{ij}$.
In case of adjacent punctures $z_i,z_j$ on the cylinder boundary, i.e.\ $j=i {\pm} 1 \ {\rm mod} \ 5$, 
the open-string field-theory limits involve a box integral $I^{(4)}$ besides the pentagon $I^{(5)}$,
\begin{align}
{\rm FT}^{\rm op}_{12345}\big[g^{(1)}_{12} \big]&= \frac{1}{2}I^{(5)}_{1,2,3,4,5}+ \frac{I^{(4)}_{12,3,4,5}}{ s_{12}} \, ,
&{\rm FT}^{\rm op}_{12345}\big[ g^{(1)}_{13}\big]&= \frac{1}{2} I^{(5)}_{1,2,3,4,5}  \label{baspin.23} \\
{\rm FT}^{\rm op}_{12345}\big[g^{(1)}_{23}\big]&=  \frac{1}{2} I^{(5)}_{1,2,3,4,5}+ \frac{I^{(4)}_{1,23,4,5}}{s_{23}} \, ,
&{\rm FT}^{\rm op}_{12345}\big[g^{(1)}_{24}\big]&= \frac{1}{2} I^{(5)}_{1,2,3,4,5} \notag
\end{align}
The field-theory limits of closed-string five-point integrals feature external propagators
$s_{ij}^{-1}$ and permutation sums of boxes $ I^{(4)}$ if the simple poles of 
$g^{(1)}_{ij}\overline{g^{(1)}_{pq}}$ match, i.e.\ if $\{ p,q\} =  \{ i,j\}$,
\begin{align}
{\rm FT}^{\rm cl}_5\big[g^{(1)}_{23}\overline{g^{(1)}_{23}} \big]&=
\frac{1}{4}\sum_{\rho \in S_{\{2,3,4,5 \}}} I^{(5)}_{1,\rho(2,3,4,5)}
 + \frac{1}{s_{23}} \sum_{\rho \in S_{\{23,4,5\}}}I_{1,\rho(23,4,5)}^{(4)}
 \notag \\
{\rm FT}^{\rm cl}_5\big[g^{(1)}_{23}\overline{g^{(1)}_{24}} \big]&=
  \frac{1}{4}\sum_{\rho \in S_{\{2,3,4,5 \}}} \sgn^\rho_{23}  \sgn^\rho_{24} I^{(5)}_{1,\rho(2,3,4,5)}
 \label{baspin.24} 
 \end{align}
 see (\ref{baspin.18}) for the sign factors $\sgn^\rho_{ij} $.
The appearance of the box integrals in (\ref{baspin.23}) and (\ref{baspin.24}) with
prefactors $s_{ij}^{-1}$ can be viewed as a one-loop
uplift of the tree-level pinching rule for open and closed strings in (\ref{baspin.03}) 
and (\ref{baspin.06}), respectively. However, the new feature at one loop is that also the pentagons 
$ I^{(5)}$, i.e.\ terms without external propagators $s_{ij}^{-1}$, contribute to the $\alpha' \rightarrow 0$ limit.

At six points, generic field-theory limits mix hexagons $I^{(6)}$ with pentagons $I^{(5)}$ and boxes~$I^{(4)}$. 
For open strings, three cyclically inequivalent examples are
\begin{align}
{\rm FT}^{\rm op}_{12\ldots 6}\big[ g^{(1)}_{12} g^{(1)}_{34} \big]&= \frac{1}{4}
I^{(6)}_{1,2,3,4,5,6}
+ \frac{I^{(5)}_{12,3,4,5,6}}{ 2s_{12} } 
+ \frac{I^{(5)}_{1,2,34,5,6}}{ 2 s_{34}} 
+ \frac{I^{(4)}_{12,34,5,6}}{ s_{12} s_{34}} 
\notag \\
{\rm FT}^{\rm op}_{12\ldots 6}\big[ g^{(1)}_{12} g^{(1)}_{23} \big]&= \frac{1}{4} 
I^{(6)}_{1,2,3,4,5,6}
+ \frac{I^{(5)}_{12,3,4,5,6}}{ 2s_{12} } 
+ \frac{I^{(5)}_{1,23,4,5,6}}{ 2s_{23} } 
+\bigg( \frac{1}{s_{12}} + \frac{1}{s_{23}} \bigg) \frac{I^{(4)}_{123,4,5,6}}{ s_{123}} 
\notag \\
{\rm FT}^{\rm op}_{12\ldots 6}\big[  g^{(1)}_{13} g^{(1)}_{23} \big]&= \frac{1}{4} 
I^{(6)}_{1,2,3,4,5,6}
+ \frac{I^{(5)}_{1,23,4,5,6}}{2 s_{23} } 
+ \frac{I^{(4)}_{123,4,5,6}}{ s_{23} s_{123}} 
 \label{baspin.25}
 \end{align} 
 whereas the non-adjacent arguments of $g^{(1)}_{12} g^{(1)}_{35}$ and $g^{(1)}_{13} g^{(1)}_{46}$ 
would lead to field-theory limits without boxes and without any boxes or pentagons, respectively. 
The propagators $(s_{23}s_{123})^{-1}$ and $(s_{12}s_{123})^{-1}$ along
with the box contributions in the last two lines of (\ref{baspin.25}) can be understood by analogy
with the tree-level pinching rules (\ref{baspin.04}) for integrals over $(z_{12} z_{23})^{-1}$ and
$(z_{13} z_{23})^{-1}$: Three-particle propagators $s_{i-1,i,i+1}^{-1}$ and the
associated boxes $I^{(4)}_{\ldots,(i-1)i(i+1),\ldots}$ are again tied to
having two out of the three simple poles of $g^{(1)}_{i-1,i}$, $g^{(1)}_{i-1,i+1} $ and $g^{(1)}_{i,i+1}$.
The two-mass box $I^{(4)}_{12,34,5,6}$ of the first line of (\ref{baspin.25}) in turn can be viewed as coming
from two independent applications of the four-point tree-level pinching rule (\ref{baspin.03}).\footnote{The direct
tree-level analogue in the ${\rm SL}(2,\mathbb R)$ frame $(z_1,z_4,z_5) \rightarrow (0,1,\infty)$ is given by $$\lim_{\ap \rightarrow 0 } (\alpha')^2\int^1_0  \dd z_3 \int^{z_3}_0  \dd z_2 \, 
 \frac{ {\cal J}^{\rm tree}_5(\alpha') }{z_{12} z_{34}} = \frac{1}{s_{12} s_{34}}
 $$}
The poles of $g^{(1)}_{ij}$ for each adjacent $j=i {\pm} 1 \ {\rm mod} \ 6$
furthermore lead to pentagons $s_{ij}^{-1} I^{(5)}_{\ldots,ij,\ldots}$.

Finally, closed-string six-point examples with contributions from hexagons, pentagons and boxes include
\begin{align}
{\rm FT}^{\rm cl}_6\big[g^{(1)}_{23} g^{(1)}_{45}
\overline{ g^{(1)}_{23}  } \overline{g^{(1)}_{45}} \big]&= \frac{1}{16} \sum_{\rho \in S_{\{2,3,4,5,6 \}}}
I^{(6)}_{1,\rho(2,3,4,5,6)}
+ \frac{1}{ 4s_{23} }   \sum_{\rho \in S_{\{23,4,5,6 \}}} I^{(5)}_{1,\rho(23,4,5,6)} \notag \\
&\quad
+ \frac{1}{  4 s_{45}}  \sum_{\rho \in S_{\{2,3,45,6 \}}}  I^{(5)}_{1,\rho(2,3,45,6)} 
+ \frac{1}{ s_{23} s_{45}}   \sum_{\rho \in S_{\{23,45,6 \}}}  I^{(4)}_{1,\rho(23,45,6)} 
  \notag \\
{\rm FT}^{\rm cl}_6\big[g^{(1)}_{23} g^{(1)}_{34}
\overline{g^{(1)}_{23} g^{(1)}_{34}} \big]&= 
\frac{1}{16} \sum_{\rho \in S_{\{2,3,4,5,6 \}}}  I^{(6)}_{1,\rho(2,3,4,5,6)}
+ \frac{1}{ 4 s_{23} }  \sum_{\rho \in S_{\{23,4,5,6 \}}}  I^{(5)}_{1,\rho(23,4,5,6)}  \notag \\
&\quad
+ \frac{1}{ 4 s_{34} }  \sum_{\rho \in S_{\{2,34,5,6 \}}} \! \! \! I^{(5)}_{1,\rho(2,34,5,6)}  
+ \bigg( \frac{1}{s_{23}} + \frac{1}{s_{34}} \bigg)  \frac{1}{ s_{234}}
\sum_{\rho \in S_{\{234,5,6 \}}} \! \! \! I^{(4)}_{1,\rho(234,5,6)} 
\notag \\
{\rm FT}^{\rm cl}_6\big[g^{(1)}_{23} g^{(1)}_{34} 
\overline{ g^{(1)}_{24} g^{(1)}_{34} } \big] &= \frac{1}{16}
\sum_{\rho \in S_{\{2,3,4,5,6 \}}} \sgn^\rho_{23}  \sgn^\rho_{24}  I^{(6)}_{1,\rho(2,3,4,5,6)}
+  \frac{1}{ 4 s_{34} }  \sum_{\rho \in S_{\{2,34,5,6 \}}}  I^{(5)}_{1,\rho(2,34,5,6)} 
\notag \\
&\quad
+ \frac{1}{ s_{34} s_{234}}  \sum_{\rho \in S_{\{234,5,6 \}}} I^{(4)}_{1,\rho(234,5,6)}
 \label{baspin.26}
 \end{align} 
 where the propagator structure of the box contributions follows
 that of the tree-level examples in (\ref{baspin.07}): Three-particle propagators $s_{ijk}^{-1}$ 
 together with one-mass boxes $I^{(4)}_{\ldots,ijk,\ldots}$
 arise when two out of the three simple poles of $g^{(1)}_{ij}, g^{(1)}_{ik} ,g^{(1)}_{jk}$
 and $\overline{ g^{(1)}_{ij} }, \overline{g^{(1)}_{ik} }, \overline{g^{(1)}_{jk}}$ in both the
 holomorphic and the antiholomorphic variables occur in the integrand. 
 The two-mass boxes $I^{(4)}_{1,\rho(23,45,6)}$
 of the second line can be viewed as coming from two applications of 
 tree-level pinching rules akin to (\ref{baspin.06}).
 Finally, each pair $g^{(1)}_{ij} \overline{ g^{(1)}_{ij} }$
 leads to permutation sums over 24 pentagons $s_{ij}^{-1} I^{(5)}_{\ldots,ij,\ldots}$.
 
Chiral correlators at higher multiplicity $n\geq 7$ feature products of up to $n{-}4$
factors of $g^{(1)}_{ij}$ which generically result in analogous cascades of 
$r=0,1,2,\ldots,n{-}4$ pinches, i.e.\ the full bandwidth of boxes, pentagons, $\ldots$, $(n{-}1)$-gons and $n$-gons in the field-theory limit. The box contributions are governed by the well-known 
combinatorics of $n$-point field-theory limits at tree level \cite{nptStringII, Zfunctions, DPellis, FTlimit, Schlotterer:2018zce, Brown:2019wna}. The $m\geq 5$ gons can be systematically assembled by adapting the tree-level pinching rules to the situation where subsets of the
$g^{(1)}_{ij}$ (or pairs with suitable $\overline{g^{(1)}_{pq}}$ in the closed-string case) contribute via (\ref{baspin.17}),
summing over all such subsets as in (\ref{baspin.25}) and~(\ref{baspin.26}).
  
%%%%%%%%%%%%%%%%%%%%%%%%%%%%%%%%%%%%%%%%%%%%%%%% 
%%%%%%%%%%%%%%%%%%%%%%%%%%%%%%%%%%%%%%%%%%%%%%%% 

\subsubsection{Summary of one-loop pinching rules}
\label{sec:2.2.4}

As a condensed way of summarizing the six-point examples
of pinching rules, we can translate (\ref{baspin.25}) and (\ref{baspin.26})
into substitution rules at the level of integrands w.r.t.\ the $z_i$. For the
open-string cases in (\ref{baspin.25}) with the color-ordering 
of ${\rm Tr}(t^{a_1}t^{a_2}\ldots t^{a_n})$, we can write
\begin{align}\label{swpara.00}
g^{(1)}_{12} g^{(1)}_{34} &\rightarrow
\frac{1}{4} \sgn(u_{12}) \sgn(u_{34} )
- \frac{\delta(z_{12}) \sgn(u_{34}) }{2 s_{12} } 
- \frac{\delta(z_{34}) \sgn(u_{12}) }{2 s_{34} } 
+\frac{\delta(z_{12})\delta(z_{34})}{s_{12} s_{34}} 
\notag \\
g^{(1)}_{12} g^{(1)}_{23} &\rightarrow \frac{1}{4} \sgn(u_{12}) \sgn(u_{23} )
- \frac{\delta(z_{12})  \sgn(u_{23}) }{2 s_{12} } 
- \frac{\delta(z_{23})  \sgn(u_{12}) }{2 s_{23} } 
+ \bigg( \frac{1}{s_{12}} {+} \frac{1}{s_{23}} \bigg) \frac{\delta(z_{12})\delta(z_{23})}{s_{123}}
\notag \\
g^{(1)}_{13} g^{(1)}_{23} &\rightarrow \frac{1}{4} \sgn(u_{13}) \sgn(u_{23} )
- \frac{\delta(z_{23})  \sgn(u_{13}) }{2 s_{23} } 
+\frac{\delta(z_{13})\delta(z_{23})}{s_{23} s_{123}}
\end{align}
where the powers of $(2\pi i)$ in the normalization factors of (\ref{attempt1.0}) are absorbed
into the $\rightarrow$ symbol. Even though the factors of $ \sgn(u_{ij})$ from (\ref{baspin.17}) are 
constant for the given color-ordering, we have kept them for clarity. The delta functions
$\delta(z_{ij})= \delta(u_{ij})/\Im \tau$ adapted to the color-ordering collapse the integrations 
over Schwinger parameters $u_i$ in the parametrization (\ref{swpara.01}) of scalar $n$-gon 
integrals and implement the pinches.

In the closed-string case, the six-point pinching rules of (\ref{baspin.26}) can be
condensed into substitution rules (which again absorb powers of $(2\pi i)$ from
(\ref{attempt1.0}) into the $\rightarrow$ symbol)
\begin{align}
g^{(1)}_{23} g^{(1)}_{45}
\overline{ g^{(1)}_{23}  } \overline{g^{(1)}_{45}}  &\rightarrow \frac{1}{16}
+ \frac{\delta^2(z_{23})}{ 4s_{23} }  
+ \frac{\delta^2(z_{45})}{  4 s_{45}}   
+ \frac{\delta^2(z_{23}) \delta^2(z_{45})}{ s_{23} s_{45}}  
\notag \\
g^{(1)}_{23} g^{(1)}_{34}
\overline{g^{(1)}_{23} g^{(1)}_{34}}&\rightarrow
\frac{1}{16} 
+ \frac{\delta^2(z_{23})}{ 4 s_{23} } 
+ \frac{\delta^2(z_{34})}{ 4 s_{34} }  
+ \bigg( \frac{1}{s_{23}} + \frac{1}{s_{34}} \bigg)  \frac{\delta^2(z_{23}) \delta^2(z_{34})}{ s_{234}} 
\notag \\
g^{(1)}_{23} g^{(1)}_{34} 
\overline{ g^{(1)}_{24} g^{(1)}_{34} } &\rightarrow \frac{1}{16}
 \sgn(u_{23})  \sgn(u_{24})  
 +  \frac{\delta^2(z_{34})}{ 4 s_{34} } 
+ \frac{\delta^2(z_{34}) \delta^2(z_{24})}{ s_{34} s_{234}} \label{swpara.02}
\end{align}
The $\sgn(u_{ij})$ factors are essential to track the relative signs
in the permutation sums of (\ref{baspin.26}) that arise from the decomposition
of the integration region $0 < u_i < 1$ (independently for all $i=2,3,\ldots,n$) 
into $(n{-}1)!$ simplices (\ref{baspin.11}).

Note that numerator factors $s_{ij}$ multiplying contributions $g^{(1)}_{ij}$
to the integrand may cancel the external propagator from the pinching rules. Nevertheless,
such factors of $s_{ij}$ do not alter the appearance of $(n{-}1)$-gons at $n$ points 
from the delta functions in (\ref{swpara.00}) and (\ref{swpara.02})
which will still be referred to as reducible diagrams.

%%%%%%%%%%%%%%%%%%%%%%%%%%%%%%%%%%%%%%%%%%%%%%%% 
%%%%%%%%%%%%%%%%%%%%%%%%%%%%%%%%%%%%%%%%%%%%%%%% 

\subsection{Explicit form of chiral correlators}
\label{sec:2.3}

In the previous section, we have reviewed the well-known pinching rules
due to simple poles $z_{ij}^{-1}$ in one-loop string integrands
adapted to the chiral-splitting formulation. Starting from six points,
however, the chiral correlators combine 
$g_{ij}^{(1)}= z_{ij}^{-1}+{\cal O}(z_{ij})$ in (\ref{baspin.16}) with higher-order
Kronecker-Eisenstein coefficients $g_{ij}^{(k\geq 2)}$ subject to new types of
pinching rules. Moreover, the interplay of $B$-cycle monodromies $z_i\rightarrow z_i {+}\tau$
and integration by parts w.r.t.\ $z_i$ introduces additional subtleties into the chiral-splitting
formulation of $(n\geq 6)$-point one-loop amplitudes. This section aims to review
the essential features of multiparticle correlators needed to complete
the pinching rules in section \ref{sec:3.1} below.

%%%%%%%%%%%%%%%%%%%%%%%%%%%%%%%%%%%%%%%%%%%%%%%% 
%%%%%%%%%%%%%%%%%%%%%%%%%%%%%%%%%%%%%%%%%%%%%%%% 

\subsubsection{Kronecker Eisenstein coefficients}
\label{sec:2.3.1}

A natural and universal function space for chiral correlators in one-loop 
amplitudes of various string theories is furnished by the Kronecker-Eisenstein 
coefficients $g_{ij}^{(k)} = g^{(k)}(z_{ij},\tau)$ with $k\geq 1$ \cite{Dolan:2007eh, Broedel:2014vla, Berg:2016wux, Tsuchiya:2017joo, Lee:2017ujn, Gerken:2018jrq, Mafra:2018pll, oneloopIII}.
Their explicit form is conveniently determined by the generating series
\beq
F(z,\eta,\tau) = \frac{\theta_1'(0,\tau) \theta_1(z{+}\eta,\tau) }{\theta_1(z,\tau) \theta_1(\eta,\tau)}
= \sum_{k=0}^{\infty} \eta^{k-1} g^{(k)}(z,\tau)
 \label{baspin.31}
\eeq
with $\theta_1'(0,\tau)= \partial_z \theta_1(z,\tau) |_{z=0}$
whose leading orders in the bookkeeping variable $\eta$ give rise to 
$g^{(0)}(z,\tau)=1$, the expression (\ref{baspin.16}) for $g^{(1)}(z,\tau)$ and for instance
\beq
g^{(2)}(z,\tau) = \frac{1}{2} \bigg[ \big( \partial_z \log \theta_1(z,\tau) \big)^2
+ \partial_z^2 \log \theta_1(z,\tau) - \frac{\theta_1'''(0,\tau) }{3 \theta_1'(0,\tau)}  \bigg]
 \label{baspin.32}
\eeq
At higher multiplicities, chiral correlators additionally feature holomorphic
Eisenstein series 
\beq
{\rm G}_k(\tau) = 
- g^{(k)}(0,\tau) = \sum_{m,n \in \mathbb Z \atop (m,n)\neq (0,0)} \frac{1}{(m\tau{+}n)^k} \, , \ \ \ \ k\geq 2
 \label{baspin.30}
\eeq
where the double sum is absolutely convergent for $k\geq 3$ and vanishes for odd $k$. 
At ${k=2}$, equivalence to the coincident limits $z_i \rightarrow z_j$ of $g_{ij}^{(k)} = g^{(k)}(z_{ij},\tau)$
selects the Eisenstein summation prescription \cite{KronEis} for the conditionally convergent double sum in (\ref{baspin.30}), resulting in the quasi-modular ${\rm G}_2(\tau) =  - \frac{\theta_1'''(0,\tau) }{3 \theta_1'(0,\tau)}$.

The multiplicative monodromies of the Kronecker-Eisenstein series (\ref{baspin.31})
\beq
F(z{+}1,\eta,\tau)  = F(z,\eta,\tau)  \, , \ \ \ \
F(z{+}\tau,\eta,\tau) =e^{-2\pi i \eta}F(z,\eta,\tau) 
 \label{baspin.33}
\eeq
translate into $A$-cycle periodicity $g^{(k)}(z{+}1,\tau) = g^{(k)}(z,\tau)$
and $B$-cycle monodromies ($k\geq 0$)
\beq
g^{(k)}(z{\pm}\tau,\tau) = g^{(k)}(z,\tau) + \sum_{\ell=1}^k \frac{(\mp 2\pi i)^\ell}{\ell!}  g^{(k-\ell)}(z,\tau) 
 \label{b.34gen}
\eeq
for instance 
\begin{align}
g^{(1)}(z{+}\tau,\tau) &= g^{(1)}(z,\tau) - 2\pi i
 \label{baspin.34} \\
g^{(2)}(z{+}\tau,\tau) &= g^{(2)}(z,\tau) - 2\pi i g^{(1)}(z,\tau) + \frac{1}{2}(2\pi i)^2 \notag
\end{align}
While $g^{(1)}(z,\tau)$ has simple poles with unit residue at any lattice point 
$z \in \mathbb Z{+} \tau \mathbb Z$,
the higher $g^{(k\geq 2)}(z,\tau)$ are regular at the origin and throughout the real axis including 
$z \in \mathbb Z$. However, the appearance of $g^{(1)}(z,\tau)$ in the
$B$-cycle monodromies (\ref{b.34gen}) causes all the $g^{(k\geq 2)}(z,\tau)$ to have
simple poles at $z=\pm \tau$ and in fact any 
lattice point $z \in  \mathbb Z{+}\tau \mathbb Z_{\neq 0}$ away from the real axis, e.g.\
\beq
g^{(k)}(z,\tau) = \left\{ \begin{array}{r} \frac{(-2\pi i)^{k-1}}{(k{-}1)!} \frac{1}{z{-}\tau} + {\cal O}\big( (z{-}\tau)^0\big)  \\
 \frac{(2\pi i)^{k-1}}{(k{-}1)!} \frac{1}{z{+}\tau} + {\cal O}\big( (z{+}\tau)^0\big)  \end{array} \right.  \label{baspin.41a}
\eeq
and in particular
\beq
 g^{(2)}(z,\tau) = \left\{ \begin{array}{r}   \frac{-2\pi i}{z{-}\tau} + {\cal O}\big( (z{-}\tau)^0\big)  \\
\frac{2\pi i}{z{+}\tau} + {\cal O}\big( (z{+}\tau)^0\big)  \end{array} \right. 
 \label{baspin.41b}
\eeq
Note that doubly-periodic completions of $g^{(k)}(z,\tau)$ denoted by $f^{(k)}(z,\tau)$
can be generated by adjoining a non-meromorphic exponential to the generating series
(\ref{baspin.31}) \cite{BrownLev, Broedel:2014vla},
\begin{align}
&\exp \bigg( 2\pi i \eta \frac{\Im z}{\Im \tau}\bigg)
\frac{\theta_1'(0,\tau) \theta_1(z{+}\eta,\tau) }{\theta_1(z,\tau) \theta_1(\eta,\tau)}
=  \sum_{k=0}^{\infty} \eta^{k-1} f^{(k)}(z,\tau)
 \label{baspin.42} \\
&\ \ f^{(k)}(z,\tau) =  g^{(k)}(z,\tau) + \sum_{\ell=1}^k \frac{1}{\ell!} \bigg( 2\pi i \frac{\Im z}{\Im \tau} \bigg)^\ell  g^{(k-\ell)}(z,\tau) 
 \notag
\end{align}
starting with $f^{(0)}(z,\tau)=1$ and $f^{(1)}(z,\tau)= g^{(1)}(z,\tau)+2\pi i \frac{\Im z}{\Im \tau}$.
The exponent $ 2\pi i \eta \frac{\Im z}{\Im \tau}$ in (\ref{baspin.42}) is tailored to compensate the multiplicative
$B$-cycle monodromies in (\ref{baspin.33}).
In the context of one-loop string amplitudes (\ref{baspin.13}), the doubly-periodic completion of 
$g^{(k)}_{ij}$ by powers of $2\pi i \frac{\Im z_{ij}}{\Im \tau}$ arises by performing the
Gaussian integral over the loop momentum \cite{Mafra:2018pll, oneloopIII}.

%%%%%%%%%%%%%%%%%%%%%%%%%%%%%%%%%%%%%%%%%%%%%%%% 
%%%%%%%%%%%%%%%%%%%%%%%%%%%%%%%%%%%%%%%%%%%%%%%% 

\subsubsection{Homology invariance}
\label{sec:2.3.2}

The integrands of one-loop open- and closed-string amplitudes (\ref{baspin.13}) with 
respect to $z_i$ and $\tau$ need to enjoy $B$-cycle periodicity under $z_i \rightarrow z_i {+} \tau$ for all of 
$i=1,2,\ldots,n$ to be well-defined on the cylinder and torus worldsheet. This
double-periodicity is tied to the loop integral in (\ref{baspin.13}) and does 
not hold pointwise in $\ell$: The $B$-cycle monodromies $z_i \rightarrow z_i {+} \tau$ of the Koba-Nielsen factor ${\cal J}_n$ in (\ref{baspin.12}) are compensated by a shift of loop momentum $\ell \rightarrow \ell {-} k_i$ which does not affect the $\int_{\RR^D} \dd^{D} \ell $ integrals of the amplitudes. The same combined transformation of $z_i$ and $\ell$ has to leave the chiral correlators ${\cal K}_n$ invariant \cite{DHoker:1989cxq, Mafra:2017ioj, Mafra:2018pll, DHoker:2020prr} which imposes non-trivial constraints on their $z_i$- and $\ell$-dependence by the $B$-cycle monodromies (\ref{b.34gen}) of the Kronecker-Eisenstein building blocks. We shall refer to any function of $\ell,k_i,z_i,\tau$ with this property and additional $A$-cycle periodicity $z_i \rightarrow z_i {+} 1$ as homology invariant,
\beq
\te{homology invariance:} \ \
E(\ell,k_i,z_i,\tau) = E(\ell{-}k_i,k_i,z_i{+}\tau,\tau)  = E(\ell,k_i,z_i{+}1,\tau)  
\label{defhi}
\eeq
The homology invariants relevant to maximally supersymmetric one-loop amplitudes at five and six points are given by \cite{Mafra:2017ioj, Mafra:2018pll}\footnote{The expressions for homology invariants in \cite{Mafra:2017ioj, Mafra:2018pll} do not feature any factors of $2\pi i$ by the change of integration variable $\ell \rightarrow \frac{\ell}{2\pi i}$ in the references. Moreover, the expression $ \partial_{z_1} g^{(1)}_{12} + s_{12} (g_{12}^{(1)})^2 - 2 s_{12} g^{(2)}_{12} $ for $ E_{1|2|3,4,5,6}$ in \cite{Mafra:2017ioj, Mafra:2018pll} is obtained from the last line of (\ref{Esat6pt}) by noting the conventions of the references where $2\alpha'=1$ and
$s_{ij}$ is normalized as $k_i\cdot k_j$ rather than $2k_i\cdot k_j$.}
\begin{align}
	E_{1\vert 23, 4,5}& = g^{(1)}_{1 2}+ g^{(1)}_{23} + g^{(1)}_{31} \nonumber\\
	E^m_{1\vert 2,3, 4,5}& =2\pi i\ell^m +\left[ k_2^m g^{(1)}_{12}+(2\leftrightarrow 3,4,5)\right]
	\label{Esat5pt}
\end{align}
as well as
\begin{align}
	E_{1\vert 234, 5,6} &=
	 g^{(1)}_{12} g^{(1)}_{23} +  g^{(1)}_{23} g^{(1)}_{34} 
	 +  g^{(1)}_{34} g^{(1)}_{41} +  g^{(1)}_{41} g^{(1)}_{12}
\nonumber\\
   &\quad + g^{(1)}_{12} g^{(1)}_{34} + g^{(1)}_{23} g^{(1)}_{41}
  +  g^{(2)}_{12}+  g^{(2)}_{23}+g^{(2)}_{34 }+  g^{(2)}_{41}\nonumber\\
   E_{1\vert 23, 4 5,6} &= (g^{(1)}_{12}+g^{(1)}_{23} + g^{(1)}_{31}) (g^{(1)}_{14}+g^{(1)}_{45} + g^{(1)}_{51})\nonumber\\
   E^m_{1\vert 23, 4 ,5,6} &=  (g^{(1)}_{12} +g^{(1)}_{23} + g^{(1)}_{31}) (2\pi i\ell^{m}+k^{m}_{4}g^{(1)}_{14} + k^{m}_{5}g^{(1)}_{15}+ k^{m}_{6}g^{(1)}_{16})\nonumber\\
   &\quad+[k^{m}_{2} (g^{(1)}_{13}
   g^{(1)}_{23}+g^{(2)}_{12}-g^{(2)}_{13}-g^{(2)}_{23})-(2\leftrightarrow 3)]\nonumber\\
   E^{mn}_{1\vert 2,3, 4 ,5,6}&=(2\pi i)^2\ell^{m} \ell^{n}+[k^{(m}_{2} k^{n)}_{3}g^{(1)}_{12} g^{(1)}_{13}+(2,3\vert 2,3,4,5,6)]\nonumber\\
   &\quad+[2\pi i \ell^{(m}_{\phantom{2}} k^{n)}_{2} g^{(1)}_{12}+2  k^{m}_{2} k^{n}_{2}g^{(2)}_{12}+(2\leftrightarrow 3,4,5,6)] \notag \\
  E_{1|2|3,4,5,6} &=
\tfrac{1}{2\ap}\partial_{z_1} g^{(1)}_{12} +\tfrac{1}{2} s_{12} (g_{12}^{(1)})^2 -  s_{12} g^{(2)}_{12} 
   \label{Esat6pt}
\end{align}
respectively, where our conventions $k^{(m}_{2} k^{n)}_{3}=k^{m}_{2} k^{n}_{3}+k^{n}_{2} k^{m}_{3}$ for (anti-)symmetrizing vector indices do not include a factor of $\frac{1}{2}$. The notation $+(i_1,\ldots,i_p | j_1,\ldots,j_q)$ instructs us to sum over all ordered subsets  $i_1,\ldots,i_p$ of $j_1,\ldots,j_q$, resulting in a total of $(\smallmatrix q \\ p \endsmallmatrix)$ terms. While the scalar examples $E_{1\vert 23, 4,5}, E_{1\vert 234, 5,6} ,E_{1\vert 23, 4 5,6} $ and $  E_{1|2|3,4,5,6} $ are elliptic functions in the conventional sense, homology invariance of the vector- and tensor examples $E^m_{1\vert 2,3, 4,5}, E^m_{1\vert 23, 4 ,5,6} ,   E^{mn}_{1\vert 2,3, 4 ,5,6}$ relies on the shift of the loop momentum in the transformation of (\ref{defhi}).

In view of the differential equation of the Koba-Nielsen factor
\beq
\ap^{-1}\partial_{z_j} \log {\cal J}_n = 4\pi i (\ell \cdot k_j) + \sum_{i=1 \atop{i \neq j}}^n s_{ij} g^{(1)}_{ji}
\label{KNder}
\eeq
the derivative $\partial_{z_1} g^{(1)}_{12} $ in the last example of (\ref{Esat6pt}) can be absorbed into 
\begin{align}
E_{1|2|3,4,5,6} {\cal J}_6 &= 
  g^{(1)}_{12} (2\pi i \ell\cdot k_2 + \tfrac{1}{2}s_{23}g^{(1)}_{23}
+ \tfrac{1}{2}s_{24}g^{(1)}_{24}  +  \tfrac{1}{2}s_{25} g^{(1)}_{25}  +  \tfrac{1}{2}s_{26} g^{(1)}_{26})  {\cal J}_6 \notag \\
&\quad -  s_{12} g^{(2)}_{12}  {\cal J}_6
- \frac{1}{2\ap}\partial_{z_2}  \big( g^{(1)}_{12}  {\cal J}_6 \big)
\label{naiveE}
\end{align}
However, as detailed in section \ref{sec:4.1}, the total Koba-Nielsen derivative $\partial_{z_2}   ( g^{(1)}_{12}  {\cal J}_6 )$ integrates to a non-trivial boundary term in the context of closed-string amplitudes (\ref{baspin.13}) and cannot be discarded. Hence, maximally supersymmetric chiral correlators 
at $n\geq 6$ points generically require refined integration by parts to eliminate
Kronecker-Eisenstein derivatives $\partial_{z_i} g^{(k)}_{ij}$ 
before integration over $\ell$, see for instance section \ref{sec:wayout}. 

After loop integration in turn, derivatives $\partial_{z_i} f^{(k)}_{ij}$ of the single-valued functions in (\ref{baspin.42}) can be removed via simpler integrations by parts. Nevertheless, we will see in
section \ref{sec:6.9} that, within the application of the chiral correlators below to ambitwistor-string theories, 
one can eliminate already $\partial_{z_i} g^{(k)}_{ij}$ via scattering equations analogous to (\ref{naiveE}).

%%%%%%%%%%%%%%%%%%%%%%%%%%%%%%%%%%%%%%%%%%%%%%%% 
%%%%%%%%%%%%%%%%%%%%%%%%%%%%%%%%%%%%%%%%%%%%%%%% 

\subsubsection{Chiral correlators up to six points}
\label{sec:2.3.4}

The homology invariants $E$ in (\ref{Esat5pt}) and (\ref{Esat6pt}) capture the dependence
of the $(n\leq 6)$-point chiral correlators ${\cal K}_n$ on the worldsheet moduli $z_i,\tau$.
The second key ingredients of the ${\cal K}_n$ are kinematic factors depending on gauge-theory
polarizations and external momenta which double copy to gravitational kinematic
factors in closed-string integrands $ \sim | {\cal K}_n|^2$. 

We shall employ the formulation
of one-loop kinematic factors in pure-spinor superspace \cite{EOMBBs,
Mafra:2014gsa, Mafra:2018nla, Berkovits:2022fth} denoted by $C_{1|23,4,5},
C^m_{1|2,3,4,5}$  at five points and $C_{1|234,5,6}$,
$C_{1|23,45,6},C^m_{1|23,4,5,6},$ $C^{mn}_{1|2,3,4,5,6},P_{1|2|3,4,5,6}$ at six
points. They all descend from particular polynomials in Gra\ss{}mann-odd
variables $\theta^\alpha$ -- Weyl spinors of ${\rm SO}(10)$ -- with gluon-polarization vectors, gaugino wavefunctions and momenta as coefficients.\footnote{In slight abuse of notation, we do not display the component prescription $\langle \ldots \rangle$ which extracts the $\lambda^3 \theta^5$ orders
of the enclosed superspace kinematic factors \cite{Berkovits:2000fe,
Mafra:2010pn}, where $\lambda^\alpha$ is a Gra\ss{}mann-even Weyl spinor of
${\rm SO}(10)$ subject to the pure-spinor constraint explained in the references.} 
In this way, the results of this work apply to any combination of external bosons and fermions in the 
supersymmetry multiplets of ten-dimensional super-Yang-Mills and type-IIA/B supergravity as well as their dimensional reductions.

The combinatorics of five- and six-point superspace kinematic factors
$C_{1|23,4,5}, C^m_{1|2,3,4,5}$ and $C_{1|234,5,6}$, $C_{1|23,45,6},C^m_{1|23,4,5,6},C^{mn}_{1|2,3,4,5,6},P_{1|2|3,4,5,6}$ mirror that of
the homology invariants in (\ref{Esat5pt}) and (\ref{Esat6pt}), respectively. In other words, the integration-by-parts relations among homology invariants (see for instance section \ref{sec:wayout} below) have a 
direct counterpart at the level of superspace kinematic factors \cite{Mafra:2014gsa}
and endow the chiral correlators with a double-copy structure \cite{Mafra:2017ioj, Mafra:2018pll, oneloopIII},\footnote{Note that the $s_{ij}$ in the references are defined as $k_i\cdot k_j$, i.e.\ without the
factor of two in the conventions (\ref{baspin.01}) of this work.}
\begin{align}
{\cal K}_4&= C_{1|2,3,4} E_{1|2,3,4}
\notag \\
{\cal K}_5&= C^{m}_{1|2,3,4,5} E^{m}_{1|2,3,4,5} 
+ \tfrac{1}{2} \big[ s_{23} C_{1|23,4,5} E_{1|23,4,5} + (2,3|2,3,4,5) \big] 
\notag \\
{\cal K}_6&=  \tfrac{1}{2} C^{mn}_{1|2,\ldots,6} E^{mn}_{1|2,\ldots,6}   +  \tfrac{1}{2} \big[ s_{23} C^m_{1|23,4,5,6} E^m_{1|23,4,5,6} + (2,3|2,3,4,5,6) \big] \notag \\
&\quad  + \tfrac{1}{4} \Big( \big[ s_{23} s_{45} C_{1|23,45,6} E_{1|23,45,6} + {\rm cyc}(3,4,5) \big] + (6\leftrightarrow 5,4,3,2) \Big)  \notag
 \\
&\quad  +  \tfrac{1}{4} \Big( \big[ s_{23}s_{34} C_{1|234,5,6}  E_{1|234,5,6} + {\rm cyc}(2,3,4) \big]    + (2,3,4|2,3,4,5,6) \Big) \notag \\
&\quad  -  \big[ P_{1|2|3,4,5,6} E_{1|2|3,4,5,6}  +(2\leftrightarrow 3,4,5,6) \big]  \label{KK6GEF}
\end{align}
The trivial homology invariant $E_{1|2,3,4}=1$ at four points has been introduced to bring ${\cal K}_4$
into the ``$C\cdot E$ form'' of ${\cal K}_5$. The four-point kinematic factor
\beq
C_{1|2,3,4} = s_{12} s_{23} A^{\rm tree}_{\rm SYM}(1,2,3,4)
\label{KK4}
\eeq
is a direct sum of color-ordered tree-level amplitudes of ten-dimensional super-Yang-Mills
involving any combination of external gauge bosons and gauginos. Similar representations
for some of the $(n\leq 6)$-point kinematic factors and compact expressions for all of their
bosonic components are reviewed in appendix \ref{app:kin}.
A collection of component expansions for kinematic factors in pure-spinor superspace can 
also be downloaded from \cite{PSSweb}. The BRST {(pseudo)}invariance of the individual 
superspace expressions $C$ and~$P$ in (\ref{KK6GEF}) implies that their bosonic and fermionic 
components are gauge invariant and supersymmetric \cite{Berkovits:2000fe}. However, the
manifest BRST properties of each kinematic factor come at the cost of obscuring the 
locality properties of the amplitudes computed from (\ref{KK6GEF}) since each $C$ and $P$
at six points has a pole structure of $s_{ij}^{-1}$, $s_{ijk}^{-1}$, $( s_{ij} s_{kl})^{-1}$ or $( s_{ij} s_{ijk})^{-1}$. 

Alternative and manifestly local representations of chiral correlators at $n\leq 7$ points can be found 
in \cite{oneloopIII}, though the superfields accompanying the individual 
homology invariants of (\ref{KK6GEF}) would then have gauge dependent components.
The computation of six-point supergravity amplitudes
in section \ref{sec:5.3} below is mostly driven by the field-theory limits of closed-string integrals over
bilinears in the homology invariants of (\ref{KK6GEF}). Hence, the methods of
this work can as well be used to obtain manifestly local expressions for
the six-point amplitudes of section \ref{sec:5.3} from the local correlator representations
of \cite{oneloopIII}.

%%%%%%%%%%%%%%%%%%%%%%%%%%%%%%%%%%%%%%%%%%%%%%%% 
%%%%%%%%%%%%%%%%%%%%%%%%%%%%%%%%%%%%%%%%%%%%%%%% 

\subsubsection{Field-theory limit only well-defined for homology invariants}
\label{sec:2.3.3}

In the same way as individual terms in the component expansion of $C$ and $P$ exhibit nonzero
gauge variations, the individual Kronecker-Eisenstein terms in the homology invariants
(\ref{Esat5pt}) and (\ref{Esat6pt}) do not have well-defined field-theory limits. We shall now
illustrate via simple counterexamples that homology invariance is essential for a well-defined
field-theory limit. 

First, the $\tau \rightarrow i \infty$ limit (\ref{baspin.17}) of $g^{(1)}_{ij}$ in planar cylinder amplitudes with
fixed $i,j$ depends on the parametrization of the cylinder boundary. For the color-ordering of 
${\rm Tr}(t^{a_1}t^{a_2}\ldots t^{a_5})$ the choice of coordinates in (\ref{baspin.11}) is as natural 
as any of its cyclic permutations 
$0 = u_i < u_{i+1}< u_{i+2}< u_{i+3}< u_{i+4}< 1$ with $u_i \cong u_{i\pm 5}$. Changing the parametrization by moving $u_1=0$ to the upper end of the integration region $u_5< u_1 < 1$ will flip the sign of  $\lim_{\tau \rightarrow i \infty} g^{(1)}_{12} = i\pi \, \sgn(u_2{-}u_1)$. The difference of $-2\pi i$ can be identified with the $B$-cycle monodromy in (\ref{baspin.34}) since the change of parametrization in the variables $u_i$ descends from $z_1 \rightarrow z_1{+}\tau$. Accordingly, any contribution to ${\cal K}_n$ with non-zero homology variation under
$z_i \rightarrow z_i{+}\tau$ and $\ell \rightarrow \ell {-}k_i$ will have a
parametrization-dependent field-theory limit.

The expressions (\ref{Esat5pt}) for $E_{1|23,4,5}$ and $E^m_{1|2,3,4,5}$ at five points 
illustrate how two different types of homology invariant completions of $g^{(1)}_{12}$ give rise to 
well-defined field-theory limits: The $B$-cycle monodromy 
$z_1 \rightarrow z_1{+}\tau$ vanishes for the $z_1$-dependent terms 
$g^{(1)}_{12} - g^{(1)}_{13}$ of $E_{1|23,4,5}$, consistent with the fact
that the contributions $ \sgn(u_2{-}u_1)-  \sgn(u_3{-}u_1)$ to the field-theory limit
change with opposite signs under change of parametrizations from $u_1=0$ to $u_5< u_1 < 1$.
Similarly, the contributions $\sum_{j=2}^5 k_j^m g^{(1)}_{1j}$ to $E^m_{1|2,3,4,5}$ have an overall $B$-cycle monodromy 
$-2\pi i k_{2345}^m = 2\pi i k_1^m$ under $z_1 \rightarrow z_1{+}\tau$ which compensates
the homology variation $\ell \rightarrow \ell{-}k_1$.

On these grounds, particular emphasis will be placed in section \ref{sec:3.2} to \ref{sec:FTE} below 
on the field-theory limits 
(\ref{attempt1.0}) of homology invariants $E$ for open strings and $E \bar E$ for closed strings.

%%%%%%%%%%%%%%%%%%%%%%%%%%%%%%%%%%%%%%%%%%%%%%%% 
%%%%%%%%%%%%%%%%%%%%%%%%%%%%%%%%%%%%%%%%%%%%%%%% 

\section{Pinching rules and field-theory limits in chiral splitting}
\label{sec:3}

In this section, the pinching rules for factors of $g^{(1)}_{ij}$ will be completed to those
for the Kronecker-Eisenstein coefficients $g^{(2)}_{ij}$ and applied to evaluate the field-theory
limits of holomology invariants of open- and closed-string genus-one amplitudes.

%%%%%%%%%%%%%%%%%%%%%%%%%%%%%%%%%%%%%%%%%%%%%%%% 
%%%%%%%%%%%%%%%%%%%%%%%%%%%%%%%%%%%%%%%%%%%%%%%% 

\subsection{New pinching rules from higher Kronecker-Eisenstein coefficients}
\label{sec:3.1}

Performing the Gaussian loop integral in one-loop string amplitudes completes 
the meromorphic but multi-valued Kronecker-Eisenstein coefficients $g^{(k)}$ 
to the single-valued but non-meromorphic ones $f^{(k)}$ \cite{Mafra:2018pll, oneloopIII}.
Since the kernels $f^{(k)}(z,\tau)$ of weight $k \geq 2$ defined in (\ref{baspin.42}) are non-singular for any $z\in \mathbb C$,
the pinching rules after loop integration only feature external propagators $s_{i\ldots j}^{-1}$
from products of $f^{(1)}_{ij}$ (as opposed to $f^{(k \geq 2)}_{ij}$). 

Before loop integration,
however, even the $g^{(k \geq 2)}(z,\tau)$ kernels acquire poles at $z= \pm \tau$ (and in fact all lattice points 
$z=m\tau {+}n$ at $m,n \in \mathbb Z$ and $m\neq 0$) that follow from the $B$-cycle monodromies
(\ref{b.34gen}) and the poles of $g^{(1)}(z,\tau)$, see (\ref{baspin.41a}). Accordingly, the singularities 
$g^{(2)}_{ij} = \frac{-2\pi i }{z_{ij}{-}\tau}+\ldots$ and $g^{(2)}_{ij} = \frac{2\pi i }{z_{ij}{+}\tau}+\ldots$
cause new pinching rules from the integration region where $z_i {-} z_j \rightarrow \pm \tau$.
More generally, chiral splitting introduces additional pinching rules for the entire tower of
$g^{(k\geq 1)}_{ij} $, even though the poles of $f^{(k)}_{ij} $ obtained from
loop integration are limited to $k=1$. Given that the simplest appearance 
of $g^{(2)}_{ij}$ in maximally supersymmetric string amplitudes occurs in the chiral 
six-point correlator ${\cal K}_6$, there is no hint of these additional pinching rules at $n\leq 5$ points.

%%%%%%%%%%%%%%%%%%%%%%%%%%%%%%%%%%%%%%%%%%%%%%%% 
%%%%%%%%%%%%%%%%%%%%%%%%%%%%%%%%%%%%%%%%%%%%%%%% 

\subsubsection{Open strings}
\label{sec:3.1.op}

The simplest pinch from the integration region $z_i {-} z_j \rightarrow \pm \tau$
for open strings occurs~for
\begin{align}
	{\rm FT}^{\rm op}_{12345}[g^{(1)}_{15}]&= \frac{1}{2} I^{(5)}_{1,2,3,4,5} -\frac{I^{(4)}_{5\underline1 ,2,3,4}}{s_{15}}
	\label{newPR.01}
\end{align}
see (\ref{baspin.20}) for the underscore notation of the one-mass box propagators $I^{(4)}_{5\underline1 ,2,3,4}$. The relative sign between the pentagon and the box is opposite to (\ref{baspin.23}) where 
the analogous box diagram originates
from $z_i {-} z_j \rightarrow 0$ rather than $z_i {-} z_j \rightarrow \pm \tau$. We shall study two types of six-point analogues of (\ref{newPR.01})
\begin{itemize}
\item iteration of the pinch $g^{(1)}_{i6} \rightarrow \frac{1}{\tau{-} z_{6i}}$:
\begin{equation}
{\rm FT}^{\rm op}_{123456}[g^{(1)}_{16}g^{(1)}_{26}]= \frac{1}{4} I^{(6)}_{1,2,3,4,5,6}-\frac{I^{(5)}_{6\underline1 ,2,3,4,5}}{2 s_{16}}+\frac{I^{(4)}_{6\underline1 2,3,4,5}}{ s_{16}s_{126}}
\label{FB6pt}
\end{equation}
\item a new type of pinch from $g^{(2)}_{16} \rightarrow \frac{2\pi i }{\tau{-}z_{61}}$:
\begin{align}
	{\rm FT}^{\rm op}_{123456}[g^{(2)}_{16}]&= \frac{1}{12} I^{(6)}_{1,2,3,4,5,6} -\frac{I^{(5)}_{6\underline1 ,2,3,4,5}}{s_{16}}
	\label{newPR.02}
\end{align}
\end{itemize}
On the one hand, the pentagon term $- s_{16}^{-1} I^{(5)}_{6\underline1 ,2,3,4,5} $ in  
${\rm FT}^{\rm op}_{123456}[g^{(2)}_{16}]$ does not have any immediate counterpart in the 
field-theory limit of the doubly-periodic and non-singular
$f^{(2)}_{16} $ arising from loop integration. On the other hand, the pentagon $ I^{(5)}_{6\underline1 ,2,3,4,5}$
is recovered as a boundary term in worldline variables when producing $f^{(2)}_{16} $ through the string-theory 
loop integral over $E_{1|2|3,4,5,6} {\cal J}_6$ before taking the field-theory limit. The
worldline limits of the resulting $f^{(2)}_{ij} $ and bilinears in $f^{(1)}_{ij} $ 
only line up with Schwinger parametrizations 
of Feynman integrals after adding these boundary terms, see appendix \ref{bdyWL}.

In the open-string integration region associated with ${\rm Tr}(t^{a_1}t^{a_2}\ldots t^{a_6})$, the remaining $g^{(2)}_{ij}$ with $\{ i,j\}\neq \{1,6\}$ do not allow for the pinch in (\ref{newPR.02}) since $(z_1,z_6)$ is the only pair of points that gets separated by $\tau$ on a boundary component of co-dimension one, i.e.\ one may write
\begin{align}
g^{(2)}_{ij} \rightarrow \left\{ \begin{array}{cl}  \tfrac{1}{12} - \tfrac{\delta(z_{ij})}{s_{ij}} &: \ |i{-}j| = n{-}1
\\ 
\tfrac{1}{12} &: \ \te{otherwise}
\end{array} \right.
	\label{newPR.03}
\end{align}
in the notation of section \ref{sec:2.2.4}.

The coefficient $\frac{1}{12}$ of the hexagon $I^{(6)}_{1,2,3,4,5,6} $ in (\ref{newPR.02}) 
and (\ref{newPR.03}) stems
from the $\tau \rightarrow i \infty$~limit
\begin{align}
	\lim_{\tau \rightarrow i \infty}g^{(2)}_{ij} = -2 \zeta_2 = \frac{(2\pi i)^2}{12}
	\label{newPR.04}
\end{align}
which no longer depends on $z_i, z_j$ and generalizes as follows to arbitrary 
Kronecker-Eisenstein coefficients (with ${\rm B}_k$ the $k^{\rm th}$ Bernoulli number):
\begin{align}
\lim_{\tau \rightarrow i \infty}g^{(k)}_{ij}  =  \left\{ \begin{array}{cl}  -2 \zeta_k = \frac{{\rm B}_k}{k!} (2\pi i)^k &: \ k\geq 2 \ \te{even}
\\ 
0 &: \ k\geq 3 \ \te{odd}
\end{array} \right.
	\label{newPR.05}
\end{align}
It would be interesting to relate the open-string pinching rules from (\ref{newPR.03}) to (\ref{newPR.05}) 
to the field-theory limits in section 2.3 of \cite{Bridges:2021ebs}.
The reference infers field-theory limits that accommodate adjustable shifts
of loop momenta by external momenta from imposing BRST invariance and manifest
color-kinematics duality of the resulting $(n\leq 7)$-point super-Yang-Mills amplitudes. 
On the one hand, the coefficients $b^{(m)}_{ij}$ and $c^{(m)}_{ij}$
in (2.14) and (2.15) of \cite{Bridges:2021ebs} have parallels with the
Bernoulli numbers in (\ref{newPR.05}) and, through the ${\rm dist}_4(i,j)$ functions in the $c^{(m)}_{ij}$,
with the new pinching rules of this section. On the other hand, the field-theory limits
of the reference are not claimed to reproduce the $\alpha' \rightarrow 0$ limits of the {\it individual} 
Koba-Nielsen integrals over Kronecker-Eisenstein coefficients in any given cylinder parametrization of
figure~\ref{figWS}. The field-theory limits of \cite{Bridges:2021ebs} can instead be understood as 
effective rules that lead to the correct 
assembly of BRST invariant (and color-kinematics dual) loop integrands.

%%%%%%%%%%%%%%%%%%%%%%%%%%%%%%%%%%%%%%%%%%%%%%%% 
%%%%%%%%%%%%%%%%%%%%%%%%%%%%%%%%%%%%%%%%%%%%%%%% 

\subsubsection{Closed strings}
\label{sec:3.1.cl}

Pinching rules from the integration 
region $z_i {-} z_j \rightarrow \pm \tau$:
\begin{itemize}
\item relative signs for $(n{-}1)$-gons:

The pinches from pairs of $g^{(1)}_{ij} \bar{g}^{(1)}_{pq} $ exemplified in  
(\ref{baspin.24}), (\ref{baspin.26}) and (\ref{swpara.02}) receive contributions
from both $z_i {-} z_j \rightarrow 0$ and $z_i {-} z_j \rightarrow \pm \tau$ when
the puncture $z_1=0$ fixed at the origin is involved, i.e.\ if $1 \in \{i,j\}$.
For instance, the boxes in the five-point case
\begin{align}
{\rm FT}^{\rm cl}_5\big[g^{(1)}_{15}\overline{g^{(1)}_{15}} \big]&=
\frac{1}{4}\sum_{\rho \in S_{\{2,3,4,5 \}}} I^{(5)}_{1,\rho(2,3,4,5)}
 +\frac{1}{2s_{15}} \sum_{\rho \in S_{\{2,3,4\}}}\big( I_{15,\rho(2,3,4)}^{(4)}
 + I_{5\underline{1},\rho(2,3,4)}^{(4)} \big) \notag \\
 &=
\frac{1}{4}\sum_{\rho \in S_{\{2,3,4,5 \}}} I^{(5)}_{1,\rho(2,3,4,5)}
 + \frac{1}{s_{15}} \sum_{\rho \in S_{\{2,3,4\}}}I_{15,\rho(2,3,4)}^{(4)}
 \label{1nex.01}
 \end{align}
receive contributions $I_{15,\rho(2,3,4)}^{(4)}$ from $z_5 {-} z_1 \rightarrow 0$ 
and $ I_{5\underline{1},\rho(2,3,4)}^{(4)}$ from $z_5 {-} z_1 \rightarrow \tau$.
In absence of loop-momentum dependent factors, we have $I_{15,\rho(2,3,4)}^{(4)}
=  I_{5\underline{1},\rho(2,3,4)}^{(4)}$ and obtain the relabeling of (\ref{baspin.24})
in the second line of (\ref{1nex.01}). However, the analogous pinches in a six-point context involve
a second pair of $g^{(1)}_{ab} \bar{g}^{(1)}_{pq} $ which may introduce relative signs between
the integration region $z_i {-} z_j \rightarrow 0$ and $z_i {-} z_j \rightarrow \pm \tau$. Indeed, the
pentagons in the field-theory limit
\begin{align}
{\rm FT}^{\rm cl}_6\big[ g^{(1)}_{16} g^{(1)}_{26} \bar g^{(1)}_{16} \bar g^{(1)}_{34} \big]
&= \frac{1}{16}
\sum_{\rho \in S_{\{2,3,4,5,6 \}}} \sgn^\rho_{26}  \sgn^\rho_{34}  I^{(6)}_{1,\rho(2,3,4,5,6)}
 \label{1nex.02} \\
&\quad +\frac{1}{8 s_{16}}
\sum_{\rho \in S_{\{2,3,4,5 \}}}  \big(  I^{(5)}_{16,\rho(2,3,4,5)}  - I^{(5)}_{6\underline1,\rho(2,3,4,5)}  \big)  \sgn^\rho_{34} 
\notag \\
&=  \frac{1}{16}
\sum_{\rho \in S_{\{2,3,4,5,6 \}}} \sgn^\rho_{26}  \sgn^\rho_{34}  I^{(6)}_{1,\rho(2,3,4,5,6)}
\notag
\end{align}
cancel since the contribution $\sgn( u_{26} )$ from $g^{(1)}_{26}$ takes opposite signs along with
$ I^{(5)}_{16,\rho(2,3,4,5)} $ (coming from the integration region $u_6 \rightarrow u_1=0 < u_2$) and $ I^{(5)}_{6\underline1,\rho(2,3,4,5)} $ (coming from $u_6 \rightarrow 1 > u_2$). The field-theory
limit ${\rm FT}^{\rm cl}_6[ g^{(1)}_{16} g^{(1)}_{12} \bar g^{(1)}_{16} \bar g^{(1)}_{34} ]$ 
with $g^{(1)}_{12}$ instead of $g^{(1)}_{26}$ in turn features both of 
$I^{(5)}_{16,\rho(2,3,4,5)} $ and $I^{(5)}_{6\underline1,\rho(2,3,4,5)}$ with
the same prefactor $-\frac{1}{8 s_{16}}$ since $\sgn( u_{12} )$ has uniform sign
for both of $u_6 \rightarrow 0$ and $u_6 \rightarrow 1$, leading to non-zero pentagon contributions.
\item pinches from $g^{(2)}_{ij} $:

There are several variants of how the poles $g^{(2)}_{ij} = \frac{\pm 2\pi i }{z_{ij}{\pm}\tau}+\ldots$
integrate to pinches multiplied by $ s_{ij}^{-1}$ in a closed-string six-point setting: 
\begin{align}
{\rm FT}^{\rm cl}_6\big[ g^{(2)}_{16}\bar{g}^{(2)}_{16} \big] &= \frac{1}{144}  \sum_{\rho \in S_{\{2,3,4,5,6 \}}}  I^{(6)}_{1,\rho(2,3,4,5,6)} + \frac{1}{2 s_{16}} \sum_{\rho \in S_{\{2,3,4,5 \}}}   I^{(5)}_{16,\rho(2,3,4,5)} 
\label{newPR.06} \\
{\rm FT}^{\rm cl}_6\big[ g^{(2)}_{16} \bar{g}^{(1)}_{16} \bar{g}^{(1)}_{ij} \big] &= 
{-} \frac{1}{48}  \sum_{\rho \in S_{\{2,3,4,5,6 \}}} \! I^{(6)}_{1,\rho(2,3,4,5,6)}   \sgn^\rho_{ij}  
  - \frac{1}{4 s_{16}} \sum_{\rho \in S_{\{2,3,4,5 \}}}   I^{(5)}_{16,\rho(2,3,4,5)}    \sgn^\rho_{ij}  
\notag \\
{\rm FT}^{\rm cl}_6\big[g^{(1)}_{16} g^{(1)}_{ij} \bar{g}^{(2)}_{16} \big] &= 
{-} \frac{1}{48}  \sum_{\rho \in S_{\{2,3,4,5,6 \}}} \! I^{(6)}_{1,\rho(2,3,4,5,6)}   \sgn^\rho_{ij}  
  - \frac{1}{4 s_{16}} \sum_{\rho \in S_{\{2,3,4,5 \}}}   I^{(5)}_{16,\rho(2,3,4,5)}   \sgn^\rho_{ij}  
\notag
\end{align}
where $\{i,j\}$ in the second and third line are distinct from $\{1,6\}$. One may have
naively expected an extra factor of two for the pentagon contributions since the poles $| z_i{-} z_j |^{-2}$
translated into $\delta^2(z_{ij})/s_{ij}$ with coefficient one in the earlier examples (\ref{swpara.02}). 
However, it is crucial that the torus integration region for $z_6$ only covers half of the disk around
$z_{61} \rightarrow \tau$ if $z_1=0$ and that the factors of $g^{(2)}_{16} $ and $\bar{g}^{(2)}_{16} $  are non-singular at
$z_{61} \rightarrow 0$. That is why the pentagon prefactors in (\ref{newPR.06}) are only half of those in
(\ref{baspin.26}) and (\ref{swpara.02}). 

More generally, closed-string pinches due to $g^{(k\geq 2)}_{1n} $ or $\bar g^{(k\geq 2)}_{1n} $ tend to come with extra factors of $\frac{1}{2}$ relative to those from pairs $g^{(1)}_{1n} \bar g^{(1)}_{1n}$ since only the latter receive contributions from both integration regions $z_{n1} \rightarrow 0$ and $z_{n1} \rightarrow \tau$.

In the notation of
section \ref{sec:2.2.4}, we may summarize (\ref{newPR.06}) as
\begin{align}
g^{(2)}_{16}\bar{g}^{(2)}_{16} &\rightarrow \frac{1}{144} +\frac{\delta^2(z_{16})}{2 s_{16}}
\label{newPR.07} \\
g^{(2)}_{16} \bar{g}^{(1)}_{16} \bar{g}^{(1)}_{ij} &\rightarrow  \bigg( \frac{1}{48}  \sgn(u_{16})  -\frac{\delta^2(z_{16})}{4 s_{16}}  \bigg) \sgn(u_{ij}) 
\notag\\
 g^{(1)}_{16} g^{(1)}_{ij}  \bar{g}^{(2)}_{16} &\rightarrow \bigg( \frac{1}{48}   \sgn(u_{16}) -\frac{\delta^2(z_{16})}{4 s_{16}}  \bigg)   \sgn(u_{ij}) 
 \notag
\end{align}
\end{itemize}
The mutual consistency of the field-theory limits in (\ref{1nex.02}) and (\ref{newPR.06}) can be cross-checked
by virtue of Fay identities among Kronecker-Eisenstein coefficients
\beq
g^{(1)}_{ij} g^{(1)}_{jk}
+ g^{(1)}_{jk} g^{(1)}_{ki}
+ g^{(1)}_{ki} g^{(1)}_{ij}
+ g^{(2)}_{ij} +g^{(2)}_{jk} + g^{(2)}_{ki} = 0
\eeq
which hold in identical form for $g^{(k)} \rightarrow f^{(k)}$ and
whose general form is most conveniently written at the level of generating functions
(\ref{baspin.31}) and (\ref{baspin.42}) \cite{BrownLev}. More specifically, Fay identities
imply the vanishing of the field-theory limits
\begin{align}
{\rm FT}^{\rm cl}_6\big[ \bar g^{(2)}_{16} (
g^{(1)}_{12}g^{(1)}_{26} - g^{(1)}_{12}g^{(1)}_{16} - g^{(1)}_{16}g^{(1)}_{26}
+g^{(2)}_{16} + g^{(2)}_{12}+ g^{(2)}_{26}) \big]
&=0
 \label{1nex.03} \\
{\rm FT}^{\rm cl}_6\big[ \bar g^{(1)}_{16} \bar g^{(1)}_{34}
(
g^{(1)}_{12}g^{(1)}_{26} - g^{(1)}_{12}g^{(1)}_{16} - g^{(1)}_{16}g^{(1)}_{26}
+g^{(2)}_{16} + g^{(2)}_{12}+ g^{(2)}_{26}) \big] &=0
\notag
\end{align}
where the first line validates the relative factors of the hexagons and pentagons of 
(\ref{newPR.06}). The second line of (\ref{1nex.03}) in turn is sensitive to the fact that the domains 
$z_i {-} z_j \rightarrow 0$ and $z_i {-} z_j \rightarrow \pm \tau$ may contribute with
opposite signs and verifies consistency of (\ref{1nex.02}) with (\ref{newPR.06}).

%%%%%%%%%%%%%%%%%%%%%%%%%%%%%%%%%%%%%%%%%%%%%%%% 
%%%%%%%%%%%%%%%%%%%%%%%%%%%%%%%%%%%%%%%%%%%%%%%% 

\subsubsection{Algorithm for multiparticle generalization}
\label{sec:3.1.sm}

Similar to the discussion in sections \ref{sec:2.2.3} and \ref{sec:2.2.4}, the pinching
rules for products of $g^{(k_1)}_{i_1j_1}g^{(k_2)}_{i_2j_2} \ldots g^{(k_r)}_{i_rj_r}$ can augment
the $n$-gons by reducible diagrams from $(n{-}1)$-gons all the way to $(n{-}r)$-gons. One again sums over the
non-trivial cascades of pinches obtained from all subsets of these $g^{(k_1)}_{i_1j_1}
g^{(k_2)}_{i_2j_2} \ldots g^{(k_r)}_{i_rj_r}$ while replacing
the remaining Kronecker-Eisenstein factors by their $\tau \rightarrow i\infty$ limits 
(\ref{baspin.17}) and (\ref{newPR.05}).
In an eight-point open-string case in the color-ordering of ${\rm Tr}(t^{a_1}t^{a_2}\ldots t^{a_8})$,
for instance, the product $g^{(2)}_{81}g^{(2)}_{71}$ induces two non-trivial pinches
$-\frac{1}{12} I^{(7)}_{8\underline{1},\ldots,6,7}/s_{81}$ (from $g^{(2)}_{81}$) and 
$ I^{(6)}_{78\underline{1},\ldots,6}/(s_{781}s_{81})$ (from both factors) which are 
added to the octagon $\frac{1}{144} I^{(8)}_{1,2,\ldots,8}$.

The same combinatorial rules apply to closed strings: the pinching rules for products
$g^{(k_1)}_{i_1j_1}g^{(k_2)}_{i_2j_2} \ldots g^{(k_r)}_{i_rj_r}$ times
$\overline{ g^{(m_1)}_{p_1q_1}}  \overline{ g^{(m_2)}_{p_2q_2} }\ldots \overline{ g^{(m_s)}_{p_sq_s}}$ 
sum over all subsets of pairings of meromorphic and antimeromorphic factors that give rise to
non-trivial pole structures by the mechanisms of section \ref{sec:3.1.cl}. In the eight-point example of
$g^{(2)}_{81}g^{(2)}_{71} \overline{ g^{(2)}_{81}g^{(1)}_{71} g^{(1)}_{34} }$, the field-theory
limit comprises heptagons $\sim s_{81}^{-1}$ from $g^{(2)}_{81} \overline{ g^{(2)}_{81}}$, heptagons $\sim s_{71}^{-1}$ from $ g^{(2)}_{71} \overline{  g^{(1)}_{71} }$, hexagons $\sim (s_{71}^{-1}{+} s_{81}^{-1}) s_{781}^{-1}$ from $g^{(2)}_{81}g^{(2)}_{71} \overline{ g^{(2)}_{81}g^{(1)}_{71} }$ and a permutation sum over octagons where each of the five factors is replaced by its $\tau \rightarrow i\infty$ limit.

%%%%%%%%%%%%%%%%%%%%%%%%%%%%%%%%%%%%%%%%%%%%%%%% 
%%%%%%%%%%%%%%%%%%%%%%%%%%%%%%%%%%%%%%%%%%%%%%%% 

\subsection{Field-theory limits of open-string homology invariants}
\label{sec:3.2}

In this section, the pinching rules of sections \ref{sec:2.2} and \ref{sec:3.1} are combined to
determine the field-theory limits of the homology invariants of section \ref{sec:2.3.2} in an open-string context.

%%%%%%%%%%%%%%%%%%%%%%%%%%%%%%%%%%%%%%%%%%%%%%%% 
%%%%%%%%%%%%%%%%%%%%%%%%%%%%%%%%%%%%%%%%%%%%%%%% 

\subsubsection{Five points}
\label{sec:3.2.a}

At five points, the pinching rules of (\ref{baspin.23}) imply the following field-theory limits
for the homology invariants $E_{1\vert 23,4,5}$ and $E_{1\vert 2,3,4,5}^m$ in
(\ref{Esat5pt}) in the color-ordering of ${\rm Tr}(t^{a_1}t^{a_2}\ldots t^{a_5})$:
\begin{align}
	{\rm FT}^{\rm op}_{12\ldots 5}\big[  E_{1\vert 23,4,5}\big]&=\frac{1}{2}I^{(5)}_{1,2,3,4,5}+\frac{1}{s_{12}}I^{(4)}_{12,3,4,5}+\frac{1}{s_{23}}I^{(4)}_{1,23,4,5} \label{5ptEex}\\
	{\rm FT}^{\rm op}_{12\ldots 5}\big[E_{1\vert 24,3,5}\big]&=\frac{1}{2}I^{(5)}_{1,2,3,4,5}+\frac{1}{s_{12}}I^{(4)}_{12,3,4,5}\nonumber\\
	{\rm FT}^{\rm op}_{12\ldots 5}\big[  E_{1\vert 25,3,4}\big]&=\frac{1}{2}I^{(5)}_{1,2,3,4,5}+\frac{1}{s_{12}}I^{(4)}_{12,3,4,5}+\frac{1}{s_{15}}I^{(4)}_{5\underline1,2,3,4}\nonumber\\
	{\rm FT}^{\rm op}_{12\ldots 5}\big[  E_{1\vert 2,3,4,5}^m\big]&=\bigg(\ell^m + \frac{1}{2}k_{2345}^m\bigg)I^{(5)}_{1,2,3,4,5}+\frac{1}{s_{12}}k_2^mI^{(4)}_{12,3,4,5}-\frac{1}{s_{15}}k_5^mI^{(4)}_{5\underline1,2,3,4}
\notag
\end{align}
The number of box integrals varies between different permutations of $E_{i \vert jk,p,q}$ -- a single box in case 
of non-adjacent $p,q$ with $|p{-}q|=2,3$ and two boxes in the adjacent case with $|p{-}q|=1,4$, consistently with
the cyclic property $E_{i \vert jk,p,q}= E_{k \vert ij,p,q}$. In the field-theory limit of $E_{1\vert 2,3,4,5}^m$, one can view the coefficient 
$\ell^m{+}\frac{1}{2}k_{2345}^m=\ell^m{-}\frac{1}{2}k_{1}^m$ of the pentagon propagators $I^{(5)}_{1,2,3,4,5}$ 
as the average of the momenta $\ell$ and $\ell{-}k_1$ adjacent to leg~1 in figure \ref{figngon}.

%%%%%%%%%%%%%%%%%%%%%%%%%%%%%%%%%%%%%%%%%%%%%%%% 
%%%%%%%%%%%%%%%%%%%%%%%%%%%%%%%%%%%%%%%%%%%%%%%% 

\subsubsection{Six points}
\label{sec:3.2.b}

Among the six-point homology invariants in (\ref{Esat6pt}), we will relegate
the case $E_{1|2|3,4,5,6}$ with double poles to section \ref{sec:FTE}.
Except for the two-tensor $E^{mn}_{1|2,3,4,5,6}$ with permutation symmetry in $2,3,4,5,6$,
the color-ordering of ${\rm Tr}(t^{a_1}t^{a_2}\ldots t^{a_6})$ gives rise to inequivalent 
field-theory limits for different permutations:  
\begin{align}
	{\rm FT}^{\rm op}_{12\ldots 6}\big[  E_{1\vert 23,45,6}\big]&=\frac{1}{4}I^{(6)}_{1,2,3,4,5,6}+\frac{1}{2s_{12}}I^{(5)}_{12,3,4,5,6}+\frac{1}{2s_{23}}I^{(5)}_{1,23,4,5,6}+\frac{1}{2s_{45}}I^{(5)}_{1,2,3,45,6} \label{newPR.09} \\
	&\quad+\frac{1}{s_{12}s_{45}}I^{(4)}_{12,3,45,6}+\frac{1}{s_{23}s_{45}}I^{(4)}_{1,23,45,6}\nonumber\\
	{\rm FT}^{\rm op}_{12\ldots 6}\big[ E_{1\vert 23,46,5}\big]&=\frac{1}{4}I^{(6)}_{1,2,3,4,5,6}+\frac{1}{2s_{12}}I^{(5)}_{12,3,4,5,6}+\frac{1}{2s_{16}}I^{(5)}_{6\underline1,2,3,4,5}+\frac{1}{2s_{23}}I^{(5)}_{1,23,4,5,6}\nonumber\\
	&\quad+\frac{1}{s_{16}s_{23}}I^{(4)}_{6\underline1,23,4,5,6}
	+\bigg( \frac{1}{s_{12}}{+}  \frac{1}{s_{16}} \bigg)  \frac{1}{s_{126}}
	I^{(4)}_{6\underline12,3,4,5}\nonumber\\
	{\rm FT}^{\rm op}_{12\ldots 6}\big[ E_{1\vert 234,5,6}\big]&=\frac{1}{3}I^{(6)}_{1,2,3,4,5,6}+\frac{1}{2s_{12}}I^{(5)}_{12,3,4,5,6}+\frac{1}{2s_{23}}I^{(5)}_{1,23,4,5,6}+\frac{1}{2s_{34}}I^{(5)}_{1,2,34,5,6}\nonumber\\
	&\quad+\frac{1}{s_{12}s_{34}}I^{(4)}_{12,34,5,6}+\!\bigg( \frac{1}{s_{12} }{+}\frac{1}{ s_{23}}\bigg) \frac{1}{s_{123}}	
	I^{(4)}_{123,4,5,6}+\!
	\bigg( \frac{1}{s_{23} }{+} \frac{1}{ s_{34} } \bigg) \frac{1}{s_{234}}
	I^{(4)}_{1,234,5,6}\nonumber\\
	{\rm FT}^{\rm op}_{12\ldots 6}\big[ E_{1\vert 235,4,6}\big]&=\frac{1}{3}I^{(6)}_{1,2,3,4,5,6}+\frac{1}{2s_{12}}I^{(5)}_{12,3,4,5,6}+\frac{1}{2s_{23}}I^{(5)}_{1,23,4,5,6}+
	\bigg(\frac{1}{s_{12}}{+}\frac{1}{s_{23}} \bigg) \frac{1}{s_{123}}
	I^{(4)}_{123,4,5,6}\nonumber\\
	{\rm FT}^{\rm op}_{12\ldots 6}\big[ E_{1\vert 23,4,5,6}^m\big]&=
\bigg(   \frac{1}{2}(\ell^m{-}\tfrac{1}{2}k_1^m) + \frac{1}{12} (k_3^m{-}k_2^m) \bigg)I^{(6)}_{1,2,3,4,5,6}
+\frac{1}{s_{12}} \big(\ell^m+\tfrac{1}{2}k_{3456}^m \big)I^{(5)}_{12,3,4,5,6}\nonumber\\
	&\quad-\frac{1}{2s_{16}}k_6^mI^{(5)}_{6\underline1,2,3,4,5}+\frac{1}{s_{23}} \big(\ell^m-\tfrac{1}{2}k_{1}^m\big)I^{(5)}_{1,23,4,5,6}-\frac{1}{s_{16}s_{23}}k_6^mI^{(4)}_{6\underline1,23,4,5}\nonumber\\ 
	&\quad+
	\bigg(
	\frac{k_2^m}{s_{23}s_{123}} + \bigg( \frac{1}{s_{12}}{+} \frac{1}{s_{23}}  \bigg) \frac{k_3^m}{s_{123}} \bigg)I^{(4)}_{123,4,5,6}
	-	\bigg( \frac{1}{s_{12}} {+} \frac{1}{s_{16}} \bigg)\frac{1}{s_{126}}k_6^mI^{(4)}_{6\underline12,3,4,5}\nonumber\\
	{\rm FT}^{\rm op}_{12\ldots 6}\big[ E_{1\vert 24,3,5,6}^m\big]&=
		\bigg( \frac{1}{2}(\ell^m{-} \tfrac{1}{2}k_1^m)+\frac{1}{12} (k_4^m{-}k_2^m) \bigg)I^{(6)}_{1,2,3,4,5,6}
	+\frac{1}{s_{12}} \big(\ell^m + \tfrac{1}{2}k_{3456}^m \big)I^{(5)}_{12,3,4,5,6}\nonumber\\
	&\quad-\frac{1}{2s_{16}}k_6^mI^{(5)}_{6\underline1,2,3,4,5}+\frac{1}{s_{12}s_{123}}k_3^mI^{(4)}_{123,4,5,6}-
	\bigg( \frac{1}{s_{12}}{+}\frac{1}{s_{16}} \bigg) \frac{1}{s_{126}} k_6^mI^{(4)}_{6\underline12,3,4,5}\nonumber\\
	{\rm FT}^{\rm op}_{12\ldots 6}\big[ E_{1\vert 26,4,5,6}^m\big]&=
	\bigg( \frac{1}{2}(\ell^m{-}\tfrac{1}{2}k_1^m) +\frac{1}{12} (k_6^m{-}k_2^m) \bigg)I^{(6)}_{1,2,3,4,5,6}
	+\frac{1}{s_{12}} \big( \ell^m + \tfrac{1}{2}k_{3456}^m \big)I^{(5)}_{12,3,4,5,6}\nonumber\\
	&\quad+\frac{1}{s_{16}} \big(\ell^m + \tfrac{1}{2}k_6^m- \tfrac{1}{2}k_1^m\big)I^{(5)}_{6\underline1,2,3,4,5}+\frac{1}{s_{12}s_{123}}k_3^mI^{(4)}_{123,4,5,6}\nonumber\\
	&\quad+
\bigg( \frac{k_2^m}{s_{16}s_{126}}- \frac{k_6^m}{s_{12}s_{126}} \bigg) 	
	I^{(4)}_{6\underline12,3,4,5}-\frac{1}{s_{16}s_{156}}k_5^mI^{(4)}_{56\underline1,2,3,4}\nonumber\\
	%%%%%
	{\rm FT}^{\rm op}_{12\ldots 6}\big[ E_{1\vert 2,3,4,5,6}^{mn}\big]&= \bigg\{ 
\ell^m_{\phantom{1}}\ell^n_{\phantom{1}} +    \frac{1}{2}\big[ k_2^{(m}\ell_{\phantom{1}}^{n)} +(2\leftrightarrow 3,4,5,6) \big]	+ \frac{1}{4} \big[ k_2^{(m}k_3^{n)}+(2,3\vert 2,3,4,5,6) \big]
	\nonumber\\
	&\quad + \frac{1}{6} \big[ k_2^mk_2^n {+}(2\leftrightarrow 3,\ldots,6) \big] \bigg\}   I^{(6)}_{1,2,3,4,5,6}+\frac{1}{s_{12}} k_2^{(m} \big( \ell_{\phantom{1}}^{n)} + \tfrac{1}{2}k_{3456}^{n)} \big)I^{(5)}_{12,3,4,5,6}\nonumber\\
	&\quad
	-\frac{1}{s_{16}}
	 k_6^{(m} (\ell_{\phantom{1}}^{n)}+ \tfrac{1}{2}k_6^{n)} - \tfrac{1}{2} k_1^{n)}) I^{(5)} _{6\underline1,2,3,4,5}
	+\frac{1}{s_{12}s_{123}} k_2^{(m}k_3^{n)}I^{(4)}_{123,4,5,6} \nonumber\\
	&\quad+\frac{1}{s_{16}s_{156}} k_5^{(m}k_6^{n)}I^{(4)}_{56\underline1,2,3,4}-
	\bigg( \frac{1}{s_{12}} {+}\frac{1}{s_{16}}\bigg) \frac{1}{s_{126}} k_2^{(m}k_6^{n)} I^{(4)}_{6\underline12,3,4,5}\nonumber
\end{align}
Note that, similar to the five-point case in the last line of (\ref{5ptEex}),\footnote{However, the cyclicity $E_{i \vert jk,p,q}= E_{k \vert ij,p,q}$ of the five-point scalar homology invariants does not carry over to the six-point vector homology invariants with the same index structure since \cite{Mafra:2018pll}
\[
E^m_{1 | 23,4,5,6}=E^m_{2| 31,4,5,6}+ \big[ k_4^m E_{1 | 324,5,6} + (4\leftrightarrow 5,6) \big]
\]
see section 5.2 of the reference for similar identities of more general homology invariants.} several 
terms in the field-theory limits of $E^m_{i \vert jk, p,q,r}$ again feature the averaged
momenta of two neighboring pentagon edges, for instance $( \ell^m {+} \tfrac{1}{2}k_{3456}^m )
I^{(5)}_{12,3,4,5,6}$ or $(\ell^m{-}\tfrac{1}{2}k_{1}^m)I^{(5)}_{1,23,4,5,6}$. A complete list of
${\rm FT}^{\rm op}_{12\ldots 6}[E]$ for six-point homology invariants can be found in the 
supplementary material of this work.

%%%%%%%%%%%%%%%%%%%%%%%%%%%%%%%%%%%%%%%%%%%%%%%% 
%%%%%%%%%%%%%%%%%%%%%%%%%%%%%%%%%%%%%%%%%%%%%%%% 

\subsection{Field-theory limits of closed-string homology invariants}
\label{sec:3.3}

In this section, we proceed to gathering field-theory limits of closed-string homology invariants,
again using the pinching rules of sections \ref{sec:2.2} and \ref{sec:3.1}. In view of the double copy
$| {\cal K}_n |^2$ of chiral correlators in the closed-string integrand of (\ref{baspin.13}), we will be interested
in arbitrary crossterms $E \bar E$ of the meromorphic homology invariants of section \ref{sec:2.3.2} and their complex conjugates. Accordingly, closed-string field-theory limits introduce considerably more inequivalent
cases than their open-string counterparts in section \ref{sec:3.2}. 

%%%%%%%%%%%%%%%%%%%%%%%%%%%%%%%%%%%%%%%%%%%%%%%% 
%%%%%%%%%%%%%%%%%%%%%%%%%%%%%%%%%%%%%%%%%%%%%%%% 

\subsubsection{Five points}
\label{sec:3.3.a}

Given the two types $E_{1\vert ij, p, q}$ and $E^m_{1\vert 2,3, 4, 5}$ of meromorphic homology invariants,
there are five permutation-inequivalent field-theory limits which we assemble from (\ref{baspin.24})
\begin{align}
	{\rm FT}^{\rm cl}_5\big[E_{1\vert 23, 4, 5}\bar E_{1\vert 45,2,3} \big]&= \frac{1}{4}  \sum_{\rho \in S_{\{2,3,4,5\}}}I_{1,\rho(2,3,4,5)}^{(5)}\sign^\rho_{23}\sign^\rho_{45} \label{newPR.10} \\
	{\rm FT}^{\rm cl}_5\big[E_{1\vert 23, 4, 5}\bar E_{1\vert 24,3,5} \big]&= \frac{1}{4} \sum_{\rho \in S_{\{2,3,4,5\}}} I_{1,\rho(2,3,4,5)}^{(5)}\sign^\rho_{23}\sign^\rho_{24} + \frac{1}{s_{12}} \sum_{\rho \in S_{\{3,4,5\}}}  I_{12,\rho(3,4,5)}^{(4)}\nonumber\\
	{\rm FT}^{\rm cl}_5\big[E_{1\vert 23, 4, 5}\bar E_{1\vert 23,4,5}\big]&= \frac{1}{4}\sum_{\rho \in S_{\{2,3,4,5\}}} I_{1,\rho(2,3,4,5)}^{(5)}+ \frac{1}{s_{23}} \sum_{\rho \in S_{\{23,4,5\}}}I_{1,\rho(23,4,5)}^{(4)}\nonumber\\
	&\quad +  \frac{1}{s_{12}} \sum_{\rho \in S_{\{3,4,5\}}} I_{12,\rho(3,4,5)}^{(4)}
	+ \frac{1}{s_{13}} \sum_{\rho \in S_{\{2,4,5\}}} I_{13,\rho(2,4,5)}^{(4)}\nonumber\\
	{\rm FT}^{\rm cl}_5\big[E_{1\vert 2,3, 4, 5}^m\bar E_{1\vert 23,4,5}\big]&= - \frac{1}{2}\big(\ell^m+ \tfrac{1}{2}k_{2345}^m \big) \sum_{\rho \in S_{\{2,3,4,5\}}} I_{1,\rho(2,3,4,5)}^{(5)}\sign^\rho_{23}\nonumber\\
	&\quad + \frac{1}{s_{12}}k_2^m \sum_{\rho \in S_{\{3,4,5\}}}I_{12,\rho(3,4,5)}^{(4)}
	-  \frac{1}{s_{13}}k_3^m\sum_{\rho \in S_{\{2,4,5\}}}I_{13,\rho(2,4,5)}^{(4)} \nonumber\\
	{\rm FT}^{\rm cl}_5\big[E_{1\vert 2,3, 4, 5}^m\bar E_{1\vert 2,3,4,5}^p\big]&=
\big( \ell^m + \tfrac{1}{2}k_{2345}^m \big)   \big( \ell^p+ \tfrac{1}{2}k_{2345}^p \big) 	
	\sum_{\rho \in S_{\{2,3,4,5\}}}I_{1,\rho(2,3,4,5)}^{(5)}\nonumber\\
	&\quad +\bigg[ \frac{1}{s_{12}}k_2^mk_2^p\sum_{\rho \in S_{\{3,4,5\}}}I_{12,\rho(3,4,5)}^{(4)}+(2\leftrightarrow 3,4,5)\bigg] \notag
\end{align}
Similar to the open-string analogues (\ref{5ptEex}), we encounter the
combined momentum $\ell^m{+}\tfrac{1}{2}k_{2345}^m$ $= \ell^m {-}\tfrac{1}{2}k_{1}^m$ along with the 
pentagon propagators which averages over the two edges with
momenta $\ell$ and $\ell{-}k_1$ in figure \ref{figngon}.

%%%%%%%%%%%%%%%%%%%%%%%%%%%%%%%%%%%%%%%%%%%%%%%% 
%%%%%%%%%%%%%%%%%%%%%%%%%%%%%%%%%%%%%%%%%%%%%%%% 

\subsubsection{Six points}
\label{sec:3.3.b}

For the pentagon contributions to closed-string field-theory limits at six points,
we have seen in section \ref{sec:3.1.cl} that there can be cancellations between
the two routings of loop momenta in $ I_{2\underline1,\rho(3,4,5,6)}^{(5)}$ and $I_{12,\rho(3,4,5,6)}^{(5)}$, see figure \ref{figngon}. We have frequently used the equivalence
\beq
\varphi(\ell) I_{2\underline1,\rho(3,4,5,6)}^{(5)} \cong \varphi(\ell{-}k_2) I_{12,\rho(3,4,5,6)}^{(5)}
\eeq
under the loop integral in simplifying the expressions for various ${\rm FT}^{\rm cl}_6[\ldots]$ 
below and in appendix \ref{app:Eex}, where $\varphi(\ell)$ is a placeholder for an arbitrary
(not necessarily scalar) function of the loop momentum.

In the present discussion of six-point field-theory limits, we shall again relegate homology
invariants involving a permutation of $E_{1|2|3,4,5,6}$ or its complex conjugates to section \ref{sec:FTE}.
Representative examples include the following (see appendix \ref{app:Eex.1} for additional examples 
and the supplementary material of this work for a comprehensive list):
\begin{align}
	{\rm FT}^{\rm cl}_6\big[E_{1\vert 23, 4 5,6}\bar E_{1\vert 23, 4 5,6}\big]&= \frac{1}{16} \sum_{\rho \in S_{\{2,3,4,5,6\}}} I_{1,\rho(2,3,4,5,6)} 	
\label{par.0} \\
	&\quad \hspace{-3.4cm}
	+\biggl[ \frac{1}{4 s_{12}} \! \sum_{\rho \in S_{\{3,4,5,6\}}} \! \! \! I_{12,\rho(3,4,5,6)}^{(5)}+(2\leftrightarrow 3,4,5)\biggr]
	+\biggl[ \frac{1}{4s_{23}}  \! \sum_{\rho \in S_{\{23,4,5,6\}}} \! \! \! I_{1,\rho(23,4,5,6)}^{(5)}+(23\leftrightarrow 45)\biggr]\nonumber\\
	&\quad \hspace{-3.4cm}
	+\biggl[\biggl( \frac{1}{ s_{14}s_{23}}  \! \sum_{\rho \in S_{\{23,5,6\}}} \! \! \! I_{14,\rho(23,5,6)}^{(4)}+(4\leftrightarrow5)\biggr)+(23\leftrightarrow 45)\biggr] + \frac{1}{ s_{23}s_{45}}  \! \sum_{\rho \in S_{\{23,45,6\}}} \! \! \! I_{1,\rho(23,45,6)}^{(4)}\nonumber\\
	&\quad \hspace{-3.4cm}+\biggl[\biggl( \frac{1}{s_{124}}\bigg(\frac{1}{s_{12}}+\frac{1}{s_{14}}\bigg)  \sum_{\rho \in S_{\{3,5,6\}}}I^{(4)}_{124,\rho(3,5,6)}+(4\leftrightarrow5)\biggr)+(2\leftrightarrow 3)\biggr]
	\nonumber \\
	&\quad \hspace{-3.4cm}+  \frac{1}{4 s_{12}} \sum_{\rho \in S_{\{3,4,5,6\}}}I_{12,\rho(3,4,5,6)}^{(5)}\sign^\rho_{34}\sign^\rho_{45}
	- \frac{1}{4 s_{14}} \sum_{\rho \in S_{\{2,3,5,6\}}}I_{14,\rho(2,3,5,6)}^{(5)} 
	\nonumber\\
	&\quad \hspace{-3.4cm} + \frac{1}{4s_{23}} \sum_{\rho \in S_{\{23,4,5,6\}}}I_{1,\rho(23,4,5,6)}^{(5)}\sign^\rho_{34}\sign^\rho_{45} - \frac{1}{s_{124}}\left(\frac{1}{s_{12}}+\frac{1}{s_{14}}\right)\sum_{\rho \in S_{\{3,5,6\}}}I^{(4)}_{124,\rho(3,5,6)}\nonumber\\
	&\quad \hspace{-3.4cm} - \frac{1}{s_{14}s_{134}} \sum_{\rho \in S_{\{2,5,6\}}}I^{(4)}_{134,\rho(2,5,6)}- \frac{1}{s_{14}s_{23}} \sum_{\rho \in S_{\{23,5,6\}}}I_{14,\rho(23,5,6)}^{(4)}
	\notag
\end{align}
as well as
\begin{align}
	{\rm FT}^{\rm cl}_6\big[E_{1\vert 23, 4 5,6}\bar E_{1\vert 23,4,5,6}^p\big]
	&=- \frac{1}{8} \big( \ell^p+ \tfrac{1}{2} k_{23456}^p\big) \sum_{\rho \in S_{\{2,3,4,5,6\}}} I_{1,\rho(2,3,4,5,6)}^{(6)}\sign_{45}^\rho\\
	&\quad \hspace{-3.4cm}   + \frac{1}{48} (k_3^p{-}k_2^p)  \sum_{\rho \in S_{\{2,3,4,5,6\}}}
	I_{1,\rho(2,3,4,5,6)}^{(6)}  \sign_{23}^\rho  \sign_{45}^\rho \notag\\
	&\quad \hspace{-3.4cm} -\bigg[ \frac{1}{2s_{12}} \sum_{\rho \in S_{\{3,4,5,6\}}}\big(\ell^p+ \tfrac{1}{2}k_{3456}^p \big) I_{12,\rho(3,4,5,6)}^{(5)}\sign_{45}^\rho+(2\leftrightarrow 3)\bigg]\nonumber\\
	&\quad \hspace{-3.4cm} +\bigg[ \frac{1}{4s_{14}}k_{4}^p \! \! \! \! \sum_{\rho \in S_{\{2,3,5,6\}}} \! \! \! \!I_{14,\rho(2,3,5,6)}^{(5)}-(4\leftrightarrow 5)\bigg]-\bigg[  \frac{1}{2s_{23}} \big(\ell_{\phantom{1}}^p{-} \tfrac{1}{2}k_{1}^p \big) \! \!  \! \! \sum_{\rho \in S_{\{23,4,5,6\}}} \! \! \! \! I_{1,\rho(23,4,5,6)}^{(5)}\sign_{45}^{\rho}\bigg]\nonumber\\
	&\quad \hspace{-3.4cm} +\bigg[\bigg( \frac{1}{s_{124}}\bigg(\frac{1}{s_{12}}{+}\frac{1}{s_{14}} \bigg)k_{4}^p\sum_{\rho \in S_{\{3,5,6\}}}I_{124,\rho(3,5,6)}^{(4)}+(2\leftrightarrow 3)\bigg)-(4\leftrightarrow 5)\bigg]\nonumber\\
	&\quad \hspace{-3.4cm} +\bigg[ \frac{1}{s_{14}s_{23}}k_4^p \sum_{\rho \in S_{\{23,5,6\}}}I_{14,\rho(23,5,6)}^{(4)}-(4\leftrightarrow 5)\bigg] \nonumber \\ 
	{\rm FT}^{\rm cl}_6\big[E_{1\vert 23, 4 5,6}\bar E_{1\vert 2,3,4,5,6}^{pq}\big]&=
	\bigg( \frac{1}{4} \ell_{\phantom{1}}^{p}\ell_{\phantom{1}}^{q} + \frac{1}{8} 
	k_{23456}^{(p}\ell_{\phantom{1}}^{q)} \bigg)
	  \sum_{\rho \in S_{\{2,3,4,5,6\}}} 
	  I_{1,\rho(2,3,4,5,6)}^{(6)}\sign_{23}^\rho\sign_{45}^\rho\nonumber\\
&\quad  \hspace{-3.4cm} + \bigg(  \frac{1}{24}
	\sum_{j=2}^6 k_j^pk_j^q
	 +\frac{1}{16} \big[ k_2^{(p}k_3^{q)}+(2,3\vert 2,3,4,5,6) \big] 
	 \bigg) 
	  \sum_{\rho \in S_{\{2,3,4,5,6\}}} I_{1,\rho(2,3,4,5,6)}^{(6)}\sign_{23}^\rho\sign_{45}^\rho\nonumber\\
&\quad \hspace{-3.4cm}-\bigg[ \frac{1}{2s_{12}} k_2^{(p} \big( \ell^{q)} _{\phantom{1}} {+} \tfrac{1}{2}k_{3456}^{q)} \big) \sum_{\rho \in S_{\{3,4,5,6\}}}I_{12,\rho(3,4,5,6)}^{(5)} \sign_{45}^\rho -(2\leftrightarrow 3)\bigg] \nonumber\\
	&\quad \hspace{-3.4cm}-\bigg[ \frac{1}{2s_{14}} k_4^{(p} \big( \ell^{q)} _{\phantom{1}} {+} \tfrac{1}{2}k_{2356}^{q)} \big) \sum_{\rho \in S_{\{3,4,5,6\}}}I_{14,\rho(2,3,5,6)}^{(5)} \sign_{23}^\rho  -(4\leftrightarrow 5)\bigg]\nonumber\\
	&\quad \hspace{-3.4cm} +\bigg[\bigg(  \frac{1}{s_{124}}\bigg(\frac{1}{s_{12}}+\frac{1}{s_{14}} \bigg)k_2^{(p}k_{4}^{q)}
	\sum_{\rho \in S_{\{3,5,6\}}} I_{124,\rho(3,5,6)}^{(4)}-(2\leftrightarrow 3)\bigg)-(4\leftrightarrow 5)\bigg]
	\nonumber
\end{align}
and finally
\begin{align}
	{\rm FT}^{\rm cl}_6\big[E_{1\vert 23,4, 5,6}^m\bar E_{1\vert 2,3,4,5,6}^{pq}\big]&=	
\frac{1}{2} \big( \ell^m + \tfrac{1}{2} k_{23456}^m \big) \ell_{\phantom{1}}^{p}\ell_{\phantom{1}}^{q}
\sum_{\rho \in S_{\{2,3,4,5,6\}}} I_{1,\rho(2,3,4,5,6)}^{(6)}
\\
&\quad \hspace{-3.4cm}
+ \frac{1}{12} (k_2^m{-} k_3^m)  \ell_{\phantom{1}}^{p}\ell_{\phantom{1}}^{q}
\sum_{\rho \in S_{\{2,3,4,5,6\}}}
 \sign_{23}^\rho I_{1,\rho(2,3,4,5,6)}^{(6)}
\notag \\
&\quad \hspace{-3.4cm}
+ \frac{1}{144} \bigg(
\big[ 2k_2^pk_2^q+6k_2^{(p}\ell_{\phantom{1}}^{q)}(2\leftrightarrow 3,4,5,6) \big]+ \big[ 3k_2^{(p}k_3^{q)}+(2,3\vert 2,3,4,5,6) \big]
 \bigg) \notag \\
 &\quad \quad \quad \hspace{-3.4cm} \times \sum_{\rho \in S_{\{2,3,4,5,6\}}}
 \bigg(
 6 \big(\ell^m+\tfrac{1}{2} k^m_{23456}\big)+ (k_2^m{-}k_3^m) \sign_{23}^\rho 
 \bigg)  I_{1,\rho(2,3,4,5,6)}^{(6)}
	  \nonumber\\
	&\quad \hspace{-3.4cm}+\bigg[
	\frac{1}{s_{12}}  \big(\ell_{\phantom{1}}^{m} + \tfrac{1}{2} k_{3456}^{m} \big) k_2^{(p}  \big(  \ell_{\phantom{1}}^{q)}+\tfrac{1}{2}k_{3456}^{q)} \big)
	\sum_{\rho \in S_{\{3,4,5,6\}}}I_{12,\rho(3,4,5,6)}^{(5)}-(2\leftrightarrow 3)\bigg]\nonumber\\
	&\quad \hspace{-3.4cm} -\bigg[  \frac{1}{2s_{14}}  k_{4}^{m} k_4^{(p} \big(\ell_{\phantom{1}}^{q)} + \tfrac{1}{2} k_{2356}^{q)}\big) \sum_{\rho \in S_{\{2,3,5,6\}}}I_{14,\rho(2,3,5,6)}^{(5)} \sign_{23}^\rho +(4\leftrightarrow 5,6)\bigg]\nonumber\\
	&\quad \hspace{-3.4cm} +\bigg[\bigg( \frac{1}{s_{124}} \bigg(\frac{1}{s_{12}}+\frac{1}{s_{14}} \bigg) k_4^m  k_2^{(p}k_{4}^{q)}\sum_{\rho \in S_{\{3,5,6\}}} I_{124,\rho(3,5,6)}^{(4)}+(4\leftrightarrow 5,6)\bigg)-(2\leftrightarrow 3)\bigg]\nonumber\\
	&\quad\hspace{-3.4cm} +\bigg[
\frac{1}{s_{123}}\bigg(\frac{k_3^m}{s_{12}}  - \frac{ k_2^m }{s_{13}} \bigg) 	 k_2^{(p}k_{3}^{q)}
	\sum_{\rho \in S_{\{4,5,6\}}}I_{123,\rho(4,5,6)}^{(4)}\bigg] \notag \\
	{\rm FT}^{\rm cl}_6\big[E_{1\vert 2,3,4, 5,6}^{mn}\bar E_{1\vert 2,3,4,5,6}^{pq}\big]&=
	\bigg(\ell^m \ell^n {+} \frac{1}{2}  k_{23456}^{(m}\ell_{\phantom{1}}^{n)}{+} \frac{1}{6} \sum_{j=2}^6 k_j^m k_j^n 
{+} \frac{1}{ 4}  \big[ k_2^{(m}k_3^{n)} {+} (2,3\vert 2,3,4,5,6)\big] \bigg) \notag \\
&\quad\quad  \hspace{-3.4cm}
\times 
\bigg(\ell^p \ell^q {+} \frac{1}{2}  k_{23456}^{(p}\ell_{\phantom{1}}^{q)}{+} \frac{1}{6} \sum_{j=2}^6 k_j^p k_j^q 
{+} \frac{1}{ 4}  \big[ k_2^{(p}k_3^{q)} {+} (2,3\vert 2,3,4,5,6)\big] \bigg)  \! \! \!
\sum_{\rho \in S_{\{2,3,4,5,6\}}}  \! \! \! \! \! \! I_{1,\rho(2,3,4,5,6)}^{(6)}\nonumber\\
	&\quad \hspace{-3.4cm} +\bigg[
	\frac{1}{s_{12}}  k_2^{(m}\big(\ell_{\phantom{1}}^{n)} + \tfrac{1}{2}  k_{3456}^{n)} \big)
	 k_2^{(p} \big( \ell_{\phantom{1}}^{q)} + \tfrac{1}{2}k_{3456}^{q)} \big)
	\sum_{\rho \in S_{\{3,4,5,6\}}}I_{12,\rho(3,4,5,6)}^{(5)}+(2\leftrightarrow 3,4,5,6)\bigg]\nonumber\\
	&\quad \hspace{-3.4cm}+\bigg[  \frac{1}{s_{123}} \bigg(\frac{1}{s_{12}}+\frac{1}{s_{13}}\bigg)k_2^{(m}k_{3}^{n)}k_2^{(p}k_{3}^{q)} \sum_{\rho \in S_{\{4,5,6\}}} I_{123,\rho(4,5,6)}^{(4)}+(2,3\vert 2,3,4,5,6)\bigg]\nonumber
\end{align}
In spite of the lengthy expressions for some of the individual field-theory limits, the
resulting six-point supergravity loop integrand in section \ref{sec:5.3} below takes a
reasonably compact form.

%%%%%%%%%%%%%%%%%%%%%%%%%%%%%%%%%%%%%%%%%%%%%%%% 
%%%%%%%%%%%%%%%%%%%%%%%%%%%%%%%%%%%%%%%%%%%%%%%% 

\subsection{Removal of double poles}
\label{sec:4}

This section is dedicated to a subtlety and its resolution in extending the
above collection of six-point field-theory limits to homology
invariants involving $E_{1|2|3,4,5,6}$ with a double pole.
As detailed in section \ref{sec:2.3.2}, the naive absorption of $\partial_{z_1} g^{(1)}_{12}$ from the 
expression (\ref{Esat6pt}) for $E_{1|2|3,4,5,6}$ into the total
Koba-Nielsen derivative in (\ref{naiveE}) may result in a 
boundary term. After a detailed description of the non-vanishing boundary terms in
section \ref{sec:4.1}, we shall provide a refined integration-by-parts strategy 
in section \ref{sec:wayout} that eliminates the double pole and reduces
the homology invariant $E_{1|2|3,4,5,6}$ to combinations of the simpler
ones. As a net result, there is no need for independent pinching rules that
address the double poles in six-point amplitudes.

%%%%%%%%%%%%%%%%%%%%%%%%%%%%%%%%%%%%%%%%%%%%%%%% 
%%%%%%%%%%%%%%%%%%%%%%%%%%%%%%%%%%%%%%%%%%%%%%%% 

\subsubsection{Boundary terms obstructing a naive integration by parts}
\label{sec:4.1}

It is tempting to absorb the combination $\partial_{z_1} g^{(1)}_{12}+ \alpha' s_{12}( g^{(1)}_{12})^2$ 
with a double pole in $E_{1|2|3,4,5,6}$ into the
total derivative of $g^{(1)}_{12} {\cal J}_6$ which would lead to
\beq
E^{\rm naive}_{1|2|3,4,5,6} = 2\pi i \ell\cdot k_2  g^{(1)}_{12}
-  s_{12} g^{(2)}_{12}+ \tfrac{1}{2} \sum_{j=3}^6  s_{2j}g^{(1)}_{2j}
\label{enaiv}
\eeq
see (\ref{naiveE}). However, the $B$-cycle monodromy $g^{(1)}(z{+}\tau)- g^{(1)}(z)= -2\pi i$
implies that the total derivative $g^{(1)}_{12} {\cal J}_6$ integrates to a non-trivial boundary
term within closed-string amplitudes and cannot be discarded: In the context of the
amplitude prescription (\ref{baspin.13}), this boundary term due to rewritings of ${\cal K}_6$ can be
understood from
\beq
{\cal B}_2(z_3,\ldots,z_6, \tau) =2i \int_{\mathbb R^D}\dd^D \ell \int_{{\cal T}_\tau} \dd^2 z_2
 \,  \partial_{z_2} \big( g^{(1)}_{12} \overline{ {\cal K}_6 } |  {\cal J}_6 |^2 \big)
\label{bterm.01}
\eeq
which enters $M^{\textrm{1-loop}}_6$ upon further integration 
over $ \tau$ and the remaining $z_i$. The $z_2$-derivative commutes with
the antimeromorphic functions
$ \overline{ {\cal K}_6 }  \overline{ {\cal J}_6 }$ (apart from delta distributions from
simple poles in $\bar z_{2i}$ that are suppressed by the local behavior
$\sim |z_{2i}|^{2\alpha' s_{2i}}$ of the Koba-Nielsen factor). 
By Stokes' theorem, the
$z_2$ integration in (\ref{bterm.01}) localizes on the boundary
$\partial {\cal T}_\tau $ of the
fundamental domain for the torus in the right panel of figure \ref{figBDY}:

\begin{figure}[t]
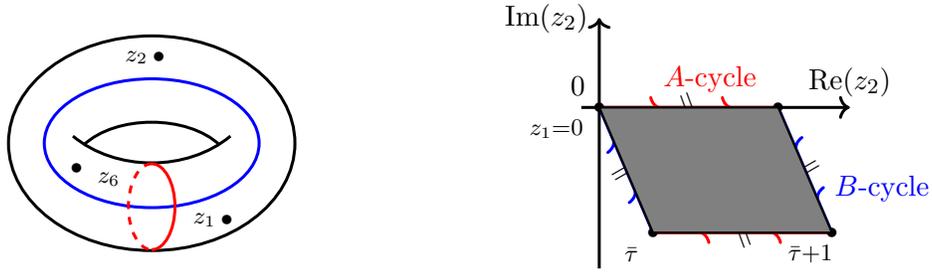

\begin{center}
\tikzpicture[scale=0.34, line width=0.40mm,scale=1.4]
\draw(0,0) ellipse  (4cm and 3cm);
\draw(-2.2,0.2) .. controls (-1,-0.8) and (1,-0.8) .. (2.2,0.2);
\draw(-1.9,-0.05) .. controls (-1,0.8) and (1,0.8) .. (1.9,-0.05);
\draw[blue](0,0) ellipse  (3cm and 1.8cm);
\draw[red] (0,-2.975) arc (-90:90:0.65cm and 1.2cm);
\draw[red,dashed] (0,-0.575) arc (90:270:0.65cm and 1.2cm);
\draw(2.1,-2.15)node{\footnotesize $\bullet$}node[left]{\footnotesize $z_1$};
\draw(0.2,2.4)node{\footnotesize $\bullet$}node[left]{\footnotesize $z_2$};
\draw(-2.1,-0.7)node{\footnotesize $\bullet$};
\draw(-1.2,-1)node{\footnotesize $z_6$};
\scope[xshift=12.5cm,yshift=1cm]
\draw[->](-0.5,0) -- (7,0) node[above]{${\rm Re}(z_2)$};
\draw[->](0,-4.5) -- (0,2.5) node[left]{${\rm Im}(z_2)$};
\draw[red](3.1,0.8)node{$A$-cycle};
\draw[blue](7.9,-2.25)node{$B$-cycle};
\draw[blue](0,0) -- (1.5,-3.5);
\draw[blue,<-](1.5*0.26,-3.5*0.26) -- (1.5*0.27,-3.5*0.27);
\draw[blue,<-](1.5*0.66,-3.5*0.66) -- (1.5*0.67,-3.5*0.67);
\draw[red](0,0) -- (5,0);
\draw[red,->](5*0.73,0) -- (5*0.74,0);
\draw[red,->](5*0.33,0) -- (5*0.34,0);
\draw[red](1.5,-3.5) -- (6.5,-3.5);
\draw[red,<-](1.5+0.66*5,-3.5) -- (1.5+0.67*5,-3.5);
\draw[red,<-](1.5+0.26*5,-3.5) -- (1.5+0.27*5,-3.5);
\draw[blue](5,0) -- (6.5,-3.5);
\draw[blue,->](5+0.33*1.5,-0.33*3.5) -- (5+0.34*1.5,-0.34*3.5);
\draw[blue,->](5+0.73*1.5,-0.73*3.5) -- (5+0.74*1.5,-0.74*3.5);
\draw(0,0)node{\footnotesize $\bullet$};
\draw(-0.6,0.6)node{$0$};
\draw(-1.2,-0.6)node{\footnotesize $z_1{=}0$};
\draw(0.75,-1.75)node[rotate=-60]{$| \; \! \!|$};
\draw (1.5,-3.5)node{\footnotesize $\bullet$} ;
\draw(0.9,-4.1)node{\footnotesize $\bar \tau$};
\draw(2.5,0)node[rotate=20]{$| \; \! \! |$};
\draw (5,0)node{\footnotesize $\bullet$};
\draw (4.4,-0.6)node{\footnotesize $1$};
\draw(4,-3.5)node[rotate=20]{$| \; \! \! |$};
\draw(5.75,-1.75)node[rotate=-60]{$| \; \! \! |$};
\draw(6.5,-3.5)node{\footnotesize $\bullet$};
\draw(5.9,-4.1)node{\footnotesize $\bar \tau{+}1$};
\draw(1.3,-1)node{\footnotesize $\bullet$}node[right]{\footnotesize $\bar z_6$};
\draw(5,-2.25)node{\footnotesize $\bullet$}node[left]{\footnotesize $\bar z_3$};
\draw(5.9/2,-4.1/2)node{\footnotesize $\ddots$};
\draw[fill=gray, opacity=0.2, line width=0.3mm] (0, 0) -- (1.5,-3.5) -- (6.5,-3.5) -- (5,0) -- cycle;
\endscope
%%%%%%%%%%%%
\endtikzpicture
\caption{Decomposition of the boundary $\partial {\cal T}_\tau$ of the fundamental domain of the torus into straight paths from $0$ to $1$ and from $\bar \tau{+}1$ to $\bar \tau$ ($A$-cycle, drawn in red) as well as straight paths from $1$ to $\bar \tau{+}1$ and from $\bar \tau$ to $0$ ($B$-cycle, drawn in blue).}
\label{figBDY}
\end{center}
\end{figure}

\begin{align}
\label{bterm.02}
{\cal B}_2(z_3,\ldots,z_6, \tau)  &=
\int_{\mathbb R^D}\dd^D \ell 
\int_{ \partial {\cal T}_\tau} \dd \bar z_2 \;  g^{(1)}_{12} \overline{ {\cal K}_6 } |  {\cal J}_6 |^2 \\
 &= \int_{\mathbb R^D}\dd^D \ell \,
  \bigg( \int_0^1  \dd \bar z_2  \, g^{(1)}_{12} \overline{ {\cal K}_6 } |  {\cal J}_6 |^2
  + \int_1^{\bar \tau+1} \dd \bar z_2  \, g^{(1)}_{12} \overline{ {\cal K}_6 } |  {\cal J}_6 |^2 \notag \\
  &\quad \quad \quad\quad \quad
   + \int_{\bar \tau+1}^{\bar \tau}  \dd \bar z_2  \, g^{(1)}_{12} \overline{ {\cal K}_6 } |  {\cal J}_6 |^2
   + \int_{\bar \tau}^{0}  \dd \bar z_2  \, g^{(1)}_{12} \overline{ {\cal K}_6 } |  {\cal J}_6 |^2\bigg) 
 \nonumber  
 \\
 &=  \int_{\mathbb R^D}\dd^D \ell \,
\bigg( \int_0^1  \dd \bar z_2  \, \overline{ {\cal K}_6 } |  {\cal J}_6 |^2 \big[ g^{(1)}(z_{12})- g^{(1)}(z_{12}{-} \tau)  \big] \notag \\
  &\quad \quad \quad\quad \quad
- \int_0^{\bar \tau}  \dd \bar z_2  \, \overline{ {\cal K}_6 } |  {\cal J}_6 |^2 \big[ g^{(1)}(z_{12})- g^{(1)}(z_{12}{-} 1)  \big]   \bigg)
 \nonumber  \\
 &= - 2\pi i  \int_{\mathbb R^D}\dd^D \ell \,
 \int_0^1  \dd \bar z_2  \, \overline{ {\cal K}_6 } |  {\cal J}_6 |^2 \notag
\end{align}
The non-vanishing of this boundary integral has been exposed by the following three steps:
\begin{itemize}
\item[(i)] decomposed the boundary $\partial {\cal T}_\tau$ into the four straight lines of figure \ref{figBDY} in passing to the second line of (\ref{bterm.02})
\item[(ii)] related the contributions $\int_1^{\bar \tau+1}$ and $\int_{\bar \tau+1}^{\bar \tau}$ to
those over $\int_0^{\bar \tau}$ and $\int_{1}^{0}$ up to shifts of the integrand by $1$ and $\tau$,
respectively, in passing to the fourth line
\item[(iii)] used the monodromies $g^{(1)}(z_{12})- g^{(1)}(z_{12}{-} 1) =0$ 
and $g^{(1)}(z_{12})- g^{(1)}(z_{12}{-} \tau) = -2\pi i$ in passing to the sixth line
\end{itemize}
Moreover, we have also used in step (ii) that the factor of $ \overline{ {\cal K}_6 } |  {\cal J}_6 |^2$ is
doubly-periodic upon integration over $\ell$ by homology invariance of both 
$ \overline{ {\cal K}_6 }$ and $ |  {\cal J}_6 |^2$. Similar types of non-vanishing boundary terms
can already be encountered at five points: One cannot trade $s_{12}g^{(1)}_{12} \overline{g^{(1)}_{12}}$
for $g^{(1)}_{12}({-}4\pi i \ell {\cdot} k_2+\sum_{j=3}^5 s_{2j} \overline{g^{(1)}_{2j}} )$ 
within Koba-Nielsen integrals since the difference is the total derivative
$\partial_{\bar z_2} ( g^{(1)}_{12}  |  {\cal J}_5 |^2 )$ of a multivalued function of $z_2$.

More generally, the boundary-term analysis in (\ref{bterm.02}) can be straightforwardly adapted to
show that total Koba-Nielsen derivatives of multi-valued functions on the torus (such as $g^{(1)}_{12}$) cannot be
discarded from closed-string loop amplitudes. Accordingly, chiral correlators ${\cal K}_n$ can only
be shifted by $z_i$-derivatives of homology invariants, for instance $g^{(1)}_{ij}{+}g^{(1)}_{jk}{+}g^{(1)}_{ki}$
or $\ell^m + \sum_{j\geq 2} k_j^m g^{(1)}_{1j}$ rather than individual $g^{(1)}_{ij}$. 
Our reasoning for this boundary term was already borrowed in the analysis of basis 
decompositions for more general Koba-Nielsen integrals at genus one
in section 5.1.2 of \cite{Rodriguez:2023qir}.

Note that this subtlety in integration-by-parts manipulations is specific to chiral splitting:
After performing the Gaussian loop integral of the string amplitudes (\ref{baspin.13}) and
completing the meromorphic $g^{(k)}_{ij}$ to doubly-periodic $f^{(k)}_{ij}$, one can discard
total Koba-Nielsen derivatives involving individual $f^{(k)}_{ij}$. However, the non-holomorphicity
$\partial_{\bar z_i} f^{(k)}_{ij} = - \frac{\pi}{\Im \tau} f^{(k-1)}_{ij}$ following from (\ref{baspin.42}) 
implies that integration by parts of the left-movers affect those of the right-movers. 
Hence, chiral splitting features closely related drawbacks and advantages: On the one hand,
discarding total derivatives before loop integration is tied to homology invariance; on the other
hand, integration-by-parts simplifications of ${\cal K}_n$
and $\overline{{\cal K}_n}$ can be performed independently at the level of the loop integrand.

It would be interesting to compare the new pinching rules of section \ref{sec:3.1} and the boundary-term discussion of this section with the monodromy relations of genus-one string amplitudes \cite{Tourkine:2016bak, Hohenegger:2017kqy, Casali:2019ihm, Casali:2020knc}. In particular, it could be worthwhile to 
investigate a connection of the pinches due to the poles of $g^{(k\geq 2)}(z_{ij})$ at $z_{ij}=\pm \tau$
with the triangle-like pinched graphs due to open-string insertions on 
the bulk cycles of the monodromy contours in the annulus in Figure 5 of \cite{Casali:2020knc}. The reference
also discusses the importance of these pinched graphs for the double copy which invites comparisons with 
the representation of the six-point supergravity amplitude in section \ref{sec:5.3} below.

%%%%%%%%%%%%%%%%%%%%%%%%%%%%%%%%%%%%%%%%%%%%%%%% 
%%%%%%%%%%%%%%%%%%%%%%%%%%%%%%%%%%%%%%%%%%%%%%%% 

\subsubsection{Integration-by-parts removal of double poles without boundary terms}
\label{sec:wayout}

The boundary-term discussion of the previous section rules out a naive integration by parts via
(\ref{naiveE}) to eliminate the double poles $\partial_{z_1} g^{(1)}_{12}+ \alpha' s_{12}( g^{(1)}_{12})^2$ 
in $E_{1|2|3,4,5,6}$. We shall now present an alternative total derivative that eliminates said double poles
in accordance with homology invariance, based on the six-point version of the five-point homology invariants~(\ref{Esat5pt}),
\begin{align}
E_{1| 23,4,5,6} &= g^{(1)}_{12}  + g^{(1)}_{23}  + g^{(1)}_{31} 
\label{bterm.07}\\
E^m_{1| 2,3,4,5,6} &= \ell^m + \big[ k_2^m g^{(1)}_{12}  + (2\leftrightarrow 3,4,5,6) \big]
\notag
\end{align}
Koba-Nielsen derivatives involving (\ref{bterm.07}) can be discarded from string amplitudes
without any boundary terms akin to (\ref{bterm.02}) and simplify
to combinations of six-point homology invariants (\ref{Esat6pt}) such as \cite{Mafra:2018pll}
\begin{align}
\partial_{z_1} \big( E_{1| 23,4,5,6}  \, {\cal J}_6\big) &= 2\alpha' ( k_1^m E^m_{1|23,4,5,6} + E_{1|2|3,4,5,6}
- E_{1|3|2,4,5,6} )  \, {\cal J}_6
\label{bterm.08}\\
\partial_{z_2} \big( E^m_{1| 2,3,4,5,6}  \, {\cal J}_6\big) &= \alpha' \big(2 k_2^p  E^{mp}_{1|2,3,4,5,6}  -2 k_2^m  E_{1|2|3,4,5,6}
+ \big[ s_{23} E^m_{1|23,4,5,6} + (3\leftrightarrow 4,5,6) \big] \big)  \, {\cal J}_6
 \notag
\end{align}
By contracting the second identity with $k_1^m$, one can solve for the homology
invariant $E_{1|2|3,4,5,6}$ involving double poles in terms of $ E^{mp}_{1|2,3,4,5,6}$
and $E^m_{1|23,4,5,6}$ with only logarithmic singularities:
\begin{align}
s_{12} E_{1|2|3,4,5,6} \cong 2 k_1^m  k_2^p  E^{mp}_{1|2,3,4,5,6} +  \big[ s_{23} k_1^m E^m_{1|23,4,5,6} + (3\leftrightarrow 4,5,6) \big] 
\label{bterm.09}
\end{align}
This equivalence up to Koba-Nielsen derivatives of homology invariants indicated by $\cong$
can be used to infer field-theory limits of $E_{1|2|3,4,5,6}$ from those of the right-hand side
which are available from sections \ref{sec:3.1} to \ref{sec:3.3}.
Field-theory limits of permutations $E_{1|j|k,p,q,r}$ can be obtained from relabelings
of (\ref{bterm.09}) or reduced to those of $E_{1|2|3,4,5,6}$ via the first line of~(\ref{bterm.08}).

%%%%%%%%%%%%%%%%%%%%%%%%%%%%%%%%%%%%%%%%%%%%%%%% 
%%%%%%%%%%%%%%%%%%%%%%%%%%%%%%%%%%%%%%%%%%%%%%%% 

\subsection{Field-theory limits involving double poles}
\label{sec:FTE}

This section (together with appendix \ref{app:Eex.2}) gathers the six-point field-theory limits 
involving the homology invariant $E_{1|2|3,4,5,6}$ with a double pole
which follow from the improved integration by parts in (\ref{bterm.09}).

%%%%%%%%%%%%%%%%%%%%%%%%%%%%%%%%%%%%%%%%%%%%%%%% 
%%%%%%%%%%%%%%%%%%%%%%%%%%%%%%%%%%%%%%%%%%%%%%%% 

\subsubsection{Open strings}
\label{sec:4.2}

Based on (\ref{bterm.09}) together with permutations of the simpler field-theory 
limits in (\ref{newPR.09}), we obtain the following samples of cyclically inequivalent cases:
\begin{align}
	{\rm FT}^{\rm op}_{12\ldots 6}\big[ E_{1\vert 2\vert 3,4,5,6}\big]&=\frac{1}{24}s_{12}I^{(6)}_{1,2,3,4,5,6}+\frac{1}{2}I^{(5)}_{12,3,4,5,6}+\frac{1}{2s_{12}}(\ell{-}k_1)^2I^{(5)}_{12,3,4,5,6}\nonumber\\
	&\quad-\frac{s_{13}}{2s_{12}s_{123}}I^{(4)}_{123,4,5,6}-\frac{s_{26}}{2s_{12}s_{126}}I^{(4)}_{6\underline12,3,4,5}\nonumber\\
	{\rm FT}^{\rm op}_{12\ldots 6}\big[ E_{1\vert 3\vert 2,4,5,6}\big]&=\frac{1}{24}s_{13}I^{(6)}_{1,2,3,4,5,6}-\frac{1}{2s_{123}}I^{(4)}_{123,4,5,6}\nonumber\\
	{\rm FT}^{\rm op}_{12\ldots 6}\big[ E_{1\vert 6\vert 2,3,4,5}\big]&=\frac{1}{24}s_{16}I^{(6)}_{1,2,3,4,5,6}+\frac{1}{2}I^{(5)}_{6\underline1,2,3,4,5}+\frac{1}{2s_{16}}\ell^2I^{(5)}_{6\underline1,2,3,4,5}\nonumber\\
	&\quad -\frac{s_{26}}{2s_{16}s_{126}}I^{(4)}_{6\underline12,3,4,5}-\frac{s_{15}}{2s_{16}s_{156}}I^{(4)}_{56\underline1,2,3,4}
	\label{bterm.10}
\end{align}
Note that, in planar open-string amplitudes, the total derivative $\partial_{z_2}g^{(1)}_{12} {\cal J}_6$ integrates to zero for all color-orderings since the boundary terms $z_2 \rightarrow z_i$ or $z_2 \rightarrow z_i{+}\tau$ are suppressed by the local behavior $| z_{2i} |^{\alpha' s_{2i}} $ and $| z_{2i} {-} \tau|^{\alpha' s_{2i}} $ of the Koba-Nielsen factor. In other words, the boundary terms (\ref{bterm.02}) obstructing the naive integration by parts (\ref{naiveE}) in $E_{1|2|3,4,5,6}$ are a peculiarity of closed strings. At any rate, manipulations of chiral amplitudes involving total derivatives of $g^{(1)}_{ij}{\cal J}_n$ undermine their universal applicability to open and closed strings, which is the original motivation for chiral splitting.

On these grounds, the open-string field-theory limits (\ref{bterm.10}) can be used to cross-check the improved integration by parts (\ref{bterm.09}): We have reproduced the expressions on the right-hand side of (\ref{bterm.10}) by applying the pinching rules of earlier sections to $E^{\rm naive}_{1|2|3,4,5,6}$ 
in (\ref{enaiv}) with logarithmic singularities instead of $E_{1|2|3,4,5,6}$.

%%%%%%%%%%%%%%%%%%%%%%%%%%%%%%%%%%%%%%%%%%%%%%%% 
%%%%%%%%%%%%%%%%%%%%%%%%%%%%%%%%%%%%%%%%%%%%%%%% 

\subsubsection{Closed strings}
\label{sec:4.3}

For closed strings, the Koba-Nielsen integrals of $E^{\rm naive}_{1|2|3,4,5,6}$ and $E_{1|2|3,4,5,6}$
differ by boundary terms with non-zero field-theory limits. That is why one has to employ the improved
integration by parts (\ref{bterm.09}) to arrive at the following results and cannot compare with any field-theory limit of (\ref{enaiv}):
\begin{align}
	{\rm FT}^{\rm cl}_6\big[E_{1\vert 2\vert 3, 4, 5,6}\bar E_{1\vert 23, 4 5,6}\big]&=
	\frac{s_{12}}{96}
	\sum_{\rho \in S_{\{2,3,4,5,6\}}} I_{1,\rho(2,3,4,5,6)}^{(6)}\sign^\rho_{23}\sign^\rho_{45}\label{bterm.11}\\
	&\quad- \frac{1}{8 s_{12}} \sum_{\rho \in S_{\{3,4,5,6\}}}\bigg((\ell-k_1)^2 I_{12,\rho(3,4,5,6)}^{(5)}-\ell^2I_{2\underline1,\rho(3,4,5,6)}^{(5)}\bigg)\sign^\rho_{45}\nonumber\\
	&\quad+\biggl[
	\bigg( \frac{s_{24}}{2s_{12}s_{124}} - \frac{1}{4s_{12}} \bigg)
	\sum_{\rho \in S_{\{3,5,6\}}} I_{124,\rho(3,5,6)}^{(4)}-(4\leftrightarrow 5)\biggr]
\notag \\
%%%
	{\rm FT}^{\rm cl}_6\big[E_{1\vert 2\vert 3, 4, 5,6}\bar E_{1\vert 234,5,6}\big]&=
	\frac{s_{12}}{288}   \sum_{\rho \in S_{\{2,3,4,5,6\}}}    I_{1,\rho(2,3,4,5,6)}^{(6)}(1{+}3\sign^\rho_{23}\sign^\rho_{34}) 	
	\nonumber\\
	&\quad+ \frac{1}{24} \sum_{\rho \in S_{\{3,4,5,6\}}} I_{12,\rho(3,4,5,6)}^{(5)}\nonumber\\
	&\quad - \frac{1}{8 s_{12}} \sum_{\rho \in S_{\{3,4,5,6\}}}\bigg((\ell-k_1)^2 I_{12,\rho(3,4,5,6)}^{(5)}-\ell^2I_{2\underline1,\rho(3,4,5,6)}^{(5)}\bigg)\sign_{34}^\rho\nonumber\\
	&\quad+\bigg(  \frac{1}{4s_{12}} -  \frac{s_{13} }{2s_{12}s_{123}}  \bigg)
	\sum_{\rho \in S_{\{4,5,6\}}} I_{123,\rho(4,5,6)}^{(4)}\nonumber\\
	&\quad+ \bigg(   \frac{1}{4s_{12}}  -  \frac{s_{24}}{2s_{12}s_{124}} \bigg)
	\sum_{\rho \in S_{\{3,5,6\}}} I_{124,\rho(3,5,6)}^{(4)}
	\notag
\end{align}
as well as
\begin{align}
{\rm FT}^{\rm cl}_6\big[E_{1\vert 2\vert 3, 4, 5,6}\bar E_{1\vert 3\vert 2, 4, 5,6}\big]&=
	\frac{s_{12}s_{13}}{576}
	\sum_{\rho \in S_{\{2,3,4,5,6\}}}I_{1,\rho(2,3,4,5,6)}^{(6)}  \label{bterm.12}\\
	&\quad \hspace{-3cm} +\bigg[ \frac{s_{13}}{48} \sum_{\rho \in S_{\{3,4,5,6\}}}I_{12,\rho(3,4,5,6)}^{(5)}+(2\leftrightarrow3)\bigg] + \frac{s_{12}{+}s_{13}}{4s_{123}}
	\sum_{\rho \in S_{\{4,5,6\}}}I_{123,\rho(4,5,6)}^{(4)} \notag \\
%%%%%%%%%
%%%%%%%%%
{\rm FT}^{\rm cl}_6\big[E_{1\vert 2\vert 3, 4, 5,6}\bar E_{1\vert 2\vert 3, 4, 5,6}\big]&=
	\frac{s_{12}^2}{576} \sum_{\rho \in S_{\{2,3,4,5,6\}}}I_{1,\rho(2,3,4,5,6)}^{(6)} 
	- \frac{s_{12}}{12}\sum_{\rho \in S_{\{3,4,5,6\}}}I_{12,\rho(3,4,5,6)}^{(5)} \notag  \\
	&\quad \hspace{-3cm}+\bigg[
	\left(
\frac{ s_{123}}{4 s_{12}}	
+ \frac{(s_{13}{+}s_{23})}{8 s_{123}}
+ \frac{ s_{13}^2{+}s_{23}^2}{8 s_{12} s_{123}}	
	-\frac{\ell {\cdot} k_{123}}{2s_{12}}\right)
	\sum_{\rho \in S_{\{4,5,6\}}} \! \! I_{123,\rho(4,5,6)}^{(4)}  +(3\leftrightarrow 4,5,6)\bigg]
\notag
\end{align}
and
\begin{align}
{\rm FT}^{\rm cl}_6\big[E_{1\vert 2\vert 3, 4, 5,6}\bar E_{1\vert 23,4,5,6}^p\big]&=
	-\frac{s_{12}}{48}  \big(\ell^p{+} \tfrac{1}{2}  k_{23456}^p \big) \sum_{\rho \in S_{\{2,3,4,5,6\}}} I_{1,\rho(2,3,4,5,6)}^{(6)}\sign_{23}^\rho 
	\label{bterm.13} \\
&\quad  +\frac{s_{12}}{288} (k_3^p{-}k_2^p) \sum_{\rho \in S_{\{2,3,4,5,6\}}}
I_{1,\rho(2,3,4,5,6)}^{(6)}  \notag \\
	&\quad+ \frac{1}{24}(2k_2^p+k_3^p)  \sum_{\rho \in S_{\{3,4,5,6\}}}I_{12,\rho(3,4,5,6)}^{(5)}\nonumber\\
	&\quad+ \frac{1}{4 s_{12}}\sum_{\rho \in S_{\{3,4,5,6\}}}\bigg( \big(\ell^p+ \tfrac{1}{2} k_{3456}^p \big)(\ell{-}k_1)^2I_{12,\rho(3,4,5,6)}^{(5)} \notag \\
	&\quad\quad\quad\quad\quad\quad\quad\quad- \big( \ell^p  + \tfrac{1}{2} k_{3456}^p+k_2^p\big)\ell^2
	I_{2\underline1,\rho(3,4,5,6)}^{(5)}\bigg)\nonumber\\
	&\quad+
	\bigg( \frac{k_2^p}{2s_{123}}  + \frac{k_3^p }{4 s_{12}}   - \frac{s_{13} k_3^p }{2 s_{12}s_{123}} \bigg) 
		\sum_{\rho \in S_{\{4,5,6\}}}I_{123,\rho(4,5,6)}^{(4)}\nonumber\\
	&\quad-\bigg[
	\bigg(  \frac{1}{4 s_{12}} -  \frac{s_{24}}{2 s_{12}s_{124}}  \bigg) k_4^p
	\sum_{\rho \in S_{\{3,5,6\}}}I_{124,\rho(3,5,6)}^{(4)}+(4\leftrightarrow 5,6)\bigg]
	\notag
\end{align}
Additional closed-string field-theory limits involving $E_{1\vert 2\vert 3, 4, 5,6}$
can be found in appendix \ref{app:Eex.2}, and the complete set of $ {\rm FT}^{\rm cl}_{ 6}[ E \bar E]$
is provided in the supplementary material. We reiterate that the
six-point supergravity loop integrand in section \ref{sec:5.3} below takes a considerably
simpler form than one might anticipate from the size of individual field-theory limits. 

%%%%%%%%%%%%%%%%%%%%%%%%%%%%%%%%%%%%%%%%%%%%%%%% 
%%%%%%%%%%%%%%%%%%%%%%%%%%%%%%%%%%%%%%%%%%%%%%%% 

\section{Six-point applications}
\label{sec:5}

The methods of the previous section determine the field-theory limits of all open-
and closed-string integrals over the homology invariants up to six points in (\ref{KK6GEF}).
In this section, we shall apply these field-theory limits to construct a new representation of the six-point one-loop 
amplitude of type-IIA/B supergravity in pure-spinor superspace. In contrast to 
earlier manifestly supersymmetric expressions in \cite{He:2016mzd, He:2017spx}, the result of this 
section is expressed in terms of Feynman propagators quadratic in the loop momentum.
Each contribution to our six-point loop integrand takes a double-copy form -- most terms 
corresponding to the cubic diagrams of the BCJ double copy 
\cite{Bern:2019prr, Bern:2022wqg, Adamo:2022dcm} and other terms
representing contact diagrams with a quintic vertex as in the \textit{generalized double copy} \cite{Bern:2017yxu}.

%%%%%%%%%%%%%%%%%%%%%%%%%%%%%%%%%%%%%%%%%%%%%%%% 
%%%%%%%%%%%%%%%%%%%%%%%%%%%%%%%%%%%%%%%%%%%%%%%% 

\subsection{Assembly of one-loop field-theory amplitudes}
\label{sec:5.1}

The field-theory amplitudes $A^{\textrm{1-loop}}_{\rm SYM}(1,2,\ldots,n) $
and $M^{\textrm{1-loop}}_{{\rm SG}, \, n} $ obtained from the $\alpha' \rightarrow 0$ limit
of one-loop superstring amplitudes take the following form upon
combining (\ref{baspin.13}) and (\ref{attempt1.0}) 
\begin{align}
A^{\textrm{1-loop}}_{\rm SYM}(1,2,\ldots,n) &=   \int_{\RR^D} \dd^{D} \ell 
\,  {\rm FT}^{\rm op}_{12\ldots n}\big[ {\cal K}_n \big]
\label{ftamp.01} \\
M^{\textrm{1-loop}}_{{\rm SG}, \, n} &=   \int_{\RR^D} \dd^{D} \ell    \,
 {\rm FT}^{\rm cl}_{ n}\big[ {\cal K}_n \overline{ {\cal K}_n  } \big]
\notag
\end{align}
with chiral correlators ${\cal K}_{n\leq 6}$ given by (\ref{KK6GEF}). The
superspace kinematic factors in ${\cal K}_{n}$ (briefly reviewed in section \ref{sec:2.3.4}
and appendix \ref{app:kin}) can be pulled out of the $ {\rm FT}[\ldots]$ prescriptions (\ref{attempt1.0}), for instance
\begin{align}
A^{\textrm{1-loop}}_{\rm SYM}(1,2,3,4,5) &=   \int_{\RR^D} \dd^{D} \ell 
\, \Big\{
C^m_{1| 2,3,4,5} {\rm FT}^{\rm op}_{12345}\big[ E^m_{1| 2,3,4,5}  \big] \label{ftamp.02} \\
&\quad
+ \tfrac{1}{2} s_{23} C_{1| 23,4,5} {\rm FT}^{\rm op}_{12345}\big[ E_{1| 23,4,5}  \big]
+ \tfrac{1}{2} s_{24} C_{1| 24,3,5} {\rm FT}^{\rm op}_{12345}\big[ E_{1| 24,3,5}  \big]
\notag \\
&\quad
+ \tfrac{1}{2} s_{25} C_{1| 25,3,4} {\rm FT}^{\rm op}_{12345}\big[ E_{1| 25,3,4}  \big]
+ \tfrac{1}{2} s_{34} C_{1| 34,2,5} {\rm FT}^{\rm op}_{12345}\big[ E_{1| 34,2,5}  \big]
\notag \\
&\quad
+ \tfrac{1}{2} s_{35} C_{1| 35,2,4} {\rm FT}^{\rm op}_{12345}\big[ E_{1| 35,2,4}  \big]
+ \tfrac{1}{2} s_{45} C_{1| 45,2,3} {\rm FT}^{\rm op}_{12345}\big[ E_{1| 45,2,3}  \big]
\Big\}
\notag
\end{align}
The translation of the leftover ${\rm FT}^{\rm op}_{12\ldots n}[ E ]$
 or $ {\rm FT}^{\rm cl}_{ n}[ E \bar E]$ into loop-momentum dependent propagators
 is discussed in section \ref{sec:3}, see the supplementary material of this work for the complete set of six-point open- and closed-string results. Upon inserting 
 these field-theory limits obtained from the new pinching rules of this work, we recover the
 manifestly supersymmetric representations of \cite{towardsOne} for the 
 super-Yang-Mills amplitudes
 \begin{align}
 A^{\textrm{1-loop}}_{\rm SYM}(1,2,\ldots,5) &=  
   \int_{\RR^D}  \! \dd^{D} \ell \, \bigg\{
 {1\over 2 } \, C_{1|23,4,5}  I^{(4)}_{1,23,4,5}
+ {1\over 2} \, C_{1|34,2,5} I^{(4)}_{1,2,34,5}
\label{ftamp.03} \\
&\quad\quad\quad\quad \ \ \ \ \ \,
  + {1\over 2} \, C_{1|45,2,3}  I^{(4)}_{1,2,3,45}
+ ( C_{1|2;3;4;5} + \ell_m C^m_{1|2,3,4,5} )  I^{(5)}_{1,2,3,4,5}
 \bigg\} \notag \\
 %%%
 %%%
  A^{\textrm{1-loop}}_{\rm SYM}(1,2,\ldots,6) &=   \int_{\RR^D} \! \dd^{D} \ell \, \bigg\{
\frac{1}{4}  \, \Big[ C_{1|234,5,6}  I^{(4)}_{1,234,5,6} + C_{1|2,345,6}  I^{(4)}_{1,2,345,6}  + C_{1|2,3,456}   I^{(4)}_{1,2,3,456} \notag \\
&\quad\quad\quad\quad \ \ \ \ \ \,
+ C_{1|23,45,6}   I^{(4)}_{1,23,45,6} + C_{1|23,4,56}  I^{(4)}_{1,23,4,56}  +C_{1|2,34,56}   I^{(4)}_{1,2,34,56} \Big]  \notag \\
%%%%%%
&\quad \hspace{-1.6cm}+ \frac{1}{2} \, \Big[  (C_{1|23;4;5;6} + \ell_m C^m_{1|23,4,5,6})   I^{(5)}_{1,23,4,5,6}  + (C_{1|2;34;5;6} + \ell_m C^m_{1|2,34,5,6})  I^{(5)}_{1,2,34,5,6} \notag \\
&\quad \hspace{-1.8cm} \ \ \  \ \ \, + (C_{1|2;3;45;6} + \ell_m C^m_{1|2,3,45,6})  I^{(5)}_{1,2,3,45,6}+ (C_{1|2;3;4;56} + \ell_m C^m_{1|2,3,4,56})  I^{(5)}_{1,2,3,4,56}\Big] \notag \\
%%%%%%%
&\quad \hspace{-1.6cm} +\big( C_{1|2;3;4;5;6} + \ell_m C^m_{1|2;3;4;5;6}  + \tfrac{1}{2} \ell_m \ell_n C_{1|2,3,4,5,6}^{mn}  \big)  I^{(6)}_{1,2,3,4,5,6}  - \frac{1}{2}   P_{1|6|2,3,4,5}   I^{(5)}_{6\underline{1},2,3,4,5}  \bigg\} \notag
 \end{align}
 and the five-point supergravity amplitude
 \begin{align}
M^{\textrm{1-loop}}_{{\rm SG}, \, 5}  &= 
 \int_{\RR^D} \! \dd^{D} \ell \, \bigg\{ \Big[   \, \big|   C_{1|2;3;4;5} + \ell_m C^m_{1|2,3,4,5}  \big|^2\,  I^{(5)}_{1,2,3,4,5} + {\rm perm}(2,3,4,5) \, \Big]    \label{ftamp.04}  \\
&\quad\quad\quad\quad\ \ \! + \frac{1}{4}  \, \Big[  s_{23} |C_{1|23,4,5}|^2  \sum_{\rho \in S_{\{23,4,5\}}}I_{1,\rho(23,4,5)}^{(4)}+ (2,3|2,3,4,5) \Big] \bigg\}
\notag
  \end{align}
also see \cite{Bjerrum-Bohr:2008vag} for earlier expressions for the bosonic components 
in generic spacetime dimension and \cite{Bern:1994zx, KLTbern, Carrasco:2011mn, 
Bjerrum-Bohr:2013iza, He:2015wgf} for
four-dimensional spinor-helicity expressions. The superspace kinematic factors
with semicolons in their subscript are composite \cite{towardsOne}:
scalar pentagon numerators 
 \begin{align}
4C_{1| 2;3;4;5} &=  s_{23} C_{1|23,4,5} + (2,3 | 2,3,4,5)
  \label{ftamp.05} 
  \\
4C_{1|23;4;5;6} &=   s_{45} C_{1|23,45,6}  
{+} s_{46} C_{1|23,46,5} {+} s_{56} C_{1|23,56,4}  + \big[ s_{34} C_{1|234,5,6} {-} s_{24} C_{1|324,5,6} {+} (4\leftrightarrow 5,6) \big]  \notag \\
  4C_{1|2;34;5;6} &= 
   s_{56} C_{1|2,34,56}  +s_{25} C_{1|25,34,6} + s_{26} C_{1|26,34,5}  \notag \\
&\quad   + \big[ s_{23} C_{1|234,5,6} + s_{45} C_{1|2,345,6} + s_{46} C_{1|2,346,5}
    - (3\leftrightarrow 4) \big]
     \notag \\
4C_{1|2;3;45;6} &=   s_{23} C_{1|23,45,6} + s_{26} C_{1|26,45,3} + s_{36} C_{1|2,36,45} \notag \\
&\quad + \big[ s_{24} C_{1|245,3,6} + s_{34} C_{1|345,2,6} + s_{56} C_{1|2,3,456} -(4\leftrightarrow 5) \big] 
 \notag \\
4C_{1|2;3;4;56} &=   s_{23} C_{1|23,4,56} {+} s_{24} C_{1|24,3,56} {+} s_{34} C_{1|2,34,56}  
+ \big[ s_{45} C_{1|2,3,456} {-} s_{46} C_{1|2,3,465} {+} (4\leftrightarrow 2,3) \big] \notag
  \end{align}
as well as scalar and vector hexagon numerators
  \begin{align}
4C^m_{1|2;3;4;5;6} &= s_{23}C^m_{1|23,4,5,6} + (2,3|2,3,4,5,6) 
\label{ftamp.00}  \\
8C_{1|2;3;4;5;6} &= {1\over2} \Big[ s_{23} s_{45} C_{1| 23,45,6} +
s_{24} s_{35} C_{1| 24,35,6} +
s_{25} s_{34} C_{1| 25,34,6} + (6\leftrightarrow 5,4,3,2)  \Big]\notag \\
& \hspace{-1.5cm}+ {1\over 3} \Big[s_{23} \big( s_{34} C_{1|234,5,6} {-}  s_{24} C_{1|324,5,6} \big)
+ s_{43} \big( s_{32} C_{1|432, 5,6}{-}  s_{24} C_{1|342,5,6} \big)  + (2,3,4|2,3,4,5,6) \Big]\notag \\
&\hspace{-1.5cm} + {1\over 3} \Big[ (k_3^m \!- \!k_2^m) s_{23} C^m_{1|23,4,5,6} + (2,3|2,3,4,5,6) \Big]
- {1\over 3} C^{mn}_{1|2,3,4,5,6} \big[ k^1_m k^1_n  + (1\leftrightarrow 2,3,4,5,6) \big] \notag
\end{align}
Following our conventions for the chiral correlators,
we employ shorthands like $|C_{1|23,4,5}|^2 =  C_{1|23,4,5} \tilde C_{1|23,4,5}$
in (\ref{ftamp.04}) and section \ref{sec:5.3} below for contributions to
supergravity amplitudes in double-copy form. The tilde on the superfields
is understood to convert the gluon polarization vectors $e_i^m$ into independent
copies $\tilde e_i^m$ and similarly for the fermion wavefunctions.

Both of (\ref{ftamp.03}) and (\ref{ftamp.04}) are free of boxes with leg 1 in a massive corner 
(say $I^{(4)}_{12,3,4,5} $,  $I^{(4)}_{5\underline{1},2,3,4} $, $I^{(4)}_{12,34,5,6} $ 
or $I^{(4)}_{6\underline{1}2,3,4,5} $). When deriving the field-theory amplitudes from (\ref{ftamp.01}) and the representation (\ref{KK6GEF}) of ${\cal K}_{n}$, these cancellations are due to superspace identities such~as
\beq 
k_2^m C^m_{1|2,3,4,5} + \tfrac{1}{2} \big[ s_{23} C_{1|23,4,5} + ( 3\leftrightarrow 4,5)\big] = 0
  \label{ftamp.06} 
  \eeq
 see sections 9 and 10 of \cite{Mafra:2014gsa} for their origin from BRST-exact expressions
 and their all-multiplicity systematics. However, the six-point amplitude in (\ref{ftamp.03}) does
 feature the pentagon $ I^{(5)}_{6\underline{1},2,3,4,5} $ with leg 1 in a massive corner which
 is crucial for the hexagon anomaly of ten-dimensional super-Yang-Mills as explained
 in section 4.5 of \cite{towardsOne}.
 
%%%%%%%%%%%%%%%%%%%%%%%%%%%%%%%%%%%%%%%%%%%%%%%% 
%%%%%%%%%%%%%%%%%%%%%%%%%%%%%%%%%%%%%%%%%%%%%%%% 

\subsection{Double-copy form of one-loop six-point amplitudes in the literature}
\label{sec:5.2}
 
 The five-point super-Yang-Mills and supergravity amplitudes in (\ref{ftamp.03}) and (\ref{ftamp.04})
 are related by the BCJ double copy \cite{Bern:2019prr, Bern:2022wqg, Adamo:2022dcm} 
 since the box- and pentagon numerators in the super-Yang-Mills
 case obey the color-kinematics duality \cite{towardsOne}. However, the six-point
 super-Yang-Mills numerators in (\ref{ftamp.03}) violate the color-kinematics duality,
 so it does not admit an immediate construction of $M^{\textrm{1-loop}}_{{\rm SG}, \, 6} $ at 
 the level of cubic graphs. Instead, the literature offers several avenues to reconcile
 maximally supersymmetric one-loop six-point amplitudes with the color-kinematics duality:
\begin{itemize}
\item restriction of the external polarizations to four-dimensional MHV helicities \cite{He:2015wgf} 
\item transition from the conventional Feynman propagators $(\ell{-}K_i)^{-2}$ to linearized
ones $(K_i^2 {-} 2 K_i {\cdot} \ell)^{-1}$ as reviewed in section \ref{sec:6.1.2} below,
where the color-kinematics duality is preserved on the resulting forward limits of tree diagrams \cite{He:2017spx}
\item admitting different shifts of loop momenta by $k_{i\ldots j}$ in the bookkeeping of
cubic graphs as in \cite{Bridges:2021ebs}, where the color-kinematics duality holds in 
superspace on quadratic propagators but does not yield the supergravity amplitude
upon double copy\footnote{As detailed in section 3.4.2 of \cite{Bridges:2021ebs}, the assignment
of separate numerators to cubic graphs with different shifts of loop momenta does
not preserve the automorphism properties of the graphs which obstructs the double
copy towards supergravity.}
\item employing the color-kinematics dual numerators of \cite{Edison:2022jln} with contributions 
$\sim (\ell{-}k_{12\ldots i})^{2}$ to the
hexagon numerator which are so far only known for bosonic external states
\end{itemize} 
Additional options to obtain the one-loop six-point supergravity integrand in superspace 
from the super-Yang-Mills amplitude in (\ref{ftamp.03}) are the one-loop field-theory KLT formula 
\cite{He:2016mzd} and the generalized double copy \cite{Bern:2017yxu}. However, both of
these options have their own drawbacks: the one-loop KLT formula for supergravity loop integrands
is only available at the level of linearized propagators,\footnote{It would be interesting to
extract a quadratic-propagator formulation of the one-loop {\it field-theory} KLT formula 
\cite{He:2016mzd} from the recent work on KLT formulae for one-loop {\it string} amplitudes
\cite{Stieberger:2022lss, Stieberger:2023nol, Bhardwaj:2023vvm, Mazloumi:2024wys}.} and the generalized double copy 
would make the construction of the contact terms from the prescription in \cite{Bern:2017yxu}
somewhat laborious. 
Hence, we will follow a different approach and compute $M^{\textrm{1-loop}}_{{\rm SG}, \, 6} $ 
from the $\alpha' \rightarrow 0$ limit of the closed-string amplitude, in particular from the field-theory
limits ${\rm FT}^{\rm cl}_{ 6}[ E \bar E]$ derived in section~\ref{sec:3}.

%%%%%%%%%%%%%%%%%%%%%%%%%%%%%%%%%%%%%%%%%%%%%%%% 
%%%%%%%%%%%%%%%%%%%%%%%%%%%%%%%%%%%%%%%%%%%%%%%% 

\subsection{The six-point one-loop supergravity amplitude}
\label{sec:5.3}

We shall now derive a new superspace representation of the
six-point one-loop supergravity amplitude from (\ref{ftamp.01}).
In order to illustrate the workflow, we introduce the following
compressed notation for the 51-term chiral correlator (\ref{KK6GEF}),
\begin{align}
{\cal K}_6 = \sum_P  {\cal C}_P E_P
  \label{ftamp.07} 
\end{align}
with collective indices $P$ gathering the subscripts and Lorentz indices of
$E_P\in \{ E^{mn}_{1|2,3,4,5,6}, $
$  \ E_{1|234,5,6},\ E_{1|2|3,4,5,6}, \ \ldots\}$
and ${\cal C}_P\in \{ \tfrac{1}{2}C^{mn}_{1|2,3,4,5,6}, \ \tfrac{1}{4} s_{23} s_{34} E_{1|234,5,6},$
$-P_{1|2|3,4,5,6}, \ldots\}$ denoting their respective coefficients in (\ref{KK6GEF}). The 
supergravity amplitude then takes the form
\begin{align}
M^{\textrm{1-loop}}_{{\rm SG}, \, 6} &=   \int_{\RR^D} \dd^{D} \ell    \,
\sum_{P,Q}  {\cal C}_P \, \tilde {\cal C}_Q \,
 {\rm FT}^{\rm cl}_{ 6 }\big[ E_P \bar E_Q\big]
  \label{ftamp.08} 
\end{align}
The closed-string field-theory limits on the right-hand side can be found in 
sections \ref{sec:3.3.b}, \ref{sec:4.3}, appendix \ref{app:Eex} and the supplementary material of this work. Each of the 51$\times$51 terms $ {\rm FT}^{\rm cl}_{ 6 }[ E_P \bar E_Q]$
contains several propagator structures $I^{(r)}$ with $r =4,5,6$, so a first step
is to gather the huge numbers of bilinears ${\cal C}_P \, \tilde {\cal C}_Q$ 
contributing to each box, pentagon and hexagon. 

In a second step, the coefficients of each $I^{(r)}$ resulting from (\ref{ftamp.08})
are simplified using kinematic identities analogous to (\ref{ftamp.06}) -- see sections 9 and 10 of \cite{Mafra:2014gsa} -- and identified as double copies. It is reasonable to expect remnants of
a double-copy form since the super-Yang-Mills numerators encoded in the open-string
field-theory limit (\ref{ftamp.03}) obey a considerable subset
of the one-loop six-point kinematic Jacobi identities \cite{towardsOne}
at the heart of the color-kinematics duality. Indeed,
the reorganization of (\ref{ftamp.08}) according to diagrams
\begin{align}
M^{\textrm{1-loop}}_{{\rm SG}, \, 6} &=   \int_{\RR^D} \dd^{D} \ell    \,
\bigg\{
\big[ |  \mN_{1 | 2,3,4,5,6} |^2 \, I^{(6)}_{1,2,3,4,5,6}  + {\rm perm}(2,3,4,5,6) \big]
 \label{ftamp.09}  \\
  &\quad +\bigg[ \frac{ 1 }{s_{23}}   \sum_{\rho \in S_{\{ 23,4,5,6\} } }  
   |  \mN_{1 | \rho(23,4,5,6)} |^2\,
  I^{(5)}_{1,\rho(23,4,5,6)} + (2,3 | 2,\ldots,6) \bigg]\notag  \\
 &\quad +\bigg[ \frac{  |  \mN_{12 | 3,4,5,6} |^2}{s_{12}} \sum_{\rho \in S_{\{ 3,4,5,6\} } }
  I^{(5)}_{12,\rho(3,4,5,6)} + (2\leftrightarrow 3,4,5,6) \bigg]\notag  \\
 &\quad +\bigg[ \frac{  |  \mN_{1 | 23,45,6} |^2}{s_{23}s_{45}} \sum_{\rho \in S_{\{ 23,45,6\} } }
  I^{(4)}_{1,\rho(23,45,6)}
  +  \frac{  |  \mN_{1 | 24,35,6} |^2}{s_{24}s_{35}} \sum_{\rho \in S_{\{ 24,35,6\} } }
  I^{(4)}_{1,\rho(24,35,6)} \notag \\
  &\quad\quad\quad\quad 
  +  \frac{  |  \mN_{1 | 25,34,6} |^2}{s_{25}s_{34}} \sum_{\rho \in S_{\{ 25,34,6\} } }
  I^{(4)}_{1,\rho(25,34,6)} + (6\leftrightarrow 5,4,3,2) \bigg] \notag  \\
 &\quad +\bigg[ \bigg(  \frac{  |  \mN_{1 | 234,5,6} |^2}{s_{23} s_{234}}
+ \frac{  |  \mN_{1 | 342,5,6} |^2}{s_{34} s_{234}}
 + \frac{  |  \mN_{1 | 423,5,6} |^2}{s_{24} s_{234}} \bigg)  \sum_{\rho \in S_{\{ 234,5,6\} } }
  I^{(4)}_{1,\rho(234,5,6)}\notag \\
  &\quad\quad\quad\quad  + (2,3,4 | 2,3,4,5,6) \bigg] \notag \\
  &\quad   {-}\frac{1}{8 }  \Big[ \big| {\cal N}_{\{123\},4,5,6}  \big|^2   \sum_{\rho \in S_{\{ 4,5,6\} } } I^{(4)}_{123,\rho(4,5,6)}  + (2,3 | 2,3,4,5,6) \Big] \bigg\}
\notag
\end{align}
has all the hexagons and pentagons in a double-copy form tailored to cubic diagrams with the following numerators:
\begin{align}
\mN_{1 | 2,3,4,5,6} &= C_{1|2;3;4;5;6}+\ell_\mu C_{1|2;3;4;5;6}^\mu+\frac{1}{2}\ell_\mu\ell_\nu C_{1|2,3,4,5,6}^{\mu\nu}
- \frac{1}{4} \big[ P_{1|2|3,4,5,6}(\ell{-}k_1)^2{+} P_{1|6|2,3,4,5}\ell^2  \big]
\nonumber\\
  \mN_{1 | 23,4,5,6}  &=  \frac{1}{2} s_{23}\big(C_{1|23;4;5;6}+\ell_\mu C^\mu_{1|23,4,5,6}\big)
\, , \ \ \ \ 
    \mN_{1 | 2,34,5,6} =   \frac{1}{2} s_{34}\big(C_{1|2;34;5;6}+\ell_\mu C^\mu_{1|2,34,5,6}\big)  \notag\\
   \mN_{1 | 2,3,45,6} &=   \frac{1}{2}  s_{45}\big(C_{1|2;3;45;6}+\ell_\mu C^\mu_{1|2,3,45,6}\big)
\, , \ \ \ \
 \mN_{1 | 2,3,4,56} =  \frac{1}{2}  s_{56}\big(C_{1|2;3;4;56}+\ell_\mu C^\mu_{1|2,3,4,56}\big)
  \notag\\
\mN_{12 | 3,4,5,6} &= \frac{1}{4} s_{12}\, P_{1|2|3,4,5,6}
  \label{ftamp.10}
\end{align}
The composite kinematic factors with semicolons in their subscripts 
can be found in (\ref{ftamp.05}) and (\ref{ftamp.00}). 

\begin{figure}[t]
\centering
	\begin{align*}
\begin{array}{c | c | c | c }
\mN_{1 | 2,3,4,5,6} &\mN_{1 | 23,4,5,6} &\mN_{1 | 234,5,6} & \mN_{1 | 23,45,6}
 \\[1.2mm]\hline
	\begin{tikzpicture}[baseline,line width=0.3mm, scale= 0.9]
        \newdimen\Q
        \newdimen\Qq
        \newdimen\R
        \newdimen\M
        \R=0.9cm
        \Q=1.3cm
        \M=0.45cm
        \Qq=1.1cm
        \draw  \foreach \x in {0,60,...,300} { (0:0) -- (\x:\Q) };
        \foreach \x/\l/\p in {60/{4}/above,120/{3}/above, 180/{2}/left, 240/{1}/below, 300/{6}/below, 0/{5}/right}
        \node[label={\p:\l}] at (\x:\Q) {};
        \draw[fill=white] (300:\R) \foreach \x in {0,60,...,300} { -- (\x:\R) };
        \draw[draw=none,postaction={decorate, decoration={markings, mark=at position 0.5 with {\arrow{<}}}}]
          (240:\R) -- (300:\R);
        \node at ($(240:\R)!0.5!(300:\R) + (0,-0.45)$) {\(\ell\)};
    \end{tikzpicture}  &
    \begin{tikzpicture}[baseline,line width=0.3mm, scale= 0.9]
        \newdimen\Q
        \newdimen\Qq
        \newdimen\R
        \newdimen\M
        \R=0.9cm
        \Q=1.3cm
        \M=0.45cm
        \Qq=1.1cm
        \draw  \foreach \x in {18,90,...,306} { (0:0) -- (\x:\Q) };
        \foreach \x/\l/\p in {90/{4}/above,234/{1}/below, 306/{6}/below, 18/{5}/right}
        \node[label={\p:\l}] at (\x:\Q) {};
        \draw[fill=white] (306:\R) \foreach \x in {18,90,...,306} { -- (\x:\R) };
        \draw (162:\Q) --++(207:\M);
        \draw (162:\Q) --++(117:\M);
        \node[label={above:3}] at ($(162:\Q) + (117:\M)$) {};
        \node[label={left:2}] at ($(162:\Q) + (207:\M)$) {};
        \draw[draw=none,postaction={decorate, decoration={markings, mark=at position 0.5 with {\arrow{<}}}}]
          (234:\R) -- (306:\R);
        \node at ($(234:\R)!0.5!(306:\R) + (0,-0.45)$) {\(\ell\)};
    \end{tikzpicture}
    &
    \begin{tikzpicture}[baseline,line width=0.3mm, scale= 0.9]
        \newdimen\Q
        \newdimen\Qq
        \newdimen\R
        \newdimen\M
        \R=0.9cm
        \Q=1.3cm
        \M=0.45cm
        \Qq=1.1cm
        \draw  \foreach \x in {45,135,...,315} { (0:0) -- (\x:\Q) };
        \foreach \x/\l/\p in {225/{1}/below, 315/{6}/below, 405/{5}/right}
        \node[label={\p:\l}] at (\x:\Q) {};
        \draw[fill=white] (315:\R) \foreach \x in {45,135,...,315} { -- (\x:\R) };
        \draw (135:\Q) --++(180:\M);
        \draw (135:\Q) --++(90:\M);
        \node[label={left:2}] at ($(135:\Q) + (180:\M)$) {};
        \node[label={above:3}] at ($(135:\Q) + (90:\M)$) {};
        \draw (135:\Qq) --++(45:\M);
        \node[label={right:4}] at ($(135:\Qq) + (45:\M)$) {};
        \draw[draw=none,postaction={decorate, decoration={markings, mark=at position 0.5 with {\arrow{<}}}}]
          (225:\R) -- (315:\R);
        \node at ($(225:\R)!0.5!(315:\R) + (0,-0.45)$) {\(\ell\)};
    \end{tikzpicture}
    & \begin{tikzpicture}[baseline,line width=0.30mm, scale= 0.9]
        \newdimen\Q
        \newdimen\R
        \newdimen\Rr
        \newdimen\M
        \R=0.9cm
        \Rr=0.636396cm
        \Q=1.3cm
        \M=0.45cm
        \draw  \foreach \x in {45,135,...,315} { (0:0) -- (\x:\Q) };
        \foreach \x/\l/\p in {225/{1}/below, 315/{6}/below}
        \node[label={\p:\l}] at (\x:\Q) {};
        \draw[fill=white] (315:\R) \foreach \x in {45,135,...,315} { -- (\x:\R) };
        \draw (135:\Q) --++(180:\M);
        \draw (135:\Q) --++(90:\M);
        \node[label={left:2}] at ($(135:\Q) + (180:\M)$) {};
        \node[label={above:3}] at ($(135:\Q) + (90:\M)$) {};
        \draw (45:\Q) --++(90:\M);
        \draw (45:\Q) --++(0:\M);
        \node[label={above:4}] at ($(45:\Q) + (90:\M)$) {};
        \node[label={right:5}] at ($(45:\Q) + (0:\M)$) {};
        \draw[draw=none, postaction={decorate, decoration={markings, mark=at position 0.5 with {\arrow{<}}}}] (225:\R) -- (315:\R);
        \node at ($(225:\R)!0.5!(315:\R) + (0,-0.45)$) {\(\ell\)};
    \end{tikzpicture}
\end{array}
\end{align*}
\caption{Cubic diagrams corresponding to the kinematic numerators $\mN_{1|\ldots}$ defined in \eqref{ftamp.10} and \eqref{ftamp.11}. 
Permutations of the external legs $2,3,4,5,6$ (i.e.\ excluding leg $1$) in the diagrams of the figure are implemented
by relabelings of the associated numerators. In case of pentagons, the figure illustrates the diagram associated with $\mN_{1 | 23,4,5,6}$ where the massive corner is adjacent to leg~1 (which always serves as the reference for the loop momentum). The numerators of the remaining pentagons (with massive corners in different positions)
can be found in \eqref{ftamp.10}, and their scalar parts $C_{1|2;34;5;6}, C_{1|2;3;45;6}, C_{1|2;3;4;56}$ given by (\ref{ftamp.05}) do not follow from a mere relabeling of the scalar part $C_{1|23;4;5;6}$ of $\mN_{1 | 23,4,5,6}$.}
\label{fig_numerat}
\end{figure}
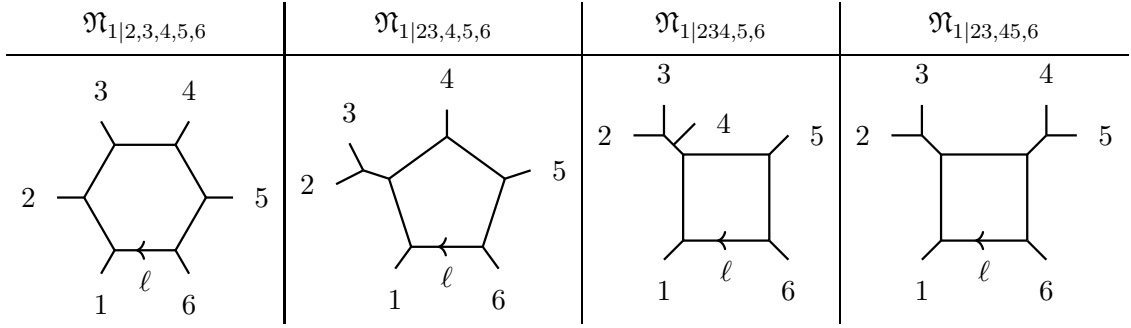

However, the box contributions to (\ref{ftamp.09}) mix different notions 
of double copy. For those boxes $I^{(4)}_{1,\ldots}$ with leg 1 in a massless
corner, the fourth to seventh line of (\ref{ftamp.09}) exhibit the six propagators of cubic
diagrams where the associated numerator factors obey the kinematic Jacobi identities:
\begin{align}
 \mN_{1 | 23,45,6}  &=  \frac{1}{4} s_{23}s_{45}\,C_{1|23,45,6}
   \label{ftamp.11}  \\
\mN_{1 | 234,5,6}  &=  \frac{1}{4}  s_{23}\big(s_{34}\,C_{1|234,5,6}-s_{24}\,C_{1|324,5,6}\big)\notag
\end{align}
The cubic diagrams associated with the numerators in (\ref{ftamp.10}) and (\ref{ftamp.11}) are drawn in
figure~\ref{fig_numerat}.

\begin{figure}[t]
\centering
	\begin{align*}
\begin{array}{c }
{\cal N}_{\{123 \},4,5,6}
\\[1.2mm]\hline
	\begin{tikzpicture}[baseline,line width=0.3mm, scale = 1]
        \newdimen\Q
        \newdimen\R
        \newdimen\M
        \R=1.1cm
        \Q=2cm
        \M=0.9cm
        \draw  \foreach \x in {45,135,315} { (0:0) -- (\x:\Q) };
        \draw (0:0) -- (225:\Q);
        \foreach \x/\l/\p in {45/{5}/right, 135/{4}/left, 225/{2}/left, 315/{6}/right}
        \node[label={\p:\l}] at (\x:\Q) {};
        \draw[fill=white] (315:\R) \foreach \x in {45,135,...,315} { -- (\x:\R) };
        \draw[draw=none,postaction={decorate, decoration={markings, mark=at position 0.5 with {\arrow{<}}}}]
          (225:\R) -- (315:\R);
        \node at ($(225:\R)!0.5!(315:\R) + (0,-0.45)$) {\(\ell\)};
        \draw (225:\R) --++(270:\M);
        \draw (225:\R) --++(180:\M);
        \node[label={below:1}] at ($(225:\R) + (270:\M)$) {};
        \node[label={left:3}] at ($(225:\R) + (180:\M)$) {};
    	\end{tikzpicture}
\end{array}
\end{align*}
\caption{Contact diagram corresponding to the numerator ${\cal N}_{{ \{ 123 \}},4,5,6}$ as defined in \eqref{weirdnum}. Its quintic vertex reflects the pole structure 
in the last line of \eqref{ftamp.09}.}
\label{fig123_box}
\end{figure}
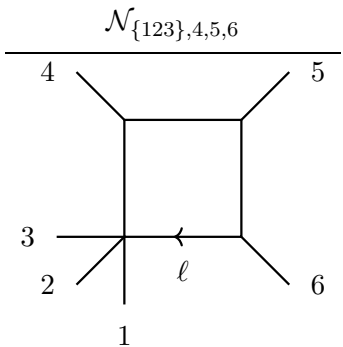

The remaining boxes in the last line of
(\ref{ftamp.09}) have leg 1 in a massive corner such as $I^{(4)}_{123,\ldots}$ and lack 
the two external propagators seen in the coefficients of $I^{(4)}_{1,\ldots}$ in the 
fourth to seventh line of (\ref{ftamp.09}).
Hence, the accompanying factors of 
\beq
{\cal N}_{\{123 \},4,5,6} = P_{1|2|3,4,5,6} - P_{1|3|2,4,5,6 }
\label{weirdnum}
\eeq
can be thought of as kinematic numerators associated with the contact diagram 
of figure~\ref{fig123_box} involving a quintic vertex. We use a different font for the quintic-vertex
numerators ${\cal N}_{\{123 \},4,5,6}$ in (\ref{weirdnum}) to distinguish them from
the cubic-diagram numerators $\mN$ in (\ref{ftamp.10}) and (\ref{ftamp.11}).

The double-copy numerators $|  {\cal N}_{\{123 \},4,5,6}|^2$ associated with the contact-term
diagrams in the last line of (\ref{ftamp.09}) are a common theme with the generalized 
double copy \cite{Bern:2017yxu}: even a gauge-theory-amplitude representation that
violates the color-kinematics duality can be taken as a starting point to construct
gravitational loop integrands if also the violations of kinematic Jacobi identities are
double copied and associated with contact diagrams. Indeed, the difference in (\ref{weirdnum})
is identified as a violation of kinematic Jacobi identities in section 6.3 of \cite{towardsOne}
associated with the representation (\ref{ftamp.03}) of the six-point one-loop amplitude of super-Yang-Mills. 
However, typical applications of the generalized double copy would first involve quartic-vertex
contact diagrams prior to the quintic diagram associated with (\ref{weirdnum}). Accordingly,
the absence of quartic-vertex diagrams in (\ref{ftamp.09}) is surprising from the viewpoint
of \cite{Bern:2017yxu}, or at least suggests that we cannot identify our representation of
$M^{\textrm{1-loop}}_{{\rm SG}, \, 6}$ with a known formulation of the generalized double copy.

%%%%%%%%%%%%%%%%%%%%%%%%%%%%%%%%%%%%%%%%%%%%%%%% 
%%%%%%%%%%%%%%%%%%%%%%%%%%%%%%%%%%%%%%%%%%%%%%%% 

\subsubsection{Comparison with the super-Yang-Mills amplitude}
\label{sec:5.4.0}

As we shall now demonstrate, the cubic-diagram numerators $\mN$ in (\ref{ftamp.10})
and (\ref{ftamp.11}) give rise to an alternative representation of the six-point
super-Yang-Mills amplitude (\ref{ftamp.03}) up to rational terms in $D=10$ dimensions. 
The general parametrization in terms of cubic diagrams without triangles, tadpoles and bubbles,
and where we furthermore impose vanishing numerators for boxes with leg 1 in a massive corner, is given by
\begin{align}
  A^{\textrm{1-loop}}_{\rm SYM}&(1,2,\ldots,6) =   \int_{\RR^D} \! \dd^{D} \ell \, \bigg\{
\frac{1}{s_{234}}  \bigg( \frac{ \mN_{1|234,5,6} }{s_{23}} + \frac{ \mN_{1|432,5,6} }{s_{34}} \bigg)  I^{(4)}_{1,234,5,6}  \label{newsym} \\
&\quad \! \! \! \! \! \!  + \frac{1}{s_{345}}  \bigg( \frac{ \mN_{1|345,2,6} }{s_{34}} + \frac{ \mN_{1|543,2,6} }{s_{45}} \bigg)  I^{(4)}_{1,2,345,6}  + \frac{1}{s_{456}}  \bigg( \frac{ \mN_{1|456,2,3} }{s_{45}} + \frac{ \mN_{1|654,2,3} }{s_{56}} \bigg)   I^{(4)}_{1,2,3,456} \notag \\
&\quad \! \! \! \! \! \!  + \frac{ \mN_{1|23,45,6}  }{s_{23} s_{45}}  I^{(4)}_{1,23,45,6} 
+ \frac{ \mN_{1|23,56,4}  }{s_{23} s_{56}}   I^{(4)}_{1,23,4,56}  
+   \frac{ \mN_{1|34,56,2}  }{s_{34} s_{56}}   I^{(4)}_{1,2,34,56}   \notag \\
%%%%%%
&\quad \! \! \! \! \! \! + \frac{ \mN_{1|23,4,5,6} }{s_{23}}  I^{(5)}_{1,23,4,5,6} 
 + \frac{ \mN_{1|2,34,5,6} }{s_{34}}  I^{(5)}_{1,2,34,5,6} 
 + \frac{ \mN_{1|2,3,45,6} }{s_{45}} I^{(5)}_{1,2,3,45,6}
 + \frac{ \mN_{1|2,3,4,56} }{s_{56}} I^{(5)}_{1,2,3,4,56} \notag \\
%%%%%%%
&\quad \! \! \! \! \! \!  +  \frac{ \mN_{12|3,4,5,6} }{s_{12}}  I^{(5)}_{12,3,4,5,6} 
  -  \frac{ \mN_{16|2,3,4,5} }{s_{61}}  I^{(5)}_{6\underline{1},2,3,4,5} 
+ \mN_{1|2,3,4,5,6}   I^{(6)}_{1,2,3,4,5,6}    \bigg\} \ \textrm{mod rat.\ terms}
\notag
\end{align}
with a minus sign in the numerator of the pentagon $I^{(5)}_{6\underline{1},2,3,4,5}$ in view of the
cubic-vertex flip that relates the cyclic image of the $I^{(5)}_{12,3,4,5,6}$ diagram to
its relabeling $2\leftrightarrow 6$. The first four lines of the 
expressions (\ref{ftamp.03}) and (\ref{newsym}) are readily seen to match
after inserting the numerators in (\ref{ftamp.10}), (\ref{ftamp.11}) and
using the shuffle property $C_{1|234,5,6} = C_{1|432,5,6}  = - C_{1|324,5,6}  - C_{1|243,5,6} $.
The integrands of the last lines of (\ref{ftamp.03}) and (\ref{newsym}) 
agree upon comparison~of
\begin{align}
&\frac{1}{4}P_{1|2|3,4,5,6} I^{(5)}_{12,3,4,5,6} 
- \frac{1}{4} P_{1|6|2,3,4,5} I^{(5)}_{6\underline{1},2,3,4,5}
- \frac{1}{4} I^{(6)}_{1,2,3,4,5,6}  \big[ (\ell{-}k_1)^2 P_{1|2|3,4,5,6}+ \ell^2 P_{1|6|2,3,4,5} \big] \notag \\
&\quad
= - \frac{1}{2} P_{1|6|2,3,4,5} I^{(5)}_{6\underline{1},2,3,4,5}\label{symagree}
\end{align}
where the first line gathers all the terms of the form $P_{1|i|j,k,l,m}$ in (\ref{ftamp.10})
that enter the hexagon and pentagon numerators.
However, the matching of loop integrands in (\ref{symagree}) makes use of
cancellations $\frac{\ell^2 }{\ell^2 } \rightarrow 1$ which are problematic upon loop
 integration in $D=10$ spacetime dimensions but readily applicable to dimensions 
 $D\leq 9$. Indeed, the integrated expressions (\ref{ftamp.03}) and (\ref{newsym}) 
 differ by rational terms $\sim ( P_{1|2|3,4,5,6}{+}P_{1|6|2,3,4,5})$ in $D=10$ that
 are known from the hexagon anomaly.

Note that (\ref{newsym}) does not feature any analogue of the boxes $| {\cal N}_{\{123\},4,5,6}  |^2 
I^{(4)}_{123,\rho(4,5,6)}$ in the last line of the supergravity integrand (\ref{ftamp.09}), 
see figure \ref{fig123_box} for their contact-term interpretation. Adding any cyclic combination of 
such boxes to (\ref{newsym}) would clearly result in departures from the super-Yang-Mills integrand
(\ref{ftamp.03}) beyond the rational terms specific to $D=10$.

%%%%%%%%%%%%%%%%%%%%%%%%%%%%%%%%%%%%%%%%%%%%%%%% 
%%%%%%%%%%%%%%%%%%%%%%%%%%%%%%%%%%%%%%%%%%%%%%%% 

\subsubsection{Crossing-symmetry properties}
\label{sec:5.4.1}

The expressions (\ref{KK6GEF}) for the chiral correlators only expose
a reduced permutation invariance in legs $2,3,\ldots,n$, i.e.\ leg 1 enters on special footing. 
Starting from six points, this is inevitable since the lack of full permutation invariance by a boundary
term in $\tau$ \cite{Mafra:2016nwr, He:2017spx, oneloopIII} is a worldsheet manifestation of the hexagon anomaly of ten-dimensional super-Yang-Mills \cite{Frampton:1983ah, Frampton:1983nr}
and individual topologies of the type-I superstring \cite{Green:1984qs, Green:1984sg, Clavelli:1986fj, anomaly}. 
As a result, the amplitude representations to be derived from (\ref{KK6GEF}) are only expected to
manifest crossing symmetry in legs $2,3,\ldots,n$ excluding leg 1. Indeed, we have seen that the
super-Yang-Mills amplitudes and five-point supergravity amplitudes in the form 
of (\ref{ftamp.03}) and (\ref{ftamp.04}) do not
feature any boxes with leg 1 in a massive corner.

 We have given separate expressions for the pentagon numerators
$\mN_{1 | 23,4,5,6}$, $  \mN_{1 | 2,34,5,6} $, $\mN_{1 | 2,3,45,6}$ and $ \mN_{1 | 2,3,4,56}$ 
in (\ref{ftamp.10}) since the functional form of their scalar parts $C_{1|23;4;5;6}$, 
$C_{1|2;34;5;6}$, $C_{1|2;3;45;6}$ and $C_{1|2;3;4;56}$ in (\ref{ftamp.05}) depends on 
the relative position of the massive corner and leg 1. For the box contributions
(\ref{ftamp.11}) and (\ref{weirdnum}) in turn, the numerators do not depend on the
orientation of the three massless legs.

Since gravitational anomalies are absent from type-II supergravity \cite{Alvarez-Gaume:1983ihn}
and type-II superstrings \cite{Kutasov:1988cm, Hayashi:1987hp, Lerche:1988np}, the lack of full permutation symmetry of ${\cal K}_6$ by a boundary 
term \cite{Mafra:2016nwr, He:2017spx, oneloopIII} does not affect $M^{\textrm{1-loop}}_{{\rm SG}, \, 6} $ 
in the field-theory limit. Hence, the six-point one-loop amplitude (\ref{ftamp.09}) must have a hidden permutation symmetry including leg 1, even though the explicit check based on permutation properties of
the BRST (pseudo)invariants \cite{Mafra:2014gsa} would be tedious.

%%%%%%%%%%%%%%%%%%%%%%%%%%%%%%%%%%%%%%%%%%%%%%%% 
%%%%%%%%%%%%%%%%%%%%%%%%%%%%%%%%%%%%%%%%%%%%%%%% 

\subsubsection{Locality properties}
\label{sec:5.4.2}

We have expressed all the dependence on the external polarizations
in terms of BRST (pseudo)invariants $C$ and $P$ which manifest spacetime
supersymmetry and gauge invariance of the bosonic components.
These manifestations are at the cost of the locality properties since
each six-point (pseudo)invariant introduces pole structures $s_{ij}^{-1}$, $s_{ijk}^{-1}$, $( s_{ij} s_{kl})^{-1}$
or $( s_{ij} s_{ijk})^{-1}$ in the external momenta. In particular, individual bilinears in the supergravity
amplitude such as $|  s_{23}(s_{34}C_{1|234,5,6}-s_{24}C_{1|324,5,6} ) |^2 / (s_{23} s_{234})$
from the box contributions (\ref{ftamp.11}) introduce propagator combinations
such as $(s_{123} s_{124})^{-1}$ that are incompatible with cubic diagrams and
cancel in the overall amplitude.

The trade-off between the manifestation of gauge invariance and locality is ubiquitous
in scattering amplitudes. An alternative representation of the supergravity amplitude that
exposes locality instead of gauge- or BRST-invariance can be derived from the
manifestly local form of the chiral correlators that are referred to as $T\cdot E$
representations in \cite{oneloopIII}. These local correlator representations can be
obtained from (\ref{KK6GEF}) through a coherent replacement of the
(pseudo)invariants $C$ and $P$ by certain local superfields denoted
by $T$ and $J$ \cite{Mafra:2014gsa, Mafra:2018nla} with a non-zero but well-controlled BRST variation. 
Inserting these local expressions for ${\cal K}_n$ into (\ref{ftamp.01}) then results
in a manifestly local form of the one-loop field-theory amplitudes.

However, the local counterparts of the BRST (pseudo)invariants no longer
obey the kinematic identities of the $C$. Instead, the local variant of the 
kinematic identity (\ref{ftamp.06}) under the formal $C\rightarrow T$ substitution
features a non-zero right-hand side for $V_1 (k^m_2 T^m_{2,3,4,5} + T_{23,4,5}+ T_{24,3,5}+ T_{25,3,4})$.
Hence, the box propagators $I^{(4)}_{12,3,4,5}$ and $I^{(4)}_{5\underline{1},2,3,4}$ would 
no longer cancel from the manifestly 
local representation of $ A^{\textrm{1-loop}}_{\rm SYM}(1,2,3,4,5) $, and the expression in (\ref{ftamp.03}) 
would grow by two terms. Thanks to the expressions for $ {\rm FT}^{\rm cl}_{ 6}[ E \bar E]$
in the supplementary material of this work, it is tedious but straightforward to assemble a
manifestly local representation of the six-point supergravity amplitude, with
a more drastic growth of terms than in the five-point super-Yang-Mills case.

%%%%%%%%%%%%%%%%%%%%%%%%%%%%%%%%%%%%%%%%%%%%%%%% 
%%%%%%%%%%%%%%%%%%%%%%%%%%%%%%%%%%%%%%%%%%%%%%%% 

\subsubsection{Ultraviolet divergences}
\label{sec:5.4.3}

As an important consistency check for our new representation
of the six-point supergravity amplitude, we have verified 
that its ultraviolet (UV) divergence in eight spacetime dimensions takes the correct form.
This UV divergence is obtained by performing the Feynman integrals
of (\ref{ftamp.09}) in $D=8{-}2\epsilon$ dimensions and isolating the
simple pole in $\epsilon$. By the power counting of one-loop supergravity amplitudes
to have at most $2m{-}8$ loop momenta in $m$-gon numerators 
\cite{Bern:1992ad, Bern:1993tz, Bjerrum-Bohr:2006xbk, Bjerrum-Bohr:2008vag},
the only contributions to the desired six-point 
UV divergence stem from
\begin{align}
&\ I^{(4)}_{A,B,C,D} \, \big|_{\rm UV} = \frac{1}{3!} 
\, , \ \ \ \ \ \ell^m \ell^n I^{(5)}_{A,B,C,D,E} \, \big|_{\rm UV} = \frac{\eta^{mn}}{ 2\cdot4!} \
 \label{ftamp.15}\\
 \ell^m &\ell^n \ell^p \ell^q I^{(6)}_{A,B,C,D,E,F} \, \big|_{\rm UV} = \frac{\eta^{mn} \eta^{pq} 
 + \eta^{mp} \eta^{nq} + \eta^{mq} \eta^{np}}{ 4\cdot 5!} \notag
\end{align}
for arbitrary massive or massless external legs $A,B,\ldots,F$,
while pentagons with $\leq 1$ power and hexagons with
$\leq 3$ powers of loop momentum yield UV-finite integrals in $D=8$.

The one-loop UV divergences of type-IIA/B supergravity in the critical dimension $D=8$
are well-known to match the matrix elements of a supersymmetrized $R^4$ counterterm 
which are also the low-energy limits of the respective string amplitudes. Up to and including six points, 
the pure-spinor superspace representations of these matrix elements are given by \cite{Mafra:2016nwr}
\begin{align}
M^{\te{1-loop}}_{{\rm SG},\, 4} \, \big|_{\rm UV}&= C_{1|2,3,4} \tilde C_{1|2,3,4}
\notag \\
2\,M^{\te{1-loop}}_{{\rm SG},\, 5} \, \big|_{\rm UV}&= C^{m}_{1|2,3,4,5} \tilde C^{m}_{1|2,3,4,5} 
+ \tfrac{1}{2} \big[ s_{23} C_{1|23,4,5} \tilde C_{1|23,4,5} + (2,3|2,3,4,5) \big] 
\notag \\
4\, M^{\te{1-loop}}_{{\rm SG},\, 6} \, \big|_{\rm UV}&=  \tfrac{1}{2} C^{mn}_{1|2,\ldots,6} \tilde C^{mn}_{1|2,\ldots,6}   +  \tfrac{1}{2} \big[ s_{23} C^m_{1|23,4,5,6} \tilde C^m_{1|23,4,5,6} + (2,3|2,3,4,5,6) \big] \notag \\
&\quad  + \tfrac{1}{4} \Big( \big[ s_{23} s_{45} C_{1|23,45,6} \tilde C_{1|23,45,6} + {\rm cyc}(3,4,5) \big] + (6\leftrightarrow 5,4,3,2) \Big)  \notag
 \\
&\quad  +  \tfrac{1}{4} \Big( \big[ s_{23}s_{34} C_{1|234,5,6}  \tilde C_{1|234,5,6} + {\rm cyc}(2,3,4) \big]    + (2,3,4|2,3,4,5,6) \Big) \notag \\
&\quad  -  \big[ P_{1|2|3,4,5,6} \tilde P_{1|2|3,4,5,6}  +(2\leftrightarrow 3,4,5,6) \big]  \label{uveqs}
\end{align}
Their formal similarity to the chiral correlators ${\cal K}_{n\leq 6}$ in (\ref{KK6GEF})
and the interpretation of this observation in the light of double-copy structures 
is discussed in \cite{Mafra:2017ioj, oneloopIII}.

According to (\ref{ftamp.15}), the UV contributions to the six-point supergravity amplitude 
(\ref{ftamp.09}) stem from boxes, pentagons and hexagons. The most delicate part is the
assembly of the $P \tilde P$ terms of (\ref{uveqs}) which receive contributions from the boxes in
the last line of (\ref{ftamp.09}) with numerators in (\ref{weirdnum})
and the double copy of the hexagon terms in (\ref{ftamp.10}) quadratic in $\ell$
\beq
\mN_{1 | 2,3,4,5,6} \, \big|_{\ell^2} = \frac{1}{2}\ell_m \ell_n C^{mn}_{1|2,3,4,5,6} -  \frac{1}{4}  \ell^2 (P_{1|2|3,4,5,6} {+} P_{1|6|2,3,4,5})
 \label{ftamp.16}
\eeq
After performing the permutation sums in (\ref{ftamp.09}), we have crosschecked in superspace
that the UV divergence of the six-point supergravity amplitude in $D=8$ dimensions\footnote{It is crucial to evaluate the contractions of metric tensors from (\ref{ftamp.15}) with those of $\ell^2 = \eta_{mn} \ell^m \ell^n $ as $\eta^{mn} \eta_{mn} = 8$ in computing eight-dimensional UV divergences. A simple crosscheck is the $D=8$ UV divergence of a pentagon with numerator $\ell^2$ whose evaluation via (\ref{ftamp.15}) and  $\eta^{mn} \eta_{mn} = 8$ matches that of a box.} 
indeed takes the desired form (\ref{uveqs}).

%%%%%%%%%%%%%%%%%%%%%%%%%%%%%%%%%%%%%%%%%%%%%%%% 
%%%%%%%%%%%%%%%%%%%%%%%%%%%%%%%%%%%%%%%%%%%%%%%% 

\section{Comparison with ambitwistor-string computations}
\label{sec:6}

An alternative worldsheet approach to loop integrands in gauge theory and (super)gravity 
descends from ambitwistor string theories which do not feature any analogue of $\alpha'$ and directly
compute field-theory amplitudes \cite{Mason:2013sva, Berkovits:2013xba, Adamo:2013tsa, Adamo:2015hoa, Geyer:2022cey}. In particular, the methods of \cite{Geyer:2015bja, Geyer:2015jch} reduce the one-loop amplitude formulae of ambitwistor strings to forward limits of tree amplitudes \cite{Baadsgaard:2015hia, He:2015yua, Cachazo:2015aol, Cardona:2016bpi, Cardona:2016wcr, He:2016mzd, He:2017spx, Geyer:2017ela, Edison:2020uzf}. We shall here compare the applications of pinching rules in conventional, $\alpha'$-dependent superstring theories with the 
forward-limit-based evaluation of ambitwistor-string integrals. 

The simplified expressions for chiral correlators ${\cal K}_n$ of conventional superstrings can be straightforwardly exported to ambitwistor string theories \cite{Gomez:2013wza, He:2017spx, Kalyanapuram:2021vjt}, possibly after taking the tensionless limit $\alpha ' \rightarrow \infty$ \cite{Huang:2016bdd, Casali:2016atr, Azevedo:2017yjy, Kalyanapuram:2021xow}. 
We will perform the comparison at the level of homology invariants $E$ 
carrying the moduli dependence of ${\cal K}_{n\leq 6}$ points instead of
the full chiral amplitudes. Already for the individual homology invariants, we 
find perfect matching for the supergravity propagators resulting from arbitrary combinations $E \bar E$ 
of five- and six-point cases in (\ref{Esat5pt}) and (\ref{Esat6pt}).

As we will see, the subtleties in the integration-by-parts treatment of the double pole of $E_{1|2|3,4,5,6}$ in section \ref{sec:4} do not carry over to the ambitwistor setup. More specifically,
the gauge-theory and supergravity propagators from worldsheet integrals over $E_{1|2|3,4,5,6}$ are
accurately reproduced by the ambitwistor-string methods when employing the naive version $E^{\rm naive}_{1|2|3,4,5,6}$ in (\ref{enaiv}), with logarithmic singularities instead of double poles.

%%%%%%%%%%%%%%%%%%%%%%%%%%%%%%%%%%%%%%%%%%%%%%%% 
%%%%%%%%%%%%%%%%%%%%%%%%%%%%%%%%%%%%%%%%%%%%%%%% 

\subsection{Basics of one-loop ambitwistor-string amplitudes}
\label{sec:6.1}

The simplification of one-loop $n$-point integrals of ambitwistor-string amplitudes using the residue theorem
for the modular parameter $\tau$ as in \cite{Geyer:2015bja, Geyer:2015jch} reproduces the tree-level measure 
\beq
\dd \mu_{n+2}^{\te{tree}} = \dd^2 \sigma_2 \, \dd^2 \sigma_3 \, \ldots \, \dd^2 \sigma_n \prod_{i=2}^n \delta^2 \bigg(2 \sum^{n+2}_{j \neq i} \frac{ k_i \cdot k_j }{\sigma_{ij}} \bigg) \, , \ \ \ \ \ \ \sigma_{ij} = \sigma_i {-} \sigma_j
\label{rev.2}
\eeq
for the moduli space of $n{+}2$ marked points on the sphere.
The total of $n{-}1$ delta distributions localizes any integral $\int_{\mathbb C^{{n-1}}} \dd \mu_{n+2}^{\te{tree}} $
to the solutions of the scattering equations $ \sum^{n+2}_{j \neq i} \frac{ k_i \cdot k_j }{\sigma_{ij}} = 0$
which are known as the backbone of CHY formulae for tree amplitudes 
\cite{Cachazo:2013gna, Cachazo:2013hca, DPellis}. The expression (\ref{rev.2}) for
the measure and later formulae of this section are adapted to the ${\rm SL}(2,\mathbb C)$ frame where three of the marked points $\sigma_i$ on the sphere are fixed at $\sigma_1 =1$ as well as
$( \sigma_{n+1}, \sigma_{n+2})  \rightarrow (0, \infty)$. We will also write
$( \sigma_+, \sigma_-) =( \sigma_{n+1}, \sigma_{n+2})$ throughout this section.

The measure (\ref{rev.2}) is a universal building block of one-loop amplitudes from
ambitwistor strings which for gauge theories and (super)gravity read
\begin{align}
A^{\te{1-loop}}_{\rm SYM}(1,2,\ldots,n) &= \int_{\mathbb R^D} \frac{ \dd^D \ell}{\ell^2} \lim_{k_{\pm} \rightarrow \pm \ell} \int_{\mathbb C^{n-1}} \dd \mu_{n+2}^{\te{tree}} \ {\cal I}_{n}(\ell) \! \!  \! \! \! \! \sum_{\gamma \in {\rm cyc}(1,2,\ldots,n)} \! \! \! \! \! \! {\rm PT}(+,\gamma(1,2,\ldots,n),-) 
 \notag \\
M^{\te{1-loop}}_{{\rm SG},\, n} &=
 \int_{\mathbb R^D} \frac{ \dd^D \ell}{\ell^2} \lim_{k_{\pm} \rightarrow \pm \ell} 
\int_{\mathbb C^{n-1}} \dd \mu_{n+2}^{\te{tree}} \ {\cal I}_n(\ell) \, \tilde {\cal I}_n(\ell)
\label{uniamps}
\end{align}
and involve the following constituents
\begin{itemize}
\item The loop momentum stems from the last two legs $k_+= k_{n+1}$ and 
$k_- = k_{n+2} $ of the $(n{+}2)$-point tree-level configuration of (\ref{rev.2}), and
$k_{\pm} \rightarrow \pm \ell$ implements a forward limit.
\item The color ordering $\sim {\rm Tr}(t^{a_1}t^{a_2}\ldots t^{a_n})$ of the planar super-Yang-Mills 
amplitudes in (\ref{uniamps})
is reflected by the cyclic ordering of legs $1,2,\ldots,n$ in the Parke-Taylor factor
\beq
{\rm PT}(+,1,2,\ldots,n,-) = \frac{1}{\sigma_{+1} \sigma_{12} \sigma_{23}\ldots \sigma_{n-1,n} }
\label{rev.3}
\eeq
\item The kinematic building blocks ${\cal I}_n(\ell)$ in (\ref{uniamps}) encode the dependence on
external polarizations and the amount of supersymmetry
of the massless spectrum in the loop. We will focus on the maximally supersymmetric case
were ${\cal I}_n(\ell)$  are the $\tau \rightarrow i\infty $ limits of the chiral correlators
${\cal K}_n(\ell)$ in (\ref{baspin.13}) at vanishing string tension
\beq
{\cal I}_{n}(\ell) =  \lim_{\alpha' \rightarrow \infty }  \lim_{\tau \rightarrow i\infty  }\frac{{\cal K}_n(\ell)}{(2\pi i)^{n-4}\sigma_1 \sigma_2\ldots \sigma_n}
\label{chy.01}
\eeq
The inverse factors of $\sigma_i$ can be understood from the change of
integration variables $\dd z_i = \frac{ \dd \sigma_i}{2\pi i \sigma_i}$ converting the
coordinate $z_i$ on the torus (with periodicities $z_i \cong z_i{+}1 \cong z_i{+}\tau$) 
to that on the sphere $\sigma_i = e^{2\pi i z_i}$. Up to and including six points, the 
tensionless limit $\alpha' \rightarrow \infty$ only affects the homology invariant $E_{1|2|3,4,5,6}$ 
by suppressing the contribution $\tfrac{1}{2\ap}\partial_{z_1} g^{(1)}_{12}$ and can otherwise
be dropped in the examples of this work.
\end{itemize}
On the support of the scattering equations in (\ref{rev.2}), any contribution
to the kinematic building block (\ref{chy.01}) can be expanded in terms of Parke-Taylor
factors (\ref{rev.3}),
\begin{align}
{\cal I}_{n}(\ell) &= \sum_{\rho \in S_n} 
N_{+|\rho(12\ldots n)| -}(\ell)  \, {\rm PT}(+,\rho(1,2,\ldots,n),-) 
\label{rev.7}
\end{align}
regardless on the amount of supersymmetry in the loop.
This follows from mathematical results of \cite{aomoto1987gauss}, led
to dramatic simplifications of string tree-level amplitudes 
\cite{Mafra:2011kj, nptStringI, Huang:2016tag, Azevedo:2018dgo} 
as reviewed in \cite{Mafra:2022wml}
and can be explicitly carried out using a variety of methods 
\cite{Baadsgaard:2015voa, Cardona:2016gon, Schlotterer:2016cxa, He:2017spx, He:2018pol, He:2019drm, Edison:2020ehu}.

%%%%%%%%%%%%%%%%%%%%%%%%%%%%%%%%%%%%%%%%%%%%%%%% 
%%%%%%%%%%%%%%%%%%%%%%%%%%%%%%%%%%%%%%%%%%%%%%%% 

\subsubsection{Evaluation of Parke-Taylor integrals}
\label{sec:6.1.1}

The sphere integrals in the one-loop amplitudes (\ref{uniamps})
are most conveniently evaluated in terms of doubly-partial amplitudes $m(\alpha | \beta)$ of bi-adjoint scalars.\footnote{The permutations $\alpha,\beta\in S_N$ refer to the coefficients of two species of traces
${\rm Tr}(t^{\alpha(1)} t^{\alpha(2)} \ldots t^{\alpha(N)})$ and ${\rm Tr}( \tilde t^{\beta(1)} \tilde t^{\beta(2)} \ldots \tilde t^{\beta(N)})$ in color-dressed $N$-point tree-level amplitudes of bi-adjoint scalars $\phi^{a| \tilde a}$ with a cubic coupling $\sim f_{abc}\tilde f_{\tilde a \tilde b \tilde c} \phi^{a| \tilde a} \phi^{b| \tilde b} \phi^{c| \tilde c}$ in their Lagrangian.} 
The latter arise from CHY integrals over two Parke-Taylor factors \cite{DPellis}
\beq
\int_{\mathbb C^{N-3}} \dd \mu_{N}^{\te{tree}}\, {\rm PT}\big(\alpha(1,2,\ldots,N)\big)  {\rm PT}\big(\beta(1,2,\ldots,N)\big) =   m(\alpha |\beta)
\label{chy.02}
\eeq
with $\alpha, \beta \in S_N$ and can be efficiently expressed in terms of products of 
tree-level propagators through the Berends-Giele recursion in \cite{FTlimit}.
Under the forward limit $k_{\pm} \rightarrow \pm \ell$ and the identification $N=n{+}2$, we obtain
the simplified form of the one-loop amplitudes (\ref{uniamps}) after inserting the Parke-Taylor
expansion (\ref{rev.7}):
\begin{align}
A^{\te{1-loop}}_{\rm SYM}(1,2,\ldots,n) &=  \int_{\mathbb R^D} \frac{ \dd^D \ell}{\ell^2} \lim_{k_{\pm} \rightarrow \pm \ell} 
\sum_{\gamma \in {\rm cyc}(1,2,\ldots,n) } \sum_{\rho \in S_n} m(+,\gamma,-|+,\rho,-) N_{+|\rho|-}(\ell)
\notag \\
M^{\te{1-loop}}_{n,{\rm SUGRA}} &=  \int_{\mathbb R^D} \frac{ \dd^D \ell}{\ell^2} \lim_{k_{\pm} \rightarrow \pm \ell} 
\sum_{\gamma, \rho \in S_n} m(+,\gamma,-|+,\rho,-) N_{+|\gamma |-}(\ell) \tilde N_{+|\rho|-}(\ell)
\label{rev.22}
\end{align}
On the one hand, (\ref{rev.22}) clarifies the interpretation of the polarization-dependent 
coefficients $N_{+|\rho(12\ldots n)| -}(\ell)$ in (\ref{rev.7}) as master numerators with
manifest color-kinematics duality for the forward limits of tree amplitudes. On the other
hand, these $N_{+|\rho(12\ldots n)| -}(\ell)$ differ from the kinematic numerators $\mN$ of 
the one-loop diagrams in section \ref{sec:5.3} as will be elaborated next.

%%%%%%%%%%%%%%%%%%%%%%%%%%%%%%%%%%%%%%%%%%%%%%%% 
%%%%%%%%%%%%%%%%%%%%%%%%%%%%%%%%%%%%%%%%%%%%%%%% 

\subsubsection{Undoing partial-fraction decompositions}
\label{sec:6.1.2}

\begin{figure}
\begin{center}
\begin{tikzpicture} [scale=0.75, line width=0.30mm]
\begin{scope}[xshift=-0.8cm]
\draw (0.5,0)--(-0.5,0);
\draw (-0.5,0)--(-0.85,-0.35);
\draw [dashed](-0.85,-0.35)--(-1.2,-0.7);
\draw (0.5,0)--(1.2,-0.7);
\draw[dashed] (-1.2,-1.7)--(-1.2,-0.7);
\draw (1.2,-1.7)--(1.2,-0.7);
\draw (1.2,-1.7)--(0.85,-2.05);
\draw[dashed] (0.85,-2.05)--(0.5,-2.4);
\draw[dashed] (-0.5,-2.4)--(0.5,-2.4);
\draw[dashed] (-0.5,-2.4)--(-1.2,-1.7);
\draw (-0.5,0)--(-0.7,0.4)node[left]{$n$};
\draw (0.5,0)--(0.7,0.4)node[right]{$1$};
\draw (1.2,-0.7)--(1.6,-0.5)node[right]{$2$};
\draw (1.2,-1.7)--(1.6,-1.9)node[right]{$3$};
\draw (0,0) node{$| \! |$};
\draw (-0.25,0.2)node{$-$};
\draw (0.25,-0.2)node{$+$};
\end{scope}
%%%%%
\draw[-> ](2.2,-1.2)  -- (3.7,-1.2);
%%%%%
\begin{scope}[xshift=2.7cm, yshift=0.5cm]
\draw(11.4,-2)node{$+ \ {\rm cyclic}(1,2,\ldots,n)$};
\draw (2.9,-2)node[left]{$+\ell$} -- (5.8,-2);
\draw (7.2,-2) -- (8.1,-2)node[right]{$-\ell$};
\draw (3.5,-2) -- (3.5,-1.5)node[above]{$1$};
\draw (4.5,-2) -- (4.5,-1.5)node[above]{$2$};
\draw (5.5,-2) -- (5.5,-1.5)node[above]{$3$};
\draw[dashed] (5.8,-2) -- (7.2,-2);
\draw (7.5,-2) -- (7.5,-1.5)node[above]{$n$};
\end{scope}
\end{tikzpicture}
\caption{$n$-gon diagram associated with the color ordering ${\rm Tr}(t^{a_1}t^{a_2}\ldots t^{a_n})$ (left panel)
and the partial-fraction decomposition (\ref{ngonPF}) relating each $n$-gon
to a cyclic orbit of the depicted $(n{+}2)$-point tree-level diagram.\label{figpfcut}}
\end{center}
\end{figure}
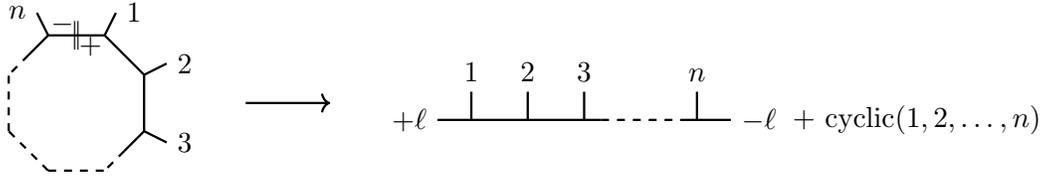

Since the tree-level building blocks $ m(+,\gamma,-|+,\rho,-)$ in the one-loop amplitudes (\ref{rev.22})
are tailored to massless external legs, there is no way of generating squares $\ell^2$ of the loop momentum
under the forward limit $k_{\pm} \rightarrow \pm \ell$. Instead, the $\ell$-dependence of their forward limits
occurs via \textit{linearized propagators} given by the inverses of
\beq
s_{i_1i_2\ldots i_r,\pm \ell} = s_{i_1i_2\ldots i_r} \pm 2 \ell\cdot k_{i_1i_2\ldots i_r} 
\label{rev.23}
\eeq
and formally obtained from discarding the $\ell^2$ summand in $(\ell{\pm}k_{i_1i_2\ldots i_r} )^{-2}$. The
linearized propagators $s_{i_1i_2\ldots i_r,\pm \ell} ^{-1}$ in the amplitudes (\ref{rev.22}) can be 
reconciled with the expected Feynman propagators quadratic in $\ell$ by performing the standard 
partial-fraction decomposition
\begin{align}
\int\limits_{\mathbb R^D} {  {\rm d}^D \ell  \, \varphi(\ell) \over \ell^2 (\ell{+}k_1)^2 (\ell{+}k_{12})^2 \ldots (\ell {+} k_{12\ldots n-1})^2}
&= \sum_{i=0}^{n-1} \ \int\limits_{\mathbb R^D} {    {\rm d}^D \ell  \, \varphi(\ell)\over (\ell {+} k_{12\ldots i})^2 } \prod^n_{j \neq i} {1\over (\ell {+} k_{12\ldots j})^2 - (\ell {+} k_{12\ldots i})^2 }\notag \\
&\hspace{-3cm}= \sum_{i=0}^{n-1}\  \int\limits_{\mathbb R^D} { {\rm d}^D \ell \over \ell^2}  \, \varphi(\ell{-}k_{12\ldots i}) \prod_{j=0}^{i-1} {1\over s_{j+1,j+2,\ldots,i,-\ell}} \prod_{j=i+1}^{n-1} {1\over s_{i+1,i+2,\ldots,j,\ell} } \label{ngonPF}
\end{align}
The numerator factor $\varphi(\ell)$ refers to an arbitrary polynomial in the loop momentum, and
each summand of the second line is interpreted as the forward limit of a tree diagram, 
see figure~\ref{figpfcut}. In the setting of section \ref{sec:5.3}, the numerators $\mN$ of 
the one-loop diagrams on quadratic propagators take the role of $\varphi(\ell)$ on the left-hand side
of (\ref{ngonPF}).
The coefficients $N_{+|\rho(12\ldots n)| -} $ in the Parke-Taylor expansion (\ref{rev.7}) 
in turn are identified with the shifted numerators $\varphi(\ell{-}k_{12\ldots i})$ on the right-hand side
of (\ref{ngonPF}), up to redefinitions that follow from the scattering equations among different permutations
of $ {\rm PT}(+,\rho(1,2,\ldots,n),-) $.

Conversely, however, it is not possible to recombine individual summands in the
second line of (\ref{ngonPF}) to Feynman integrals with quadratic propagators 
$(\ell{\pm}k_{i_1i_2\ldots i_r} )^{-2}$. In particular, the shifts in the argument of the
polynomial $\varphi(\ell{-}k_{12\ldots i})$ between different summands
in (\ref{ngonPF}) require even more care in recombining linearized 
propagators to quadratic ones. It is not obvious from the general
expression (\ref{rev.22}) for one-loop super-Yang-Mills and supergravity
amplitudes that they are eventually expressible in terms of quadratic Feynman propagators.

Note that a variety of strategies to recombine linearized propagators in the second line of (\ref{ngonPF})
to quadratic ones and to directly obtain quadratic propagators from ambitwistor strings can be found in \cite{Gomez:2016cqb, Gomez:2017lhy, Gomez:2017cpe, Ahmadiniaz:2018nvr, Agerskov:2019ryp, Farrow:2020voh, Feng:2022wee, Dong:2023stt, Xie:2024pro}.

%%%%%%%%%%%%%%%%%%%%%%%%%%%%%%%%%%%%%%%%%%%%%%%% 
%%%%%%%%%%%%%%%%%%%%%%%%%%%%%%%%%%%%%%%%%%%%%%%% 

\subsection{Matching pinching rules with ambitwistor integrals}
\label{sec:6.9}

When extracting ambitwistor-string integrands ${\cal I}_n(\ell)$ from chiral correlators ${\cal K}_n(\ell)$ on the torus via (\ref{chy.01}), one finds that individual terms $g^{(1)}_{ij}$ or $\ell^m$ of ${\cal K}_n(\ell)$ do not yield amplitude contributions to (\ref{rev.22}) that can be recombined to quadratic propagators. This is the ambitwistor-string counterpart of the closely related issue in section \ref{sec:2.3.3} that Koba-Nielsen integrals over
individual terms $g^{(1)}_{ij}$ or $\ell^m$ of chiral correlators do not have a well-defined field-theory limit. In the same was as homology invariant combinations of $g^{(k)}_{ij}$ and $\ell^m$ resolve all ambiguities in the field-theory limit of string integrals, 
it was conjectured in section 5.2 of \cite{Edison:2021ebi} that homology invariance guarantees the availability of quadratic-propagator representations. More specifically, the reference claims that the loop integrals
\begin{align}
  \int_{\RR^D} \dd^{D} \ell \  {\rm AT}^{\rm op}_{12\ldots n}\big[E_P \big]  &=
 \int_{\mathbb R^D} \frac{ \dd^D \ell}{\ell^2}
\int  _{\mathbb C^{n-1}} \dd \mu_{n+2}^{\te{tree}}  \lim_{ \tau \rightarrow i\infty\atop{ \alpha' \rightarrow \infty}} \,  \frac{  E_P(\sigma_j,\ell,\tau) }{(2\pi i)^{n-4}\sigma_1 \sigma_2 \ldots \sigma_n} 
\label{prop.79} \\
&\quad\quad\quad\quad\quad\quad \times
\sum_{\gamma \in {\rm cyc}(1,2,\ldots,n)} \! \! \! \! \! \! {\rm PT}(+,\gamma(1,2,\ldots,n),-)  \notag \\
  \int_{\RR^D} \dd^{D} \ell \ {\rm AT}^{\rm cl}_n\big[ E_P  E_Q \big] &=
 \int_{\mathbb R^D} \frac{ \dd^D \ell}{\ell^2}
\int_{\mathbb C^{n-1}} \dd \mu_{n+2}^{\te{tree}}  \lim_{ \tau \rightarrow i\infty\atop{ \alpha' \rightarrow \infty}}  \, \frac{ E_P(\sigma_j,\ell,\tau )\,
E_Q(\sigma_j,\ell,\tau ) }{ (2\pi)^{2n-8} | \sigma_1 \sigma_2 \ldots \sigma_n |^2} \notag 
\end{align}
adapted to super-Yang-Mills and supergravity amplitudes can be
combined to those over quadratic propagators for arbitrary homology invariants 
$E_P(z_j,\ell,\tau) $ and $E_Q(z_j,\ell,\tau) $. The quadratic-propagator representations
of these ${\rm AT}^{\rm op}_{12\ldots n}[\ldots]$ and ${\rm AT}^{\rm cl}_n[\ldots]$ 
can only be attained after shifts of loop momentum as in (\ref{ngonPF}), that is
why incorporated the loop integrations into the definition (\ref{prop.79}).

%%%%%%%%%%%%%%%%%%%%%%%%%%%%%%%%%%%%%%%%%%%%%%%% 
%%%%%%%%%%%%%%%%%%%%%%%%%%%%%%%%%%%%%%%%%%%%%%%% 

\subsubsection{Comparison at the level of homology invariants}
\label{sec:6.1.3}

The combinations of loop integrals in (\ref{prop.79}) are the direct ambitwistor-string analogues of
the notation (\ref{attempt1.0}) for field-theory limits ${\rm FT}^{\rm op}_{12\ldots n}[\ldots]$ 
and ${\rm FT}^{\rm cl}_n[\ldots]$ which are also defined only up to shifts of loop momenta. 
Since ambitwistor strings and field-theory limits of conventional superstrings are expected
to yield the same super-Yang-Mills and supergravity amplitudes, a necessary requirement is to have
\beq
{\rm FT}^{\rm op}_{12\ldots n}\big[ {\cal K}_n  \big] = {\rm AT}^{\rm op}_{12\ldots n}\big[ {\cal K}_n  \big] \, , \ \ \ \ 
{\rm FT}^{\rm cl}_n\big[ {\cal K}_n \overline{{\cal K}_n} \big] = {\rm AT}^{\rm cl}_n\big[ {\cal K}_n \tilde{\cal K}_n\big]
\, , \ \ \ \ \ {\rm mod} \ (\ell \rightarrow \ell{\pm} k_j)
\label{prop.80}
\eeq
at the level of the full-fledged chiral amplitudes. However, we shall here make the stronger claim
that the FT$[\ldots]$ and AT$[\ldots]$ prescriptions in (\ref{attempt1.0}) and (\ref{prop.79}) already 
match at the level of individual homology invariants
\beq
{\rm FT}^{\rm op}_{12\ldots n}\big[ E_P \big] = {\rm AT}^{\rm op}_{12\ldots n}\big[ E_P  \big] \, , \ \ \ \ 
{\rm FT}^{\rm cl}_n\big[ E_P \overline{E_Q} \big] = {\rm AT}^{\rm cl}_n\big[ E_P E_Q\big]
\, , \ \ \ \ \ {\rm mod} \ (\ell \rightarrow \ell{\pm} k_j)
\label{prop.81}
\eeq
which implies (\ref{prop.80}) without any further information on the kinematic coefficients of $E_P$
within the chiral correlators. This is plausible since both the FT$[\ldots]$ and AT$[\ldots]$
prescriptions share the 
consistency condition of homology invariance for the expression in the square brackets.
Moreover, we have explicitly verified (\ref{prop.81}) for all the homology invariants in
the five- and six-point correlators of section \ref{sec:2.3} (including all the $51\times 51$
combinations $E_P  \overline{E_Q}$ for ${\rm FT}^{\rm cl}_6$ and $E_P E_Q$ for ${\rm AT}^{\rm cl}_6$). Note that $ E_P$ and
$ \overline{E_Q} $ in (\ref{prop.81}) need to obey the homology-invariance condition
(\ref{defhi}) already before integration over $z_i$, that is why $E^{\rm naive}_{1|2|3,4,5,6}$ 
in (\ref{enaiv}) with Koba-Nielsen derivatives as its homology variation is not a valid choice.

The advantage of (\ref{prop.81}) over (\ref{prop.80}) is that different worldsheet approaches to
loop amplitudes can be compared at a refined level. Given that ${\cal K}_6$
in the form of (\ref{KK6GEF}) comprises 51 homology invariants, (\ref{prop.81}) allows us to study the interplay 
of conventional superstrings and ambitwistor strings on the basis of considerably smaller expressions than
the complete six-point one-loop amplitudes. By the intricacies of the pinching rules 
in presence of multivalued $g^{(k\geq 2)}$ as described in section \ref{sec:3.1}, it is valuable to have
(\ref{prop.81}) as an alternative evaluation method of field-theory limits, either as cross-checks
or to optimize runtimes in computer-algebra implementations. Conversely, it may be useful that
direct calculations of $\alpha' \rightarrow 0$ limits anticipate the results of recombining linearized-propagator
expressions in intermediate steps of AT$[\ldots]$ to quadratic propagators.

%%%%%%%%%%%%%%%%%%%%%%%%%%%%%%%%%%%%%%%%%%%%%%%% 
%%%%%%%%%%%%%%%%%%%%%%%%%%%%%%%%%%%%%%%%%%%%%%%% 

\subsubsection{Naive integration by parts in ambitwistor strings: six-point case study}
\label{sec:6.1.7}

The second main result of this section besides (\ref{prop.81}) is that the
homology invariants with double poles or other non-logarithmic singularities
admit more direct simplification methods in an ambitwistor setting.
More specifically, the boundary term of section \ref{sec:4.1} which obstructed
a naive integration-by-parts simplification of $E_{1|2|3,4,5,6}$ via (\ref{naiveE})
does not have any analogue on the support of the scattering equations
in the ambitwistor measure (\ref{rev.2}). This can be seen by applying
the degenerations (recall that $\sigma_i = e^{2\pi i z_i}$)
\beq
\frac{1}{2\pi i}\lim_{\tau \rightarrow i \infty} g^{(1)}_{ab} = \frac{\sigma_a {+} \sigma_b}{2 \sigma_{ab}} \, , \ \ \ \
\frac{1}{(2\pi i)^2}\lim_{\tau \rightarrow i \infty} g^{(2)}_{ab} = \frac{1}{12}
\label{prop.83}
\eeq
to the limits on the right-hand side of (\ref{prop.79}) relevant to the AT$[\ldots]$ prescriptions:
\begin{align}
 \lim_{ \tau \rightarrow i\infty\atop{ \alpha' \rightarrow \infty}}  \frac{ E_{1|2|3,4,5,6} }{(2\pi i)^2 \sigma_1  \sigma_2\ldots  \sigma_6 } &= 
\frac{k_1 \cdot k_2}{(2\pi i)^2 \sigma_1  \sigma_2\ldots  \sigma_6 }   \lim_{ \tau \rightarrow i\infty }  \big[ (g_{12}^{(1)})^2 -  2 g^{(2)}_{12}   \big] \notag\\
&= \frac{k_1 \cdot k_2}{  \sigma_3  \sigma_4  \sigma_5  \sigma_6 }
\bigg\{
\frac{1}{\sigma_{12}^2} + \frac{1}{12 \sigma_1 \sigma_2}
\bigg\}  \label{prop.84} 
\end{align}
This needs to be compared with the AT$[\ldots]$ images of $E^{\rm naive}_{1|2|3,4,5,6}$ in (\ref{enaiv}).
The latter does not depend on $\alpha'$, so we only need to implement the
$\tau \rightarrow i \infty$ limit via (\ref{prop.83}) and obtain
\begin{align}
 \lim_{ \tau \rightarrow i\infty}  \frac{ E^{\rm naive}_{1|2|3,4,5,6} }{(2\pi i)^2 \sigma_1  \sigma_2\ldots  \sigma_6 } &=
 \frac{1}{\sigma_1 \sigma_2 \ldots \sigma_6} \bigg\{ \ell {\cdot} k_2 \frac{\sigma_1{+}\sigma_2}{ 2 \sigma_{12}} -
 \frac{k_1 {\cdot} k_2}{6} +
   \frac{\sigma_1{+}\sigma_2}{ 2 \sigma_{12}}\sum_{j=3}^6 k_2 {\cdot} k_j \frac{\sigma_2{+}\sigma_j}{ 2 \sigma_{2j}}
  \bigg\}
   \label{prop.85} 
\end{align}
On the support of the scattering equations and momentum conservation, the rational functions
on the right-hand sides of (\ref{prop.84}) and (\ref{prop.85}) match since their difference is given by
\beq
 \lim_{ \tau \rightarrow i\infty\atop{ \alpha' \rightarrow \infty}} 
  \frac{ E^{\rm naive}_{1|2|3,4,5,6} - E_{1|2|3,4,5,6} }{(2\pi i)^2 \sigma_1  \sigma_2\ldots  \sigma_6 }
 = \frac{1}{ \sigma_3  \sigma_4  \sigma_5  \sigma_6}
 \bigg\{ \frac{1}{\sigma_{12}} - \frac{1}{2\sigma_1} \bigg\} \bigg( \frac{\ell \cdot k_2}{\sigma_2} + \sum_{j=1 \atop{j\neq 2}}^6  \frac{ k_2 \cdot k_j }{\sigma_{2j}} \bigg) \cong 0
    \label{prop.86}
 \eeq
More precisely, the expression in the round parenthesis vanishes on the support of the scattering equation 
at $i=2$ in (\ref{rev.2}) in the ${\rm SL}(2,\mathbb C)$ frame with $(\sigma_{n+1},\sigma_{n+2}) \rightarrow (0,\infty)$
and the forward limit $(k_{n+1},k_{n+2}) \rightarrow (\ell,{-}\ell)$. 

%%%%%%%%%%%%%%%%%%%%%%%%%%%%%%%%%%%%%%%%%%%%%%%% 
%%%%%%%%%%%%%%%%%%%%%%%%%%%%%%%%%%%%%%%%%%%%%%%% 

\subsubsection{Naive integration by parts in ambitwistor strings: general argument}
\label{sec:6.1.0}

The viability of the naive integration-by-parts simplification (\ref{naiveE})
in an ambitwistor-string context is not surprising since 
$E^{\rm naive}_{1|2|3,4,5,6}$ is obtained from $E_{1|2|3,4,5,6} $ by discarding a total derivative
\beq
\partial_{z_a} \big( h(z_i,\tau,\ell)   {\cal J}_n\big) =
\bigg\{  \partial_{z_a}  h(z_i,\tau,\ell) 
+ \alpha' h(z_i,\tau,\ell) 
\bigg( 4\pi i   (\ell \cdot k_a) + \sum_{b=1\atop{b\neq a}}^n s_{ab} g^{(1)}_{ab}\bigg) \bigg\} \,
 {\cal J}_n
  \label{prop.99}
\eeq
in the special case of $n=6$, $a=2$ and $ h(z_i,\tau,\ell) = g^{(1)}_{12}$. 
For a general choice of  $n$ and $ h(z_i,\tau,\ell)$, the
$\alpha' \rightarrow \infty$ limit relevant to
the integrand of ambitwistor strings
suppresses the first term $\partial_{z_a}  h(z_i,\tau,\ell) $ in the curly bracket.
The second term in turn is proportional to $\partial_{z_a}\log {\cal J}_n$ and vanishes on
the support of the scattering 
equations on the torus \cite{Adamo:2013tsa}. By this general argument, one can discard arbitrary total
derivatives (\ref{prop.99}) when exporting chiral correlators to the ambitwistor-string setting via (\ref{chy.01}), 
even if the function $h(z_i,\tau,\ell) $ is not homology invariant. That is why the boundary term in
section \ref{sec:4.1} obtained from integrating (\ref{prop.99}) at $h(z_i,\tau,\ell) = g^{(1)}_{12}$ does not 
have any ambitwistor analogue, and one can identify $E_{1|2|3,4,5,6} \rightarrow E^{\rm naive}_{1|2|3,4,5,6}$ 
within the AT$[\ldots]$ prescription (\ref{prop.79}). However, we reiterate that it is not possible to
replace $E_{1|2|3,4,5,6} \rightarrow E^{\rm naive}_{1|2|3,4,5,6}$ within the prescription
${\rm FT}^{\rm cl}_n[\ldots]$ of conventional strings
since the boundary terms of section \ref{sec:4.1} generically have non-vanishing field-theory limits.

Note that the degeneration $\tau \rightarrow i \infty$ reduces the
scattering equations on the torus $\partial_{z_a}\log {\cal J}_n=0$ 
to those on the sphere enforced by the delta distributions in (\ref{rev.2}) \cite{Geyer:2015bja, Geyer:2015jch}.
Accordingly, (\ref{prop.86}) is nothing but the  $\tau \rightarrow i \infty$ limit of the relation
\beq
E^{\rm naive}_{1|2|3,4,5,6}  = \lim_{\alpha' \rightarrow \infty}E_{1|2|3,4,5,6} 
+ \frac{1}{2\alpha'}  g^{(1)}_{12} \partial_{z_2}\log {\cal J}_6
  \label{prop.87}
\eeq
which explains why the curly parenthesis in (\ref{prop.86}) is given by 
the $\tau \rightarrow i\infty$ limit of $\frac{g^{(1)}_{12}}{2\pi i \sigma_1}$.

%%%%%%%%%%%%%%%%%%%%%%%%%%%%%%%%%%%%%%%%%%%%%%%% 
%%%%%%%%%%%%%%%%%%%%%%%%%%%%%%%%%%%%%%%%%%%%%%%% 

\section{Conclusion and outlook}
\label{sec:9}

In this work, we have developed advanced tools to perform the field-theory
limit of multiparticle one-loop string amplitudes in their chiral-splitting formulation
\cite{Verlinde:1987sd, DHoker:1988pdl, DHoker:1989cxq}.
The latter employs loop momenta to express closed-string integrands as 
double copies of meromorphic chiral amplitudes. Starting from six external legs, the 
modern formulation of chiral amplitudes in terms of Kronecker-Eisenstein coefficients is shown
to introduce technical pitfalls in the pinching rules that determine the reducible
diagrams in the field-theory limit. We close a gap in the literature by discussing
and overcoming subtleties in multiparticle pinching rules and restrictions on integration-by-parts
simplifications of chiral amplitudes.

The goal of this work is to pave the way for practical calculations. Accordingly, our
extended pinching rules for chiral splitting are applied to six-point one-loop supergravity
amplitudes: we obtain a new representation (\ref{ftamp.09}) of their loop integrand
from the field-theory limit of type-IIA/B superstrings. As an echo of chiral splitting,
 the kinematic factors of the box-, pentagon- and hexagon diagrams in the loop 
 integrand (\ref{ftamp.09}) are presented in a double-copy form. However, instead of
 the initial cubic-diagram formulation of the gravitational double copy \cite{loopBCJ}, 
 our supergravity integrand also involves contact terms akin to the generalized double 
copy \cite{Bern:2017yxu}.
We leave it as an open problem for the future to identify the 
incarnation of the generalized double copy at higher multiplicity and loop order
which results from the extended pinching rules of chiral splitting.

Our representation (\ref{ftamp.09}) of six-point one-loop supergravity amplitudes for the first time 
reconciles manifest spacetime supersymmetry via pure-spinor methods with the conventional
Feynman propagators quadratic in the loop momentum. This standard form of the
propagators is sometimes obscured in ambitwistor-string constructions
and closely related all-multiplicity realizations of one-loop double copy in field theory \cite{He:2016mzd, He:2017spx}.
The outcomes of our pinching rules in terms of quadratic propagators
match the results of ambitwistor-string computations, not only at the level of the full
six-point amplitudes but already for considerably smaller portions that arise from
so-called homology invariants on the torus. Hence, our findings identify homology invariants
as the optimal organization level of amplitude computations for the comparison 
of pinching rules with ambitwistor methods and for a broader exploration of
their synergies. A notable example of such synergies between ambitwistor and conventional
string theories is the construction of chiral amplitudes from color-kinematics
dual numerators in \cite{Geyer:2021oox, Geyer:2024}.

Another open problem for the future is the all-multiplicity systematics of the refined
integration-by-parts relations that preserve the homology-invariance condition
required by chiral splitting. At six points, chiral splitting necessitates the more
elaborate integration-by-parts relation (\ref{bterm.09}) to eliminate double poles
as compared to the simpler identity (\ref{bdsec.04}) of the
 manifestly doubly-periodic setup after integrating the loop momentum. 
A comprehensive integration-by-parts elimination of non-logarithmic singularities
after loop integration is performed in
\cite{Rodriguez:2023qir, Zhang:2024yfp}. It remains to pinpoint the total derivatives 
of homology invariants that eliminate the analogous non-logarithmic singularities 
in chiral splitting:
a key step is already taken in the all-multiplicity relations in section 5.1.3 of \cite{Mafra:2018pll},
but the residual task is to pinpoint convenient tensor contractions of the tensor-valued identities 
in the reference.

It would furthermore be interesting to adapt the pinching rules of this work to
string- and field-theory amplitudes with reduced supersymmetry. On the one hand,
the parity-even parts of $n$-point amplitudes with half- and
quarter-maximal supersymmetry share the combinatorial structure and
the types of Kronecker-Eisenstein coefficients with maximally supersymmetric
$(n{+}2)$-point amplitudes \cite{Berg:2016wux, Berg:2016fui}. On the other hand,
it remains to analyze the interplay of the infrared-regularization scheme of the references
for $s_{12\ldots n-1}^{-1}$ dubbed {\it Minahaning} \cite{Minahan:1987ha} with the kinematic poles introduced
by our pinching rules in non-maximally supersymmetric cases. Upcoming work
of Berg, Haack and Zimmerman \cite{Berg:2024} will describe new facets of Minahaning in gravitational
four-point amplitudes with 16 supercharges which correspond to the six-point
amplitude (\ref{ftamp.09}) under the combinatorial dictionary of \cite{Berg:2016wux, Berg:2016fui}.

At higher genus, the worldsheet integrands of multiparticle string amplitudes
are expected to have natural representations in terms of the integration
kernels for higher-genus polylogarithms. Particularly promising candidates
are the single-valued but non-holomorphic $f$-tensors of the polylogarithms 
in \cite{DHoker:2023vax} -- see \cite{DHoker:2023khh} for their appearance from fermion Green functions
and \cite{DHoker:2024ozn} for their functional identities. Their analogues for the loop integrand
in chiral splitting are the meromorphic but 
multivalued Enriquez kernels \cite{Enriquez:2011} which were recently investigated from the
viewpoint of Schottky parametrizations \cite{Baune:2024biq} and functional identities \cite{DHoker:2024ozn, Baune:2024ber}. An important open problem is to formulate pinching rules for these types of higher-genus integration kernels and the associated homology invariants as done at the two-loop five-point level in \cite{DHoker:2020prr}.

%%%%%%%%%%%%%%%%%%%%%%%%%%%%%%%%%%%%%%%%%%%%%%%% 
%%%%%%%%%%%%%%%%%%%%%%%%%%%%%%%%%%%%%%%%%%%%%%%% 
%%%%%%%%%%%%%%%%%%%%%%%%%%%%%%%%%%%%%%%%%%%%%%%% 
%%%%%%%%%%%%%%%%%%%%%%%%%%%%%%%%%%%%%%%%%%%%%%%% 
%%%%%%%%%%%%%%%%%%%%%%%%%%%%%%%%%%%%%%%%%%%%%%%% 

\acknowledgments
We are grateful to Marcus Berg, Eric D'Hoker, Jin Dong, Michael Haack, Cristhiam Lopez-Arcos, Carlos Mafra, Christian Schubert, Piotr Tourkine, Yong Zhang and Yonatan Zimmerman for combinations of inspiring discussions, correspondence, comments on the draft and collaboration on related topics. We especially thank Ricardo Monteiro and Carlos Rodriguez for many helpful discussions. 
The research of AE was supported by the US DOE under contract DE-SC0015910 and by Northwestern University via the Amplitudes and Insight Group, Department of Physics and Astronomy, and Weinberg College of Arts and Sciences.
The research of OS is supported by the European Research Council under ERC-STG-804286 UNISCAMP and the strength area ``Universe and mathematical physics'' which is funded by the Faculty of Science and Technology at Uppsala University.

%%%%%%%%%%%%%%%%%%%%%%%%%%%%%%%%%%%%%% 
%%%%%%%%%%%%%%%%%%%%%%%%%%%%%%%%%%%%%% 
%%%%%%%%%%%%%%%%%%%%%%%%%%%%%%%%%%%%%% 
%%%%%%%%%%%%%%%%%%%%%%%%%%%%%%%%%%%%%% 

\appendix
\section*{Appendix}

%%%%%%%%%%%%%%%%%%%%%%%%%%%%%%%%%%%%%%%%%%%%%%%% 
%%%%%%%%%%%%%%%%%%%%%%%%%%%%%%%%%%%%%%%%%%%%%%%% 

\section{Reducible diagrams from worldline boundary terms}
\label{bdyWL}

As detailed in section \ref{sec:3.1}, the pinching rule (\ref{newPR.02}) producing
a pentagon from six-point integrals over $g^{(2)}_{16}$ is specific to 
meromorphic Kronecker-Eisenstein kernels with simple poles at 
$z_{61}{=}\pm\tau$ and does not arise for its doubly-periodic and non-singular
counterpart $f^{(2)}_{16}$. We shall here describe an alternative mechanism that
generates reducible diagrams in field-theory limits from 
representations of homology invariants
in terms of $f^{(k)}_{ij}$. 
As we will see, analyzing field-theory limits of $E_{1|2|3,4,5,6}$
in terms of $f^{(k)}_{ij}$ introduces boundary terms with respect to worldline variables 
that reproduce the pentagon in (\ref{newPR.02}) from a different perspective.

Throughout this appendix, the discussion is tailored to the contribution to one-loop
open-string amplitudes from planar cylinder worldsheets in the parametrization of
figure \ref{figWS}. The closed-string counterpart
of the analysis is similar but combinatorially more involved.
The key steps in this appendix also appeared in section 6.2 of \cite{Berg:2016fui}
in the context of four-point one-loop open-string amplitudes with reduced supersymmetry.

%%%%%%%%%%%%%%%%%%%%%%%%%%%%%%%%%%%%%%%%%%%%%%%% 
%%%%%%%%%%%%%%%%%%%%%%%%%%%%%%%%%%%%%%%%%%%%%%%% 

\subsection{Non-chiral Koba-Nielsen factor and $f^{(k)}$ representations}
\label{bdyWL.1}

The first step is to provide a manifestly doubly-periodic representation of 
the homology invariant $E_{1|2|3,4,5,6}$ in (\ref{Esat6pt}) with only simple poles.
For this purpose, we rewrite the chiral Koba-Nielsen factor (\ref{baspin.12}) as
\begin{align}
{\cal J}_n &= \exp\bigg( 2\pi i  \alpha'  \tau \bigg[ \ell + \sum_{j=1}^n \frac{z_j}{\tau} k_j \bigg]^2 
- \frac{ 2\pi i  \alpha' }{ \tau } \sum_{i=1}^n  z_i k_i \cdot  \sum_{j=1}^n z_j k_j 
 + \alpha' \sum_{1\leq i<j}^n s_{ij} \log \theta_1(z_{ij},\tau)  \bigg)  \notag \\
&= \exp\bigg( 2\pi i \tau \alpha' p^2  + \alpha' \sum_{1\leq i<j}^n s_{ij} {\cal G}_B(z_{ij},\tau) \bigg)
\label{bdsec.01}
\end{align}
with the shifted loop momentum $p$ and the $B$-cycle Green function ${\cal G}_B$ in the second line
\begin{align}
p^ m = \ell^m + \sum_{j=1}^n \frac{z_j}{\tau} k^m_j  \, , \ \ \ \ 
{\cal G}_B(z,\tau) =  \log \theta_1(z,\tau) +\frac{ i\pi  z^2}{\tau}
\label{bdsec.02}
\end{align}
For chiral correlators ${\cal K}_n$ independent on $\ell$, the Gaussian loop integral over
the first factor $e^{ 2\pi i  \tau \alpha' p^2 }$ with purely imaginary $\tau$ can be readily 
performed to yield a non-chiral version $\Pi_n$ of the Koba-Nielsen factor (\ref{baspin.12})
\begin{align}
\int_{\mathbb R^D} \dd^D \ell\, 
{\cal J}_n  =  \frac{  \Pi_n }{  (2 \ap \Im \tau)^{D/2} }  \, , \ \ \ \
\Pi_n = \exp\bigg( \alpha' \sum_{1\leq i<j}^n s_{ij} {\cal G}_B(z_{ij},\tau)  \bigg)
\label{bdsec.00}
\end{align}
The Green function ${\cal G}_B$ in (\ref{bdsec.02}) is meromorphic, and its
$z$-derivative coincides with the restriction of the non-meromorphic $f^{(1)}$ to the boundary 
of the cylinder worldsheet in figure~\ref{figWS}, i.e.\ to the 
$B$-cycle $[0,1)\times \tau$ where $\frac{z}{\tau} = \frac{\Im z}{\Im \tau}$,
\beq
\partial_z {\cal G}_B(z,\tau)  = \partial_z  \log \theta_1(z,\tau) + 2\pi i \frac{z}{\tau} 
=  \partial_z  \log \theta_1(z,\tau) + 2\pi i \frac{\Im z}{\Im \tau} \, \bigg|_{[0,1)\tau}
= f^{(1)}(z,\tau) \, \big|_{[0,1)\tau} 
\label{bdsec.03}
\eeq
Conversely, with the understanding of $z$ and $\tau $ as open-string 
moduli with one real (instead of complex) degree of freedom each, the second
derivative of ${\cal G}_B$ yields $ \partial_z f^{(1)}(z,\tau)  \big|_{[0,1)\tau} 
= \partial_z g^{(1)}(z,\tau){+} \frac{2\pi }{\Im \tau} $
upon restriction of $z$ to the $B$-cycle and $\tau \in i \mathbb R_+$.\footnote{In a closed-string context where $ \partial_z$ is understood as a complex derivative distinct from $\partial_{\bar z}$, the analogous differential equation is $ \partial_z f^{(1)}(z,\tau)   = \partial_z g^{(1)}(z,\tau){+} \frac{\pi }{\tau} $ without the factor of two.}
As a consequence, the homology invariant $E_{1|2|3,4,5,6}$ admits the following rewriting
within cylinder amplitudes
\begin{align}
E_{1|2|3,4,5,6} &= - \frac{\pi}{\alpha' \Im \tau} + \frac{1}{2\alpha'} \partial_{z_1}f^{(1)}_{12} + \frac{1}{2} s_{12} ( f^{(1)}_{12} )^2 - s_{12} f^{(2)}_{12}
\label{bdsec.04} \\
&\cong  - \frac{\pi}{\alpha' \Im \tau} - s_{12} f^{(2)}_{12} + \frac{1}{2}   f^{(1)}_{12} ( s_{23}   f^{(1)}_{23}
+ s_{24}   f^{(1)}_{24} + s_{25}   f^{(1)}_{25} + s_{26}   f^{(1)}_{26})
\notag
\end{align}
where all the singularities of the second line are logarithmic.
In passing to the second line, we have shifted $E_{1|2|3,4,5,6} \Pi_n$ by a total $z_2$ derivative of 
$ \frac{1}{2\alpha'} f^{(1)}_{12} \Pi_n$, where the non-chiral Koba-Nielsen factor $\Pi_n $ in 
(\ref{bdsec.01}) gives rise to
\beq
\partial_{z_a} \log \Pi_n = \alpha' \sum_{b=1 \atop{b \neq a}}^n s_{ab} f^{(1)}_{ab}
\label{bdsec.05} 
\eeq
maintaining the restriction of all the $z_i$ to the $B$-cycle. Nevertheless, 
the integration by parts in (\ref{bdsec.04}) is also valid for closed strings since
the function $f^{(1)}_{12}$ in the primitive is doubly periodic, and there is no analogue 
of the boundary term (\ref{bterm.02}).

%%%%%%%%%%%%%%%%%%%%%%%%%%%%%%%%%%%%%%%%%%%%%%%% 
%%%%%%%%%%%%%%%%%%%%%%%%%%%%%%%%%%%%%%%%%%%%%%%% 

\subsection{Worldline limits}
\label{bdyWL.2}

This section gathers the worldline limits of the non-chiral Koba-Nielsen factor (\ref{bdsec.00}) and the
representation (\ref{bdsec.04}) of $E_{1|2|3,4,5,6}$. It will be 
convenient to introduce the shorthands
\begin{align}
t = 2\pi \alpha' \Im \tau \, , \ \ \ \ u_j = \frac{\Im z_j}{\Im \tau} 
\label{bdsec.06} 
\end{align}
and all the subsequent field-theory limits $\alpha' \rightarrow 0$ and $\tau \rightarrow i\infty$
 are taken at fixed values of the
worldline length $t$ and the comoving coordinates~$u_j$. By the degeneration 
(\ref{baspin.15}) of the $\theta_1$-function,
the $\tau \rightarrow i\infty$ asymptotics of the non-chiral Koba-Nielsen factor is given by
\begin{align}
\Pi_n  \rightarrow \Pi^{i\infty}_n = \exp\bigg( {-} t  \sum_{1\leq i<j}^n k_{i}\cdot k_j \big(  u_{ij}^2 - | u_{ij}| \big) \bigg)
\label{bdsec.07} 
\end{align}
which reproduces the exponential in the Schwinger parametrization (\ref{swpara.01}) of Feynman integrals.
The first step towards the analogous worldline limit of the homology invariant $E_{1|2|3,4,5,6}$ is
to identify
\begin{align}
\frac{1}{2\pi i}\lim_{\tau \rightarrow i\infty} f^{(1)}_{ij} = u_{ij}  - \frac{1}{2}\sgn(u_{ij}) \, ,  \ \ \ \
\frac{1}{(2\pi i)^2}\lim_{\tau \rightarrow i\infty} f^{(2)}_{ij} = \frac{1}{2} u_{ij}^2 - \frac{1}{2} | u_{ij} | + \frac{1}{12}
\label{bdsec.08} 
\end{align}
based on the degenerations (\ref{baspin.17}) and (\ref{newPR.03}) of $g^{(k)}$, where we again keep
the comoving coordinates $u_j$ fixed.
In the first place, (\ref{bdsec.08}) brings the degeneration of (\ref{bdsec.04}) into the form
\begin{align}
\lim_{\tau \rightarrow i\infty} \frac{E_{1|2|3,4,5,6}}{(2\pi i)^2} &= \frac{1}{2t}
+ \frac{1}{2} \Big[ s_{23} \big( u_{12} u_{13} 
+ \tfrac{1}{2}\sgn(u_{12})(u_{21}{+} u_{31}) 
+ \tfrac{1}{2}\sgn(u_{23}) u_{21}
\label{bdsec.09}  \\
&\hspace{3.6cm}+ \tfrac{1}{4} \sgn(u_{12}) \sgn(u_{23})+ \tfrac{1}{6} \big) + (3\leftrightarrow 4,5,6) \Big]
\notag
\end{align}
The comoving coordinates $u_i$ are interpreted as proper times 
of the worldline that arises from the tropical degeneration of both torus and
cylinder worldsheets. In order to find a momentum-space representation of
(\ref{bdsec.09}), it remains to match the polynomials in $u_i$ with the
results of the Schwinger parametrizations (with $u_1=0$)
\begin{align}
&\int_{\mathbb R^D} \frac{\dd^D \ell}{\pi^{D/2}} \frac{\alpha + \beta_m \ell^m + \gamma_{mn} \ell^m \ell^n}{\ell^2 (\ell{-}K_1)^2 \ldots  (\ell{-}K_{12\ldots n-1})^2}
= \int^\infty_0 \frac{\dd t}{t}  \, t^{n-D/2} \,  \int_{0<u_2<\ldots <u_n<1}  \! \!  \dd u_2\, \dd u_3\,\ldots \, \dd u_n
\notag  \\
&\ \ \ \times 
 \bigg(
\alpha + \beta_m L^m(u_i) + \gamma_{mn} \, \bigg[ L^m(u_i) L^n(u_i) {+} \frac{\eta^{mn}}{2t} \bigg] \,
 \bigg)  \, \exp \bigg[ {-}t \sum_{1\leq i<j}^n K_i \cdot K_j \big( u_{ij}^2 - |u_{ij}| \big)\bigg]
\label{bdsec.10}
\end{align}
generalizing (\ref{swpara.01}) to vector and tensor Feynman integrals. In intermediate
steps of the derivation of (\ref{bdsec.10}), the loop integral is brought into 
Gaussian form whose evaluation necessitates a shift of the loop momentum by
\begin{align}
L^m(u_i) = - \sum^n_{j=1} K_j^m u_j
\label{bdsec.10a}
\end{align}
which is equivalent to $\ell^m = p^m+L^m(u_i)$ in (\ref{bdsec.02}) for $z_j$ on the $B$-cycle
and massless momenta $K_j^m \rightarrow k_j^m$.
In the form of (\ref{bdsec.09}), it is not possible to match the degree-two
polynomial in $u_i$ with a Schwinger-parametrized hexagon (\ref{bdsec.10}), regardless of the
choice of $\alpha,\beta_m,\gamma_{mn}$. Instead, we absorb the
quadratic terms in $u_i$ of (\ref{bdsec.09}) into the total worldline derivative of (\ref{bdsec.07})
 \begin{align}
\partial_{u_a} \log \Pi_n^{i\infty} = -t \sum_{b=1 \atop{b\neq a}}^n s_{ab} \, \big( u_{ab} - \tfrac{1}{2}\sgn(u_{ab}) \big)
\label{bdsec.11}  
\end{align}
such that we can recombine all the residual factors of $u_i$
into the shift of loop momentum in (\ref{bdsec.10a}) via
$s_{23} u_{31} + (3\leftrightarrow 4,5,6) = -2 k_2\cdot L$ with the massless
version $K_j^m \rightarrow k_j^m$ of the momenta in (\ref{bdsec.10a}):
\begin{align}
\lim_{\tau \rightarrow i\infty} \frac{E_{1|2|3,4,5,6}}{(2\pi i)^2}  \, \Pi_6&=  \frac{1}{2t} \partial_{u_2} \big( u_{21} \Pi^{i\infty}_6 \big) +  \frac{1}{2} \,  \Pi^{i\infty}_6\, \sgn(u_{21}) \, k_2 \cdot L 
\label{bdsec.12}  \\
&\quad
+ \Pi^{i\infty}_6 \bigg( \frac{1}{8}\sgn(u_{12}) \big[ s_{23} \sgn(u_{23})+ (3\leftrightarrow 4,5,6) \big]
- \frac{s_{12}}{12}\bigg)
\notag
\end{align}
One can also view this result as the worldline limit of (\ref{naiveE}) whose
total worldsheet derivative $- \frac{1}{2\ap}\partial_{z_2}  ( g^{(1)}_{12}  {\cal J}_6 )$ 
in the second line corresponds to the $u_2$-derivative in (\ref{bdsec.12}). Note, however,
that the worldline analogue (\ref{bdsec.07}) of the Koba-Nielsen factor 
does not share the short-distance behavior $| z_{ij}|^{2\alpha' s_{ij}}$
of ${\cal J}_n$. Instead, the limits $u_{i} \rightarrow u_j$ of $\Pi^{i\infty}_6$ are non-zero 
and give rise to the composite $K_a = k_{ij}$ in the exponentials of (\ref{bdsec.10}) reducible diagrams. 
Hence, while boundary terms $z_{i} \rightarrow z_j$ drop out from open-string 
worldsheet integrals, this is not the case for the worldline boundary terms
from $ \partial_{u_2} ( u_{21} \Pi^{i\infty}_6 )$ in (\ref{bdsec.12}).

%%%%%%%%%%%%%%%%%%%%%%%%%%%%%%%%%%%%%%%%%%%%%%%% 
%%%%%%%%%%%%%%%%%%%%%%%%%%%%%%%%%%%%%%%%%%%%%%%% 

\subsection{Contributions before pinching}
\label{bdyWL.3}

With the manifestly doubly-periodic representation (\ref{bdsec.04}) and its worldline
limit (\ref{bdsec.12}) at hand, we shall now present an alternative computation of the
field-theory limit of open-string integrals over $E_{1|2|3,4,5,6}$. Similar to the assembly of field-theory limits from
the pinching rules of $g^{(k)}_{ij}$ in the main body of this work, we evaluate the
pinch contributions to the representation (\ref{bdsec.04}) in terms of $f^{(k)}_{ij}$
in a separate section \ref{bdyWL.4} below. The present section is
dedicated to the first term in
\beq
{\rm FT}^{\rm op}_{12\ldots 6}\big[ E_{1|2|3,4,5,6} \big] 
= {\rm FT}^{i\infty}_{12\ldots 6}\big[ E_{1|2|3,4,5,6} \big]  
+ {\rm FT}^{\rm pinch}_{12\ldots 6}\big[ E_{1|2|3,4,5,6} \big] 
\label{bdsec.13} 
\eeq
which gathers the contributions from the worldline limit (\ref{bdsec.12}) before accounting 
for any pinches of (\ref{bdsec.04}). From the translation (\ref{bdsec.10}) between
momentum space and Schwinger parameters, the target
expression  ${\rm FT}^{i\infty}_{12\ldots 6}\big[ E_{1|2|3,4,5,6} \big]  $
will be determined from
\begin{align}
\int_{\mathbb R^D} \frac{\dd^D \ell}{\pi^{D/2}} \, &{\rm FT}^{i\infty}_{12\ldots 6}\big[ E_{1|2|3,4,5,6} \big]  
=  \int^\infty_0 \frac{\dd t}{t} \, t^{6-D/2}  \! \! \! \! \!  \int \limits_{0<u_2<\ldots <u_6<1}  \! \! \! \! \!  \dd u_2\,\ldots \, \dd u_6 \,
 \Pi_6^{i\infty} \lim_{\tau \rightarrow i\infty} \frac{E_{1|2|3,4,5,6}}{(2\pi i)^2} 
\label{bdsec.14}
\end{align}
Upon insertion of (\ref{bdsec.12}) on the right-hand side, we find
\beq
{\rm FT}^{i\infty}_{12\ldots 6}\big[ E_{1|2|3,4,5,6} \big]
= \bigg( \frac{1}{2} \ell \cdot k_2 - \frac{5}{24} s_{12} \bigg) I^{(6)}_{1,2,3,4,5,6} - \frac{1}{2} I^{(5)}_{1,23,4,5,6}[u_{13}]
\label{bdsec.15}
\eeq
 where $ I^{(5)}_{1,23,4,5,6}[u_{13}]$ refers to a one-mass pentagon with an additional insertion of
$u_{13}$ into its Schwinger representation (\ref{bdsec.10}) due to the total derivative on the right-hand side
of (\ref{bdsec.12}). Since $ I^{(5)}_{1,23,4,5,6}[u_{13}]$ will eventually drop out from (\ref{bdsec.13}) within worldline computations, we do not need the cumbersome momentum-space description of this 
quantity.\footnote{A momentum-space realization of the spurious quantity 
$ I^{(5)}_{1,23,4,5,6}[u_{13}]$ in (\ref{bdsec.15}) can be found by noticing that an extra factor of
$u_{13}$ in the worldline integrand of (\ref{bdsec.10}) results from the
loop integral of $ I^{(5)}_{1,23,4,5,6} \ell^m$: when expanding this vector integral
in the independent momenta
$k_{23}^m, k_4^m, k_5^m, k_6^m$ of the problem and projecting to the component along $k_{23}^m$,
the remnant $L^m$ of the loop momentum in (\ref{bdsec.10a}) for this one-mass pentagon
reduces to the desired factor of $u_{13}$ in the worldline integrand (with a renaming of the proper times
in (\ref{bdsec.10}) to $u_1=0< u_3<u_4< u_5< u_6 < 1$).} 
 
After the rewriting $2 \ell {\cdot} k_2 =  (\ell{-}k_1)^2
- (\ell{-}k_{12})^2+s_{12}$ of the hexagon numerator in (\ref{bdsec.15}), 
one arrives at the following alternative form:
\beq
{\rm FT}^{i\infty}_{12\ldots 6}\big[ E_{1|2|3,4,5,6} \big]
=   \frac{ s_{12} }{24} I^{(6)}_{1,2,3,4,5,6}
+ \frac{1}{4} I^{(5)}_{12,3,4,5,6}
-\frac{1}{4} I^{(5)}_{1,23,4,5,6}
 - \frac{1}{2} I^{(5)}_{1,23,4,5,6}[u_{13}]
\label{bdsec.16}
\eeq
Upon extending (\ref{bdsec.13}) and (\ref{bdsec.14}) to the permutations
$E_{1|3|2,4,5,6}$ and $E_{1|6|2,3,4,5}$ of the above homology invariant, similar
methods result in
\begin{align}
{\rm FT}^{i\infty}_{12\ldots 6}\big[ E_{1|3|2,4,5,6} \big]
&=   \frac{ s_{13} }{24} I^{(6)}_{1,2,3,4,5,6}
{+} \frac{1}{4} ( I^{(5)}_{1,23,4,5,6}
{-} I^{(5)}_{1,2,34,5,6} )
 {+}\frac{1}{2} ( I^{(5)}_{1,23,4,5,6}[u_{13}]
 {-} I^{(5)}_{1,2,34,5,6}[u_{13}] )\notag \\
 {\rm FT}^{i\infty}_{12\ldots 6}\big[ E_{1|6|2,3,4,5} \big]
&=   \frac{ s_{16} }{24} I^{(6)}_{1,2,3,4,5,6}
+ \frac{1}{4} I^{(5)}_{6\underline{1},2,3,4,5}
+ \frac{1}{4} I^{(5)}_{1,2,3,4,56}
+ \frac{1}{2} I^{(5)}_{1,2,3,4,56}[u_{15}] 
\label{bdsec.17}
\end{align}
where the coefficient $ \frac{1}{4} $ of $I^{(5)}_{6\underline{1},2,3,4,5}$ receives contributions
from both the boundary term
\begin{align}
 \int^\infty_0 \frac{\dd t}{t} \, t^{6-D/2}  \! \! \! \! \!  \!  \int \limits_{0<u_2<\ldots <u_6<1} \!  \! \! \! \! \!  \dd u_2\,\ldots \, \dd u_6 \, \partial_{u_6}\bigg( \frac{u_{61}}{ 2t} \, \Pi^{i\infty}_6 \bigg) =  \int_{\mathbb R^D} \frac{\dd^D \ell}{\pi^{D/2}} \,
  \frac{1}{2} \, \big( I^{(5)}_{6\underline{1},2,3,4,5} + I^{(5)}_{1,2,3,4,56}[u_{15}]  \big)
 \label{bdsec.18}
\end{align}
and the rewriting of a hexagon numerator $2 \ell {\cdot} k_6 = (\ell{+}k_6)^2 - \ell^2$ in intermediate steps.

%%%%%%%%%%%%%%%%%%%%%%%%%%%%%%%%%%%%%%%%%%%%%%%% 
%%%%%%%%%%%%%%%%%%%%%%%%%%%%%%%%%%%%%%%%%%%%%%%% 

\subsection{Adding pinched terms}
\label{bdyWL.4}

Our last step in the assembly of the field-theory limits of open-string integrals over $E_{1|i|j,p,q,r}$
is to combine (\ref{bdsec.16}) and (\ref{bdsec.17}) with the pinch contributions to (\ref{bdsec.13}). Given that the
doubly-periodic $f^{(k)}_{ij}$ with $k\geq 2$ are non-singular throughout the torus, 
the only source of reducible diagrams is $f^{(1)}_{ij}$. Its pinching rules are identical to those
of $g^{(1)}_{ij}$ and formally obtained from those in (\ref{swpara.00}) by replacing $\sgn(u_{ij}) \rightarrow \sgn(u_{ij})-2u_{ij}$. The terms in (\ref{swpara.00}) without any delta distributions 
(with two factors of $\sgn(u_{ij})$) need to be dropped since they are already accounted
for in section \ref{bdyWL.3}. For the permutations discussed
in the previous section, we find
\begin{align}
 {\rm FT}^{\rm pinch}_{12\ldots 6}\big[ E_{1|2|3,4,5,6} \big]  &= \frac{1}{4} I^{(5)}_{1,23,4,5,6} + \frac{1}{2} I^{(5)}_{1,23,4,5,6}[u_{13}]  - \frac{1}{4} I^{(5)}_{12,3,4,5,6} + \frac{\ell \cdot k_2}{s_{12}} I^{(5)}_{12,3,4,5,6} 
   \label{bdsec.19} \\
   &\quad - \frac{s_{26}}{2 s_{12} s_{612}} I^{(4)}_{6\underline{1}2,3,4,5}
 +  \frac{s_{23}}{2   s_{123}}  \bigg(  \frac{1}{s_{12}} + \frac{1}{s_{23}}\bigg) I^{(4)}_{123,4,5,6}
\notag \\
  {\rm FT}^{\rm pinch}_{12\ldots 6}\big[ E_{1|3|2,4,5,6} \big]  &= \frac{1}{4} I^{(5)}_{1,2,34,5,6} + \frac{1}{2} I^{(5)}_{1,2,34,5,6}[u_{13}]  - \frac{1}{4} I^{(5)}_{1,23,4,5,6} - \frac{1}{2} I^{(5)}_{1,23,4,5,6}[u_{13}] - \frac{I^{(4)}_{123,4,5,6}}{2 s_{123}}
  \notag \\
   {\rm FT}^{\rm pinch}_{12\ldots 6}\big[ E_{1|6|2,3,4,5} \big] &= \frac{1}{4}  I^{(5)}_{6\underline{1},2,3,4,5}
   - \frac{\ell \cdot k_6}{s_{16}}  I^{(5)}_{6\underline{1},2,3,4,5}
   - \frac{1}{4} I^{(5)}_{1,2,3,4,56} - \frac{1}{2} I^{(5)}_{1,2,3,4,56}[u_{15}] 
   \notag \\
   &\quad - \frac{s_{26}}{2 s_{16} s_{612}} I^{(4)}_{6\underline{1}2,3,4,5}
 +  \frac{s_{56}}{2   s_{561}}  \bigg(  \frac{1}{s_{56}} + \frac{1}{s_{16}}\bigg) I^{(4)}_{56\underline{1},2,3,4} \notag
\end{align}
By adding this to the $ {\rm FT}^{i\infty}_{12\ldots 6}\big[ E_{1|i|j,p,q,r} \big]$ in (\ref{bdsec.16}) and (\ref{bdsec.17}), we arrive at the final results for the field-theory limits
\begin{align}
	{\rm FT}^{\rm op}_{12\ldots 6}\big[ E_{1\vert 2\vert 3,4,5,6}\big]&=\frac{1}{24}s_{12}I^{(6)}_{1,2,3,4,5,6}
	+\frac{\ell\cdot k_2}{s_{12}}I^{(5)}_{12,3,4,5,6}\nonumber\\
	&\quad-\frac{s_{26}}{2s_{12}s_{126}}I^{(4)}_{6\underline12,3,4,5}
	+  \frac{s_{23}}{2   s_{123}}  \bigg(  \frac{1}{s_{12}} + \frac{1}{s_{23}}\bigg)  I^{(4)}_{123,4,5,6} \nonumber\\
	{\rm FT}^{\rm op}_{12\ldots 6}\big[ E_{1\vert 3\vert 2,4,5,6}\big]&=\frac{1}{24}s_{13}I^{(6)}_{1,2,3,4,5,6}-\frac{1}{2s_{123}}I^{(4)}_{123,4,5,6}\nonumber\\
	{\rm FT}^{\rm op}_{12\ldots 6}\big[ E_{1\vert 6\vert 2,3,4,5}\big]&=\frac{1}{24}s_{16}I^{(6)}_{1,2,3,4,5,6}+\bigg( \frac{1}{2} - \frac{\ell\cdot k_6}{s_{16}} \bigg)I^{(5)}_{6\underline1,2,3,4,5} \nonumber\\
	&\quad -\frac{s_{26}}{2s_{16}s_{126}}I^{(4)}_{6\underline12,3,4,5} 
	+ \frac{s_{56}}{2   s_{561}}  \bigg(  \frac{1}{s_{56}} + \frac{1}{s_{16}}\bigg)  I^{(4)}_{56\underline1,2,3,4}
	\label{bdsec.20}
\end{align}
which match the expressions in (\ref{bterm.10}) obtained from the pinching rules of $g^{(1)}_{ij}$
and $g^{(2)}_{ij}$ after minor rearrangements such as $(\ell{-}k_1)^2 = (\ell{-}k_{12})^2{+}2 \ell {\cdot}k_2{-}s_{12}$. Note that all the pentagons with insertions of $u_{1i}$ in the Schwinger integrand (such as
$I^{(5)}_{1,23,4,5,6}[u_{13}] $ in the case of $E_{1|2|3,4,5,6}$)
cancel between the pinch contributions (\ref{bdsec.19}) and those from the boundary terms in the
worldline limits of (\ref{bdsec.16}) and (\ref{bdsec.17}).

%%%%%%%%%%%%%%%%%%%%%%%%%%%%%%%%%%%%%%%%%%%%%%%% 
%%%%%%%%%%%%%%%%%%%%%%%%%%%%%%%%%%%%%%%%%%%%%%%% 

\section{Components of superspace kinematic factors}
\label{app:kin}

This appendix gathers further information on the components 
of the kinematic factors
in pure-spinor superspace that enter the chiral correlators
(\ref{KK6GEF}) at $n\leq 6$ points.

%%%%%%%%%%%%%%%%%%%%%%%%%%%%%%%%%%%%%%%%%%%%%%%% 
%%%%%%%%%%%%%%%%%%%%%%%%%%%%%%%%%%%%%%%%%%%%%%%% 

\subsection{Scalar kinematic factors in terms of SYM trees}
\label{app:kin.1}

We shall here generalize the four-point relation (\ref{KK4}) between
scalar BRST invariants $C_{1|A,B,C}$ and color-ordered tree-level amplitudes
$ A^{\rm tree}_{\rm SYM}(1,\ldots,n)$ of ten-dimensional super-Yang-Mills
to five and six points. By the all-multiplicity dictionary between
$C_{1|A,B,C}$ and $ A^{\rm tree}_{\rm SYM}(1,\ldots,n)$ in
appendix B.2 of \cite{EOMBBs}, we have
\beq
C_{1| 23, 4, 5} =  s_{45} \big[ s_{34} A^{\rm tree}_{\rm SYM}(1,2,3,4,5) - s_{24}   A^{\rm tree}_{\rm SYM}(1,3,2,4,5) \big]
\label{CtoAYM.1}
\eeq
as well as
\begin{align}
C_{1| 23 4, 5, 6} &= s_{45} s_{56} A^{\rm tree}_{\rm SYM}(1,2,3,4,5,6) + s_{25}
s_{56} A^{\rm tree}_{\rm SYM}(1,4,3,2,5,6)
\label{CtoAYM.2} \\ 
&\quad - s_{35} s_{56}  \big[ A^{\rm tree}_{\rm SYM}(1,2,4,3,5,6) + A^{\rm tree}_{\rm SYM}(1,4,2,3,5,6)  \big] \notag \\
C_{1| 23, 45, 6} &= s_{36} s_{46} A^{\rm tree}_{\rm SYM}(1,2,3,6,4,5) 
- s_{36} s_{56} A^{\rm tree}_{\rm SYM}(1,2,3,6,5,4) \notag \\
&\quad
- s_{26} s_{46} A^{\rm tree}_{\rm SYM}(1,3,2,6,4,5)  
+ s_{26} s_{56} A^{\rm tree}_{\rm SYM}(1,3,2,6,5,4)   \notag
\end{align}
The bosonic and fermionic components of the super-Yang-Mills tree amplitudes in (\ref{CtoAYM.1})
and (\ref{CtoAYM.2}) can be downloaded from \cite{PSSweb}, also see \cite{BGBCJ} for a description in terms
of Berends-Giele currents with a superspace origin in \cite{nptFT}. The gluon-polarization vectors
and gaugino wavefunctions in the component expressions can be straightforwardly dimensionally
reduced and yield kinematic factors of maximally supersymmetric Yang-Mills theories in $D< 10$ dimensions.

%%%%%%%%%%%%%%%%%%%%%%%%%%%%%%%%%%%%%%%%%%%%%%%% 
%%%%%%%%%%%%%%%%%%%%%%%%%%%%%%%%%%%%%%%%%%%%%%%% 

\subsection{Bosonic components of six-point kinematic factors}
\label{app:kin.2}

Generic contributions to chiral correlators ${\cal K}_n$ cannot be expressed
in terms of super-Yang-Mills tree amplitudes. This is obvious for kinematic factors
$C^m$ and $C^{mn}$ with free vector indices, but also the scalar quantity
$P_{1|2|3,4,5,6}$ in the last line of (\ref{KK6GEF}) features tensor structures
including parity-odd terms that do not arise in $ A^{\rm tree}_{\rm SYM}$.

Accordingly, we cannot describe the components of all the kinematic factors
in ${\cal K}_{n\leq 6}$ in terms of $ A^{\rm tree}_{\rm SYM}$. Instead, we shall here review
the Berends-Giele construction of the bosonic components for all kinematic factors 
beyond the reach of $C_{1|A,B,C}$.

In a first step, we combine the polarization vectors $e_i^m$ of the $i^{\rm th}$ external gluon
(subject to transversality $k_i \cdot e_i = 0$) to
Berends-Giele currents \cite{BerendsME} of rank two 
\begin{align}
\mathfrak{e}_{12}^m &= \frac{1}{s_{12}} \big[  2e_2^m (k_2\cdot e_1) 
-  2e_1^m (k_1\cdot e_2) + (k_1^m - k_2^m)(e_1\cdot e_2) \big]
 \label{CtoAYM.11} \\
 \mathfrak{f}_{12}^{mn} &= k_{12}^m \mathfrak{e}_{12}^n -  k_{12}^n \mathfrak{e}_{12}^m
- e_1^m e_2^n +  e_1^n e_2^m
 \notag
\end{align}
and rank three
\begin{align}
\mathfrak{e}_{123}^m &= \frac{1}{s_{123}} \big[ 
 \mathfrak{e}_3^m (k_3\cdot \mathfrak{e}_{12})
 + \mathfrak{e}_3^p \mathfrak{f}^{mp}_{12}
 -  \mathfrak{e}_{12}^m (k_3\cdot \mathfrak{e}_3)
 -  \mathfrak{e}_{12}^p \mathfrak{f}^{mp}_3   \label{CtoAYM.12} \\
 &\quad \quad +\mathfrak{e}_{23}^m (k_3\cdot \mathfrak{e}_{1})
 + \mathfrak{e}_{23}^p \mathfrak{f}^{mp}_{1}
 -  \mathfrak{e}_{1}^m (k_3\cdot \mathfrak{e}_{23})
 -  \mathfrak{e}_{1}^p \mathfrak{f}^{mp}_{23}
 \big]
\notag \\
 \mathfrak{f}_{123}^{mn} &= k_{123}^m \mathfrak{e}_{123}^n -  k_{123}^n \mathfrak{e}_{123}^m
- \mathfrak{e}_{12}^m \mathfrak{e}_3^n +  \mathfrak{e}_{12}^n \mathfrak{e}_3^m
- \mathfrak{e}_1^m \mathfrak{e}_{23}^n +  \mathfrak{e}_1^n \mathfrak{e}_{23}^m
 \notag
\end{align}
We have introduced the notation $ \mathfrak{e}_i^m = e_i^m$ and
$ \mathfrak{f}_i^{mn} = k_i^m e_i^n - k_i^n e_i^m$ for the sake of a unified representation.

In a second step, the above Berends-Giele currents are contracted with the famous $t_8$-tensor
and its generalizations with free vector indices and parity-odd terms \cite{He:2017spx}
\begin{align}
t_{A,B,C,D} &= {\rm tr}( \mathfrak{f}_A \mathfrak{f}_B \mathfrak{f}_C \mathfrak{f}_D ) 
- \frac{1}{4} {\rm tr}( \mathfrak{f}_A \mathfrak{f}_B  )  {\rm tr}(  \mathfrak{f}_C \mathfrak{f}_D ) 
+ {\rm cyc}(B,C,D)    \label{CtoAYM.13} \\
 t_{A | B,C,D,E}^m &= \big[ \mathfrak{e}_A^m t_{B,C,D,E} + (A\leftrightarrow
 B,C,D,E) \big] + \frac{i}{16}
 \varepsilon_{10}^{m}(\mathfrak{e}_A, \mathfrak{f}_B,\mathfrak{f}_C,\mathfrak{f}_D,\mathfrak{f}_E)  \notag \\
 t_{A | B,C,D,E,F}^{mn} &= \big[ \mathfrak{e}_A^{(m}  \mathfrak{e}_B^{n)} t_{C,D,E,F} + (A,B | A, B,C,D,E,F) \big]  \notag \\
 &\quad + \frac{i}{16} \big[ \mathfrak{e}_B^{(m}
 \varepsilon_{10}^{n)}(\mathfrak{e}_A, \mathfrak{f}_C,\mathfrak{f}_D,\mathfrak{f}_E,\mathfrak{f}_F)
 + (B\leftrightarrow C,D,E,F) \big]  \notag
\end{align}
where  the traces in the first line refer to Lorentz indices, e.g.\ $ {\rm tr}(
\mathfrak{f}_A \mathfrak{f}_B \mathfrak{f}_C \mathfrak{f}_D ) =
\mathfrak{f}_A^{mn} \mathfrak{f}_B^{np} \mathfrak{f}_C^{pq}
\mathfrak{f}_D^{qm}$, and $\varepsilon_{10}^{m}(\mathfrak{e}_A,
\mathfrak{f}_B,\ldots) = \varepsilon_{10}^{mnpq\ldots}\mathfrak{e}_A^n
\mathfrak{f}_B^{pq}\ldots $. We reiterate that the symmetrization conventions in
$\mathfrak{e}_A^{(m}  \mathfrak{e}_B^{n)}= \mathfrak{e}_A^{m}
\mathfrak{e}_B^{n}+ \mathfrak{e}_A^{n}  \mathfrak{e}_B^{m}$ and elsewhere do not
include a factor of $\frac{1}{2}$.

In a third step, the bosonic components of the kinematic factors in
chiral correlators are assembled from combinations of the $t$-tensors
in (\ref{CtoAYM.13}) of different rank \cite{Mafra:2014gsa,
  He:2017spx},
\begin{align}
C^m_{1| 2,3, 4,5} &= t^m_{1|2,3,4,5}+ \big[  k_2^m  t_{12,3,4,5}+ (2\leftrightarrow 3,4,5) \big]
 \label{CtoAYM.3} \\
C^m_{1| 23, 4,5, 6} &= t^m_{1|23,4,5,6} + t^m_{12|3,4,5,6} - t^m_{13|2,4,5,6}
+ k_3^m t_{123,4,5,6} - k_2^m t_{132,4,5,6} \notag \\
&\quad + \big[  k_4^m ( t_{14,23,5,6} - t_{214,3,5,6} +  t_{314,2,5,6}  ) +(4\leftrightarrow 5,6)\big]
\notag \\
C^{mn}_{1| 2,3, 4,5, 6} &= t^{mn}_{1| 2,3,4,5,6}
+  \big[  k_2^{(m}  t^{n)}_{12|3,4,5,6}+ (2\leftrightarrow 3,4,5,6) \big]
\notag \\
&\quad- \big[  k_2^{(m} k_3^{n)}  t_{213,4,5,6}+ (2,3|2,3,4,5,6) \big]
\notag \\
P_{1| 2|3, 4,5, 6} &= (e_1\cdot e_2) t_{3,4,5,6} 
-   \frac{i}{16}   \varepsilon_{10}(e_1,e_2,
\mathfrak{f}_3,\mathfrak{f}_4,\mathfrak{f}_5,\mathfrak{f}_6)
 + \frac{1}{2} \big[ (e_2\cdot e_3) t_{1,4,5,6}
+ (3\leftrightarrow 4,5,6) \big]\notag \\
&\quad  + k_2^m  t^m_{12|3,4,5,6}   + \big[ (k_2\cdot k_3) t_{123,4,5,6}
+ (3\leftrightarrow 4,5,6) \big]
\notag 
\end{align} 
The local terms in the first line of $P_{1| 2|3, 4,5, 6} $ are inferred from the components of
the closely related kinematic factor $P_{1| 2|3, 4} $ relevant to half-maximally supersymmetric
one-loop amplitudes~\cite{Berg:2016fui}.
The bosonic components of $C^m_{1| 2,3, 4,5},C^m_{1| 23, 4,5,6}  $ and all
the $C_{1|A,B,C}$ enjoy linearized gauge invariance under $e_i^m \rightarrow k_i^m$
for all of $i=1,2,\ldots,n$. The six-point kinematic factors $C^{mn}_{1| 2,3, 4,5, 6} $ and
$P_{1| 2|3, 4,5, 6}$ are only gauge invariant in legs $2,3,\ldots,6$ but exhibit
anomalous variations
\begin{align}
C^{mn}_{1| 2,3, 4,5, 6} \, \big|_{e_1^m \rightarrow k_1^m} &= \frac{i}{16} \eta^{mn} 
\varepsilon_{10}(\mathfrak{f}_2,\mathfrak{f}_3,\mathfrak{f}_4,\mathfrak{f}_5,\mathfrak{f}_6)
 \label{CtoAYM.7} \\
P_{1| 2 | 3, 4,5, 6} \, \big|_{e_1^m \rightarrow k_1^m} &= \frac{i}{16}
\varepsilon_{10}(\mathfrak{f}_2,\mathfrak{f}_3,\mathfrak{f}_4,\mathfrak{f}_5,\mathfrak{f}_6)
\notag
\end{align}
in the first leg. These gauge variations descend from the notion of BRST pseudo-invariance 
in pure-spinor superspace \cite{Mafra:2014gsa} and give rise to the hexagon gauge anomaly
of the six-point one-loop amplitude of ten-dimensional super-Yang-Mills \cite{Frampton:1983ah, Frampton:1983nr}, see in particular section 4.5 of \cite{towardsOne}.

%%%%%%%%%%%%%%%%%%%%%%%%%%%%%%%%%%%%%%%%%%%%%%%% 
%%%%%%%%%%%%%%%%%%%%%%%%%%%%%%%%%%%%%%%%%%%%%%%% 

\section{Field-theory limits of closed-string six-point homology invariants}
\label{app:Eex}

In this appendix, we display additional examples of the closed-string field-theory limits
(\ref{attempt1.0}) at $n=6$ points. However, we do not attempt to cover all
the permutation-inequivalent instances of ${\rm FT}^{\rm cl}_6[E_P \overline E_Q]$
in this appendix that are missing in sections \ref{sec:3.3.b} and \ref{sec:4.3}. The reader
is referred to the supplementary material of this work for a comprehensive 
list of ${\rm FT}^{\rm cl}_6[E_P \overline E_Q]$ relevant to ${\cal K}_6$ in (\ref{KK6GEF}).

%%%%%%%%%%%%%%%%%%%%%%%%%%%%%%%%%%%%%%%%%%%%%%%% 
%%%%%%%%%%%%%%%%%%%%%%%%%%%%%%%%%%%%%%%%%%%%%%%% 

\subsection{Cases without factors of $E_{1|2|3,4,5,6}$}
\label{app:Eex.1}

Section \ref{sec:3.3.b} gathers field-theory limits of six-point
closed-string homology invariants without the double poles of 
$E_{1|2|3,4,5,6}$. We shall here give additional 
permutation-inequivalent samples, namely two scalars
\begin{align}
	{\rm FT}^{\rm cl}_6\big[E_{1\vert 23, 4 5,6}\bar E_{1\vert 23, 4 6,5}\big]&=
	\frac{1}{16}\sum_{\rho \in S_{\{2,3,4,5,6\}}} I_{1,\rho(2,3,4,5,6)}^{(6)}\sign^\rho_{46}\sign^\rho_{45}
	\\
&\quad+\biggl[\bigg( \frac{1}{4 s_{12}} \sum_{\rho \in S_{\{3,4,5,6\}}}I_{12,\rho(3,4,5,6)}^{(5)}+(2\leftrightarrow 3,4)\bigg)\sign^\rho_{46}\sign^\rho_{45}\biggr]\nonumber\\
	&\quad+ \frac{1}{4s_{23}} \sum_{\rho \in S_{\{23,4,5,6\}}}I_{1,\rho(23,4,5,6)}^{(5)}\sign^\rho_{46}\sign^\rho_{45}\nonumber\\
	&\quad+ \frac{1}{ s_{14}s_{23}}  \sum_{\rho \in S_{\{23,5,6\}}}I_{14,\rho(23,5,6)}^{(4)}\nonumber\\	
	&\quad+\bigg[ \frac{1}{s_{124}}\left(\frac{1}{s_{12}}+\frac{1}{s_{14}}\right)  \sum_{\rho \in S_{\{3,5,6\}}}I^{(4)}_{124,\rho(3,5,6)}+(2\leftrightarrow 3)\bigg] \notag \\
	{\rm FT}^{\rm cl}_6\big[E_{1\vert 23 4, 5,6}\bar E_{1\vert 256,3, 4}\big]&= \frac{1}{144}  \!    \sum_{\rho \in S_{\{2,3,4,5,6\}}} \! \!  I_{1,\rho(2,3,4,5,6)}^{(6)}(1{+}3\sign^\rho_{23}\sign^\rho_{34})(1{+}3\sign^\rho_{25}\sign^\rho_{56})\nonumber\\
	&\quad+\frac{1}{4 s_{12}} \sum_{\rho \in S_{\{3,4,5,6\} }} I_{12,\rho(3,4,5,6)}^{(5)}\sign^\rho_{34}\sign^\rho_{56}
	\notag
\end{align}
as well as one vector and two-tensor each: 
\begin{align}
	{\rm FT}^{\rm cl}_6\big[E_{1\vert 23, 4 5,6}\bar E_{1\vert 26,3,4,5}^p\big]&=
- \frac{1}{8} \big(\ell^p+\tfrac{1}{2}  k_{23456}^p \big)
	\sum_{\rho \in S_{\{2,3,4,5,6\}}}  I_{1,\rho(2,3,4,5,6)}^{(6)}  \sign_{23}^\rho\sign_{45}^\rho \sign_{26}^\rho \\
	&\quad	 +\frac{1}{48} (k_6^p{-}k_2^p ) \sum_{\rho \in S_{\{2,3,4,5,6\}}} 
	 I_{1,\rho(2,3,4,5,6)}^{(6)}  \sign_{23}^\rho\sign_{45}^\rho 
	\nonumber\\
	&\quad+ \frac{1}{2s_{12}} \big(\ell^p+\tfrac{1}{2}k_{3456}^p \big)  \sum_{\rho \in S_{\{3,4,5,6\}}}I_{12,\rho(3,4,5,6)}^{(5)}\nonumber\\
	&\quad- \frac{1}{4s_{13}}k_{3}^p  \sum_{\rho \in S_{\{2,4,5,6\}}}I_{13,\rho(2,4,5,6)}^{(5)}\sign_{45}^\rho\sign_{26}^\rho\nonumber\\
	&\quad+\bigg[ \frac{1}{4s_{14}}k_{4}^p \sum_{\rho \in S_{\{2,3,5,6\}}} I_{14,\rho(2,3,5,6)}^{(5)}\sign_{23}^\rho\sign_{26}^\rho-(4\leftrightarrow 5)\bigg]\nonumber\\
	&\quad+\bigg[ \frac{1}{s_{124}} \bigg(\frac{1}{s_{12}}+\frac{1}{s_{14}} \bigg)k_{4}^p \sum_{\rho \in S_{\{3,5,6\}}} I_{124,\rho(3,5,6)}^{(4)}-(4\leftrightarrow 5)\bigg] \notag \\
	{\rm FT}^{\rm cl}_6\big[E_{1\vert 234, 5,6}\bar E_{1\vert 2,3,4,5,6}^{pq}\big]&=
	\frac{1}{12} \bigg(\ell^p \ell^q {+} \frac{1}{2}  k_{23456}^{(p}\ell_{\phantom{1}}^{q)}{+} \frac{1}{6} \sum_{j=2}^6 k_j^p k_j^q 
{+} \frac{1}{ 4}  \big[ k_2^{(p}k_3^{q)} {+} (2,3\vert 2,3,4,5,6)\big] \bigg)  \notag \\
&\quad \quad \quad \times
\sum_{\rho \in S_{\{2,3,4,5,6\}}}
	 I_{1,\rho(2,3,4,5,6)}^{(6)}(1{+}3\sign^\rho_{23}\sign^\rho_{34})\nonumber\\
	&\quad+\bigg[
	\frac{1}{2s_{12}} k_2^{(p} \big(\ell_{\phantom{1}}^{q)}  {+} \tfrac{1}{2}k_{3456}^{q)}\big)
	\! \! \sum_{\rho \in S_{\{3,4,5,6\}}} \! \!  \! I_{12,\rho(3,4,5,6)}^{(5)}\sign_{23}^\rho\sign_{34}^\rho +(2\leftrightarrow 4)\bigg]\nonumber\\
	&\quad+\bigg[ \frac{1}{s_{12}s_{123}}k_2^{(p}k_{3}^{q)}  \sum_{\rho \in S_{\{4,5,6\}}}I_{123,\rho(4,5,6)}^{(4)}+(2\leftrightarrow 4)\bigg]\nonumber\\
	&\quad-\bigg[ \frac{1}{s_{124}} \bigg(\frac{1}{s_{12}}+\frac{1}{s_{14}} \bigg)k_2^{(p}k_{4}^{q)}  \sum_{\rho \in S_{\{3,5,6\}}}I_{124,\rho(3,5,6)}^{(4)}\bigg]\notag
\end{align}

%%%%%%%%%%%%%%%%%%%%%%%%%%%%%%%%%%%%%%%%%%%%%%%% 
%%%%%%%%%%%%%%%%%%%%%%%%%%%%%%%%%%%%%%%%%%%%%%%% 

\subsection{Cases with factors of $E_{1|2|3,4,5,6}$}
\label{app:Eex.2}

%%%%%%%%%%%%%%%%%%%%%%%%%%%%%%%%%%%

This section extends the examples in section \ref{sec:4.3} by additional 
instances of ${\rm FT}^{\rm cl}_6[E_{1|2|3,4,5,6} \overline E_Q]$
\begin{align}\label{appB.1}
	{\rm FT}^{\rm cl}_6\big[E_{1\vert 2\vert 3, 4, 5,6}\bar E_{1\vert 34,56,2}\big]&=
	\frac{s_{12}}{96} \sum_{\rho \in S_{\{2,3,4,5,6\}}} I_{1,\rho(2,3,4,5,6)}^{(6)}\sign^\rho_{34}\sign^\rho_{56}\\
	&\quad+ \frac{1}{8} \sum_{\rho \in S_{\{3,4,5,6\}}} I_{12,\rho(3,4,5,6)}^{(5)}\sign^\rho_{34}\sign^\rho_{56}
	\nonumber\\
%%%
	{\rm FT}^{\rm cl}_6\big[E_{1\vert 2\vert 3, 4, 5,6}\bar E_{1\vert 345,2,6}\big]&=
	\frac{s_{12}}{288} \sum_{\rho \in S_{\{2,3,4,5,6\}}} I_{1,\rho(2,3,4,5,6)}^{(6)}(1+3\sign^\rho_{34}\sign^\rho_{45})\nonumber\\
	&\quad+\frac{1}{24}  \sum_{\rho \in S_{\{3,4,5,6\}}}I_{12,\rho(3,4,5,6)}^{(5)}(1+3\sign^\rho_{34}\sign^\rho_{45})
\notag \\
%%%
{\rm FT}^{\rm cl}_6\big[E_{1\vert 2\vert 3, 4, 5,6}\bar E_{1\vert 2,3,4,5,6}^{pq}\big]&=
	\frac{s_{12}}{24} \bigg(\ell^p \ell^q + \frac{1}{2}  k_{23456}^{(p}\ell_{\phantom{1}}^{q)}+ \frac{1}{6} \sum_{j=2}^6 k_j^p k_j^q 
	\notag \\
	&\quad \quad \quad \quad + \frac{1}{ 4}  \big[ k_2^{(p}k_3^{q)} {+} (2,3\vert 2,3,4,5,6)\big] \bigg) 	\sum_{\rho \in S_{\{2,3,4,5,6\}}}
	I_{1,\rho(2,3,4,5,6)}^{(6)}\nonumber\\
	&\quad+ 
	\bigg( \frac{1}{2}  \ell^p\ell^q + \frac{1}{4}  k^{(p}_{3456} \ell_{\phantom{1}}^{q)}
	+ \frac{1}{12} \sum_{j=2}^6   k^{p}_{j} k^{q}_{j}
	\nonumber\\
	&\quad\quad  \quad \quad +\frac{1}{8}\big[  k^{(p}_{3} k^{q)}_{4}+(3,4\vert3,4,5,6) \big]\bigg) 
\sum_{\rho \in S_{\{2,4,5,6\}}} I_{12,\rho(3,4,5,6)}^{(5)} \nonumber\\
	&\quad+\frac{1}{4 s_{12}} \sum_{\rho \in S_{\{3,4,5,6\}}}\bigg( k^{(p}_{2} \big(  \ell^{q)}_{\phantom{1}} + \tfrac{1}{2} k^{q)}_{3456}\big)(\ell-k_1)^2 I_{12,\rho(3,4,5,6)}^{(5)}\nonumber\\
	&\quad\quad\quad\quad
	\quad\quad\quad\quad \ - k^{(p}_{2} \big(   \ell^{q)}_{\phantom{1}}+\tfrac{1}{2}k^{q)}_{3456} + k_{2}^{q)} \big)\ell^2I_{2\underline1,\rho(3,4,5,6)}^{(5)}\bigg)\nonumber\\	
	&\quad-\bigg[
	\bigg(  \frac{1}{4 s_{12} } - \frac{s_{23}}{2 s_{12} s_{123}}\bigg)
	 k^{(p}_{2} k^{q)}_{3}
	\sum_{\rho \in S_{\{4,5,6\}}}I_{123,\rho(4,5,6)}^{(4)}+(3\leftrightarrow 4,5,6)\bigg]\nonumber
\end{align}

\bibliographystyle{JHEP}
\bibliography{Pinching_and_supergravity}{}

\end{document}